\documentclass{aa}  

\usepackage{graphicx}
\usepackage{txfonts}
\usepackage{setspace}
\usepackage{lscape}
\usepackage{caption}
\usepackage[english]{babel}
\usepackage{graphicx}
\usepackage{subfigure}
\usepackage{txfonts} 
\usepackage{textcomp}
\usepackage{amssymb}
\usepackage{graphicx}
\usepackage{tabularx}
\usepackage[english]{babel}
\usepackage[T1]{fontenc}
\usepackage{multirow}
\usepackage{natbib}
\bibpunct{(}{)}{;}{a}{}{,}

\usepackage{color}

\def\HI{\ion{H}{I}}

\newcommand{\kms}{$\,$km~s$^{-1}$}

\newcommand{\mJy}{mJy}
\newcommand{\whz}{W~Hz$^{-1}$}

\newcommand{\nhi}{N$_{\rm\HI}$}
\newcommand{\cmsq}{cm$^{-2}$}

\newcommand{\um }{$\mu$m}
\def\emph#1{{\sl #1}}
\newcommand{\eg}{\mbox{e.g.}}
\newcommand{\ie}{\mbox{i.e.}}

\newcommand{\atlas}{{ATLAS$^{\rm 3D}$}} 
\newcommand{\pks}{\mbox PKS B1718--649}

\newcommand{\flux}{{$S_{\rm 1.4\, GHz}$}}
\newcommand{\zint}{{$0.02<z<0.25$}}
\newcommand{\flint}{{\mbox{$30$ mJy $< S_{\rm 1.4\, GHz}< 50$ mJy}}}
\newcommand{\rp}{{\mbox{$\log\,P_{1.4 \,\rm GHz}>24$~W Hz$^{-1}$}}}
\newcommand{\rpl}{{\mbox{$\log\, P_{1.4 \, \rm GHz}<24$~W Hz$^{-1}$}}}
\newcommand{\pow}{\mbox{{$\log\,P_{1.4 \, \rm GHz}$}}}

\defcitealias{gereb2015}{G15}
\defcitealias{gereb2014}{G14}

\newcommand{\ps}{\citetalias{gereb2014}}
\newcommand{\pz}{\citetalias{gereb2015}}

\begin{document}

   \title{Kinematics and physical conditions of \HI\ in nearby radio sources}

   \subtitle{The last survey of the old Westerbork Synthesis Radio Telescope}

   \author{F. M. Maccagni\inst{1}$^,$\inst{2}
   \and R. Morganti\inst{1}$^,$\inst{2}
   \and T. A. Oosterloo\inst{1}$^,$\inst{2} 
   \and K. Ger\'eb \inst{3}
   \and N. Maddox\inst{2}}

   \institute{ Kapteyn Astronomical Institute, University of Groningen, Postbus 800, 9700 AV Groningen, The Netherlands
   \and ASTRON, Netherlands Institute for Radio Astronomy, Postbus 2, 7990 AA, Dwingeloo, The Netherlands
   \and Centre for Astrophysics and Supercomputing, Swinburne University of Technology, Hawthorn, Victoria 3122, Australia\\
              \email{maccagni@astro.rug.nl}             }


  \abstract {We present an analysis of the properties of neutral hydrogen (\HI) in $248$ nearby (\zint) radio galaxies with \flux$>30$ mJy and for which optical spectroscopy is available. The observations were carried out with the Westerbork Synthesis Radio Telescope as the last large project before the upgrade of the telescope with phased array feed receivers (Apertif). The sample covers almost four orders of magnitude in radio power from \pow$=22.5$~\whz and $26.2$~\whz. We detect \HI\ in absorption in $27\% \pm 5.5\%$ of the objects. The detections are found over the full range of radio power. However, the distribution and kinematics of the absorbing \HI\ gas appear to depend on radio power, the properties of the radio continuum emission, and the dust content of the sources. Among the sources where \HI\ is detected, gas with kinematics deviating from regular rotation is more likely found as the radio power increases. In the great majority of these cases, the \HI\ profile is asymmetric with a significant blue-shifted component. This is particularly common for sources with \rp, where the radio emission is small, possibly because these radio sources are young. The same is found for sources that are bright in the mid-infrared, i.e. sources rich in heated dust. In these sources,  the \HI\ is outflowing likely under the effect of the interaction with the radio emission. Conversely, in dust-poor galaxies, and in sources with extended radio emission, at all radio powers we only detect \HI\ distributed in a rotating disk. Stacking experiments show that in sources for which we do not detect \HI\ in absorption directly, the \HI\ has a column density that is lower than $3.5\times10^{17} \,(T_\mathrm{spin}/c_f)$~\cmsq. We use our results to predict the number and type of \HI\ absorption lines that will be detected by the upcoming surveys of the Square Kilometre Array precursors and pathfinders (Apertif, MeerKAT, and ASKAP).}
  

   \maketitle
%

\section{Introduction}
\label{sec:intro}

The interaction between the energy released by the active central supermassive black hole (SMBH) and the surrounding interstellar medium (ISM) plays an important role in galaxy evolution. This interaction affects the star formation history of the host galaxy and the observed relation between the masses of the bulge and central black hole (\eg~\citealt{silk1998,bower2005,ciotti2010,faucher2012,faucher2013,king2015a,king2015b}).

The actual effect of this interplay depends on the properties of the ISM and the type of active galactic nucleus (AGN) in the host galaxy. Among the diverse families of active AGN, radio AGN show radio jets expanding through the ISM. This facilitates a detailed study of the effects of the nuclear activity on the kinematics of the different components of the  gas, i.e. the ionised gas (e.g.~\citealt{tadhunter2000,holt2003,mcnamara2007,holt2008,labiano2008,reeves2009,fabian2012}), the molecular gas (\eg~\citealt{garciaburillo2005,feruglio2010,alatalo2011,dasyra2011,guillard2012,morganti2013b,morganti2015,mahony2015,maccagni2016,morganti2016}) and the atomic gas (e.g.~\citealt{morganti2005b,morganti2005a,morganti2013a,lehnert2011,allison2015}). In radio AGN it is possible to estimate the age of the activity directly from the peaked radio continuum spectrum and from the extent and expansion velocity of its radio jets. Hence, it is possible to study the effects of the nuclear activity on the ISM at different stages of its evolution. 

Observations of the neutral hydrogen (\HI) in early-type galaxies, which are the typical hosts of a radio AGN, allow us to study the cold gas in relation to their star formation and nuclear activity. A number of studies (\eg~\citealt{morganti2006,oosterloo2010a,emonts2010,serra2012,serra2014}) show that a significant fraction of early-type galaxies ($\sim40\%$) have \HI\ detected in emission, which can be found in settled rotating disks and in unsettled configurations. The \HI\ may represent the reservoir of cold gas for the formation of new stars in early-type galaxies, and, likely, also for the fuelling of the central SMBH. When the AGN are radio loud, \HI\ seen against the radio continuum has been used to study the morphology and kinematics of the cold gas in their very centre (e.g.~\citealt{vangorkom1989,morganti2001,vermeulen2003,morganti2005a,gupta2006}). \HI\ absorption lines can trace a wide variety of morphologies and phenomena. In some radio AGN, the detection of \HI\ in emission and absorption has allowed us to associate the \HI\ absorption lines detected at the systemic velocity of the galaxy with a circumnuclear rotating disk (\eg~\citealt{vanderhulst1983,conway1995,beswick2002,beswick2004,struve2010a,struve2010b}). Narrow ($\lesssim 50$\kms) \HI\ absorption lines offset with respect to the systemic velocity have been in some cases associated with the fuelling of the radio sources (\eg~\citealt{morganti2009,maccagni2014}). In a group of radio galaxies, the shallow optical depth ($\tau<0.05$) and the blue-shifted wings of the \HI\ absorption lines trace a fast outflow of neutral hydrogen drive by the expansion of the radio jets (\eg~\citealt{morganti2005b,kanekar2008,morganti2013a,mahony2013,gereb2015,allison2015}).

Absorption studies also have the advantage of allowing us to detect the gas using relatively short radio observations and reaching higher redshifts than emission studies, since the detection in absorption does not depend on the redshift of the source but only on the strength of the radio background continuum. Thus, \HI\ absorption studies allow us to study the impact of the radio nuclear activity on the ISM in a variety of radio sources (\eg~\citealt{vermeulen2003,pihlstrom2003,gupta2006,emonts2010,allison2012,chandola2013,allison2014,gereb2014,gereb2015,chandola2017,glowacki2017,curran2017}). Among several results, these studies show that in compact steep spectrum sources and gigahertz peaked sources, (\ie\ sources that possibly represent young radio AGN;~\citealt{fanti1995,readhead1996,odea1998}) the \HI\ is detected more often. Collecting a large sample also allows us to perform stacking experiments on the sources where the \HI\ is not detected, thus providing a complete characterisation of the content of \HI\ in radio AGN. Through this technique, \cite{gereb2014} show that in the direct non-detections, the \HI\, if present, it must have very low optical depths ($\tau \lesssim 0.002$) compared to the detections, suggesting a dichotomy in the properties of the neutral hydrogen in radio-loud early-type galaxies.

Until now, to obtain a good trade-off between sensitivity and observing time, \HI\ absorption surveys have been focussing on the high-power radio sources (\rp), but low-power sources (\rpl) in early-type galaxies form the bulk of the radio AGN population~\citep{bahcall1997,best2005,sadler2007}. Hence, to fully understand the importance of radio AGN in galaxy evolution scenarios it is crucial to investigate the interplay between the radio activity and ISM among sources of all radio powers. 

Here, we expand the work of \citealt{gereb2014,gereb2015} (hereinafter \ps\ and \pz) by extending the sample to low radio powers, \ie~\pow\ $=22.5$~\whz. 
The final sample combines the sample of \ps\ and \pz\ with the sample presented here and includes 248 sources. The survey presented here was constructed to observe all the objects in a uniform way and to reach, even for the weakest sources of the sample (\flux$\sim30$~\mJy), optical depths of a few percent. The survey also provides a statistically significant sample in preparation of future \HI\ absorption surveys,\ which are about to start with the SKA pathfinders and precursors.  

We carried out the observations of our sample with the Westerbork Synthesis Radio Telescope (WSRT) between December 2013 and February 2015. Over the same period, the dishes of the telescope were refurbished and the receivers upgraded to the phased array feed system, Apertif~\citep{oosterloo2010b}. Since the observations of this survey  were carried out up to the very last hours before the telescope closed for the final upgrade, this survey is the last one undertaken with the old Westerbork Synthesis Radio Telescope. 

This paper is structured as follows. In Sect.~\ref{sec:obs} we outline the selection of the sample and describe the $1.4$ GHz observations, while in Sect.~\ref{sec:sample} we discuss how we can classify our sources depending on their mid-infrared (MIR) colours and the extension of their radio continuum emission. In Sect.~\ref{sec:res} we report the results of the survey. In particular, we analyse the occurrence of \HI\ in the sample (Section~\ref{sec:full}), we determine the main properties of the detected absorption lines (Sect.~\ref{sec:det}), and we perform stacking experiments to search for low column density \HI, which is undetectable by single observations (Sect.~\ref{sec:stacking}). Section~\ref{sec:disc0} discusses the results of this survey focussing on the impact of the radio source on the \HI\ of the host galaxies. In Sect.~\ref{sec:surveys} we use 
the results from this work to predict how many and what type of HI absorbers the upcoming surveys 
with  the Square Kilometre Array pathfinders may produce.
Section~\ref{sec:conc} summarises our results and conclusions. In the Appendix, we show the new \HI\ absorption lines we detected in the survey (see Fig.~\ref{fig:Profiles1},~\ref{fig:Profiles2},~~\ref{fig:Profiles3}) and the main properties of all sources of the sample, such as redshift, radio continuum flux, and radio power (see Table~\ref{tab:long}). 

Throughout this paper we use a $\Lambda$CDM cosmology, with Hubble constant $H_0$ = 70\kms\ Mpc$^{-1}$ and $\Omega_\Lambda = 0.7$ and $\Omega_{\rm M} = 0.3$.

\section{Description of the sample}
\subsection{Sample selection and observations}
\label{sec:obs}

We expand the sample of radio sources presented in \ps\ and \pz\ to lower radio fluxes and radio powers. As in those studies, we selected the sources by cross-correlating the seventh data release of the Sloan Digital Sky Survey catalogue (SDSS DR7;~\citealt{york2000}) with the Faint Images of the Radio Sky at Twenty-cm catalogue (FIRST;~\citealt{Becker1995}). 
The sources lie above declination $\delta >10^\circ$ and between  $07^{\rm h}51^{\rm m}00^{\rm s}$ and $17^{\rm h}22^{\rm m}25^{\rm s}$ in right ascension. The sources are restricted to the redshift range \zint, which is the redshift interval covered by the WSRT observing band, $1150$ MHz -- $1400$ MHz.
The sample of \ps\ and \pz\ was limited to sources brighter than $50$ mJy and consists of $101$ sources. In the present study, we selected all sources that have radio core flux density \flint. This includes $219$ sources, of which $183$ were successfully observed before the telescope was switched off for the upgrade of the receivers.

In $37$ objects the observed band was affected by strong radio interference making the data unusable. One source of $353$~mJy (J082133.60+470237.3) was not included in SDSS DR7, but its spectrum became available with the DR9 data release. This source falls in the field of view of the observation of J082209.54+470552.9, and we include it in the final sample of $147$ sources.

The observations were carried out in the period December 2013 -- Feb. 2015 (proj. num. \mbox{R14A019} and \mbox{R14B006}).
We used a similar set-up of the telescope as in \ps\ and \pz.  We also did not use a full synthesis (12 hrs.) in order to observe as many objects as possible.
However, given that fewer WSRT dishes were available at the time of our observations and to maintain the same sensitivity as in \ps\ and \pz, in all observations we increased the integration time to six hours per source. The observational set-up consists of $1024$ channels covering a bandwidth of $20$ MHz.

We reduced the data following a similar procedure to that presented in \ps\ and \pz\ via the $\tt{MIRIAD}$ package~\citep{sault1995}. The final \HI\ data cubes have a velocity resolution of $16$\kms. 
The median noise of the final spectra is $0.81$~\mJy~beam$^{-1}$ and $\sim90\%$ of the observations have a noise level lower than $1.3$~\mJy~beam$^{-1}$. We created continuum images using the line-free channels to measure the continuum flux density of the sources. However, because of the limited uv coverage of the observations, the beam of the observations is very elongated, typically about $45\arcsec \times 12\arcsec$. In Table~\ref{tab:long} we report a full summary of the main radio properties of the sources.

For the analysis that follows, we combine the newly observed $147$ low-power sources (\flint) with the $101$ sources with \flux$>50$~\mJy\ presented in \ps\ and \pz, obtaining a sample of $248$ sources.  


\begin{figure}[tbh]
\begin{center}
\includegraphics[trim = 0 0 0 0, clip,width=.49\textwidth]{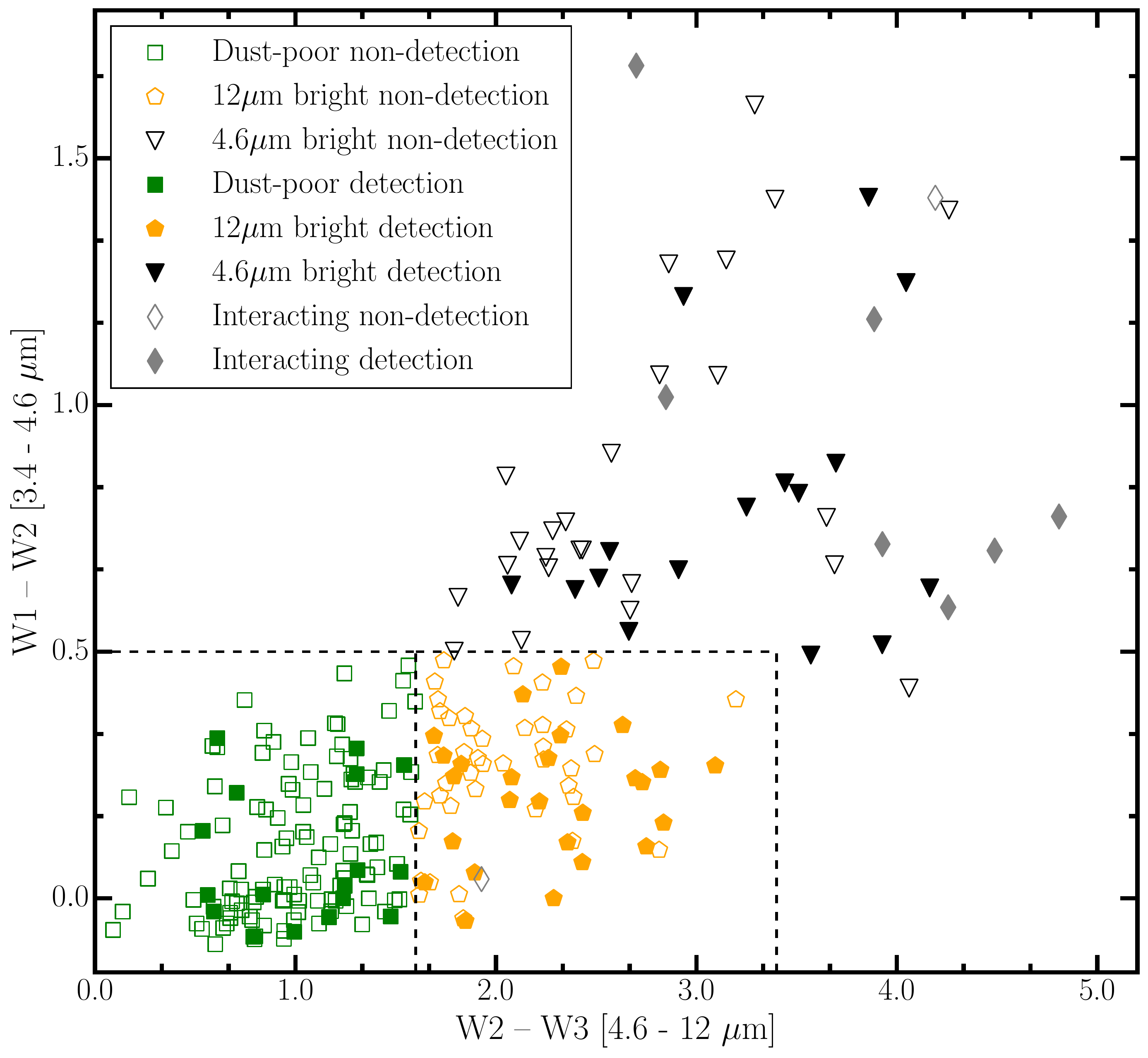}
\caption{WISE colour-colour plot for the sources of the sample, separated into \HI\ detections (filled symbols) and \HI\ non-detections (empty symbols) Dust-poor galaxies are green squares, $12$~\um\ bright sources are orange pentagons and $4.6$~\um\ bright sources are black triangles. Interacting sources are grey diamonds. The dashed lines indicate the cut-offs of the WISE colours we used to classify our sources. Further details on the classification of the sources are shown in Table~\ref{tab:stats}.}
\label{fig:wise}
\end{center}
\end{figure}

\subsection{Characterisation of the AGN sample}
\label{sec:sample}
One goal of this survey is to determine the occurrence and properties of \HI\ for the different types of radio sources present in our sample. The sources of our sample lie in the massive end of the red sequence ($-24<{M_{\rm r}}<-21$, \emph{u -- r} $\sim2.7\pm0.6$). Nine sources are an exception ({\em u--r} $\lesssim 2$). Their SDSS images show that they are undergoing a merger with a spiral galaxy or that they have large tails of gas and stars, suggesting a recent or ongoing interaction with a companion. These sources are different from the bulk of our sample of early-type radio galaxies, hence, from here on, we classify them as interacting and we denote them with grey diamonds in all figures.

The mid-infrared (MIR) colours allow us to classify the galaxies according to their dust content. For such a purpose, different studies (e.g.~\citealt{wright2010,stern2012,jarrett2013,mingo2016}) have used data from the all-sky Wide-field Infrared Survey Explorer (WISE), which observed in four MIR bands, i.e. $3.4\mu$m (W1), $4.6$~\um\ (W2), $12$~\um\ (W3), and $22\mu$m (W4). The W3 band is sensitive to the dust continuum and the presence of polycyclic-aromatic hydrocarbons (PAHs), whose emission lines peak at $11.3$~\um\ and may trace the star formation activity in a galaxy~\citep{lee2013,cluver2014}. The luminosity at $12$~\um\ of dust-poor red-sequence galaxies is dominated by the old stellar population and is similar to the luminosity at $4.6$~\um, whereas galaxies rich in PAHs and dust have enhanced luminosity at $12$~\um, which increases their W2--W3 colour. Starburst galaxies typically have W2--W3$>3.4$ (\eg~\citealt{rosario2013}). The W1--W2 colour is sensitive to heated dust.  When an AGN is present, as in the sources of our sample, galaxies bright at  $4.6$~\um\ are likely to have a dust-rich circumnuclear region that is heated by the nuclear activity.

In this study, we cross-matched the sky coordinates of each source with the WISE catalogue to extract the WISE magnitudes, making use of the VizieR catalogue access tool~\citep{vizier2000}. We followed \cite{mingo2016} to distinguish between so-called dust-poor sources (\mbox{W1--W2 $<0.5$} and \mbox{W2--W3 $<1.6$}) and galaxies with MIR emission enhanced by the dust continuum and PAHs, which we call $12$~\um\ bright sources \mbox{(W1--W2 $<0.5$} and \mbox{$1.6<$ W2--W3$<3.4$}); AGNs with hot dust in the circumnuclear regions (\mbox{W1--W2 $>0.5$}); and dust-rich starburst galaxies (\mbox{W1--W2 $<0.5$} and \mbox{W2--W3 $>3.4$}). Given that there are only three starburst galaxies in our sample, with W1--W2 $\sim0.5$, we include them in the sample of AGN rich of heated dust, which we name $4.6$~\um\ bright sources because of their enhanced flux at this wavelength. In the following sections, we sometimes use the generic name MIR bright sources to refer to $12$~\um\ bright and to $4.6$~\um\ bright sources altogether. 

\begin{table*}
\caption{Statistics of the sample}             
\label{tab:stats}      
\centering                          
\begin{tabular}{l c c c c}        
\hline\hline                 
  & Number of sources & Non-detections & Detections & Detection rate ($\%$)   \\
\hline                        

All sources  &   248    &       182     &       66       & $27\pm5.5$ \\

\\
\hline
Radio continuum classification & & & &  \\
Compact sources         &               131 &           89 &    42      &               $32\pm 7.9$\\
Extended sources        &               108 &           91       &      17 &               $16\pm 6.8$\\
\hline
WISE colour classification & & & &   \\
Dust-poor       sources &       129     &       112     &       17      & $13\pm 5.8$ \\
$12$~\um\ bright sources &       68              &       42              &       26      & $38\pm 11$\\
$4.6$~\um\ bright sources&       42              &       26              &       16      & $38\pm 16$\\
\hline
Interacting sources      &              9 &             2 &     7       &$78\pm27$ \\
\hline                           
\end{tabular}
\tablefoot{Number of observed sources (1), number of \HI\ absorption non-detections (2), number of \HI\ detections (3), and detection rates (4) for all sources and different subsamples of compact and extended sources, based on the radio-continuum classification (see Fig.~\ref{fig:ce}), dust-poor, $12$~\um\ bright and $4.6$ \um\ bright sources, according to the WISE colour-colour plot (Fig.~\ref{fig:wise}), and interacting sources.}
\end{table*}

As shown in Fig.~\ref{fig:wise} and Table~\ref{tab:stats}, half of the sources are classified as dust-poor ($52\%$, indicated by green squares in this and the following figures) while $27\%$ are $12$~\um\ bright sources (indicated by orange pentagons). Forty-two sources ($17\%$) are classified as $4.6$~\um\ bright galaxies (indicated by black triangles). 

\begin{figure}[tbh]
\begin{center}
\includegraphics[trim = 0 0 0 0, clip,width=.49\textwidth]{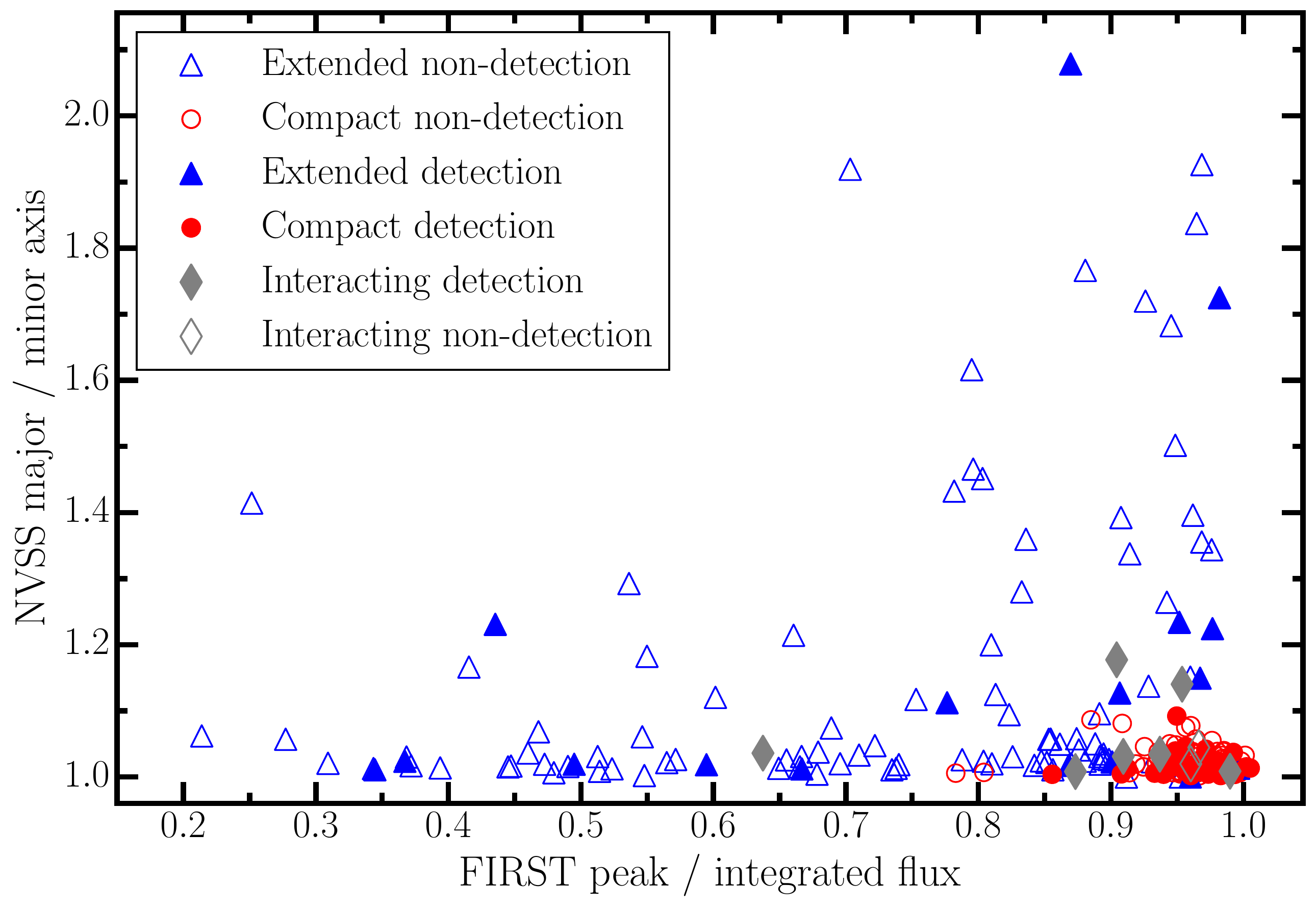}
\caption{Radio morphological classification of the sample. Red circles indicate compact sources, extended sources are indicated by blue triangles, interacting sources are shown in grey diamonds. \HI\ detections are indicated by filled symbols, while empty symbols represent non-detections.}
\label{fig:ce}
\end{center}
\end{figure}

Following \ps, we classified the radio continuum emission of the sample depending on the NVSS major-to-minor axis ratio versus the FIRST peak-to-integrated flux ratio. Figure~\ref{fig:ce} and Table~\ref{tab:stats} show the classification for the sample of $248$ sources, where we designate $52\%$ as compact sources (in red in the figure) and $48\%$ as extended sources (in blue). Compact sources typically have the radio continuum embedded in the host galaxy at sub-galactic scales ($\lesssim$ few kilo-parsec), while extended sources have radio continuum at super-galactic scales. For a group of radio AGN, the extent of their radio continuum emission can be related to the age of the nuclear activity. Compact steep spectrum sources (CSS) and gigahertz peaked sources (GPS) are the youngest family of radio AGN~\citep{odea1998,murgia2003,fanti2009,sadler2016,orienti2016}, and in general, they can be considered younger radio AGN than extended sources. A fraction of the compact sources of our sample belong to this group of young radio AGN (see \ps\ and \pz). 

\begin{figure*}[tbh]
\begin{center}
\includegraphics[trim = 0 0 0 0, clip,width=.495\textwidth]{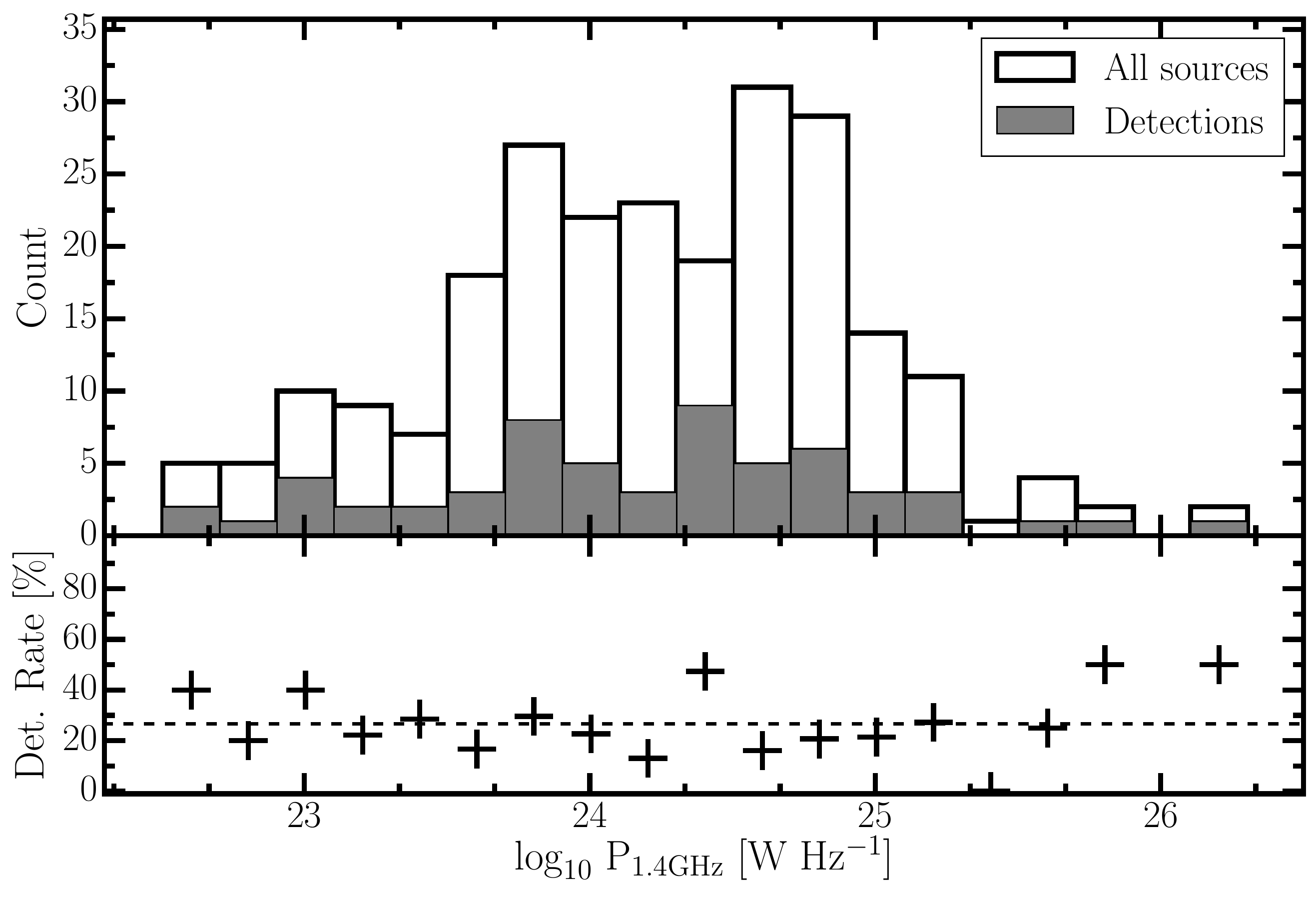}
\includegraphics[trim = 0 0 0 0, clip,width=.48\textwidth]{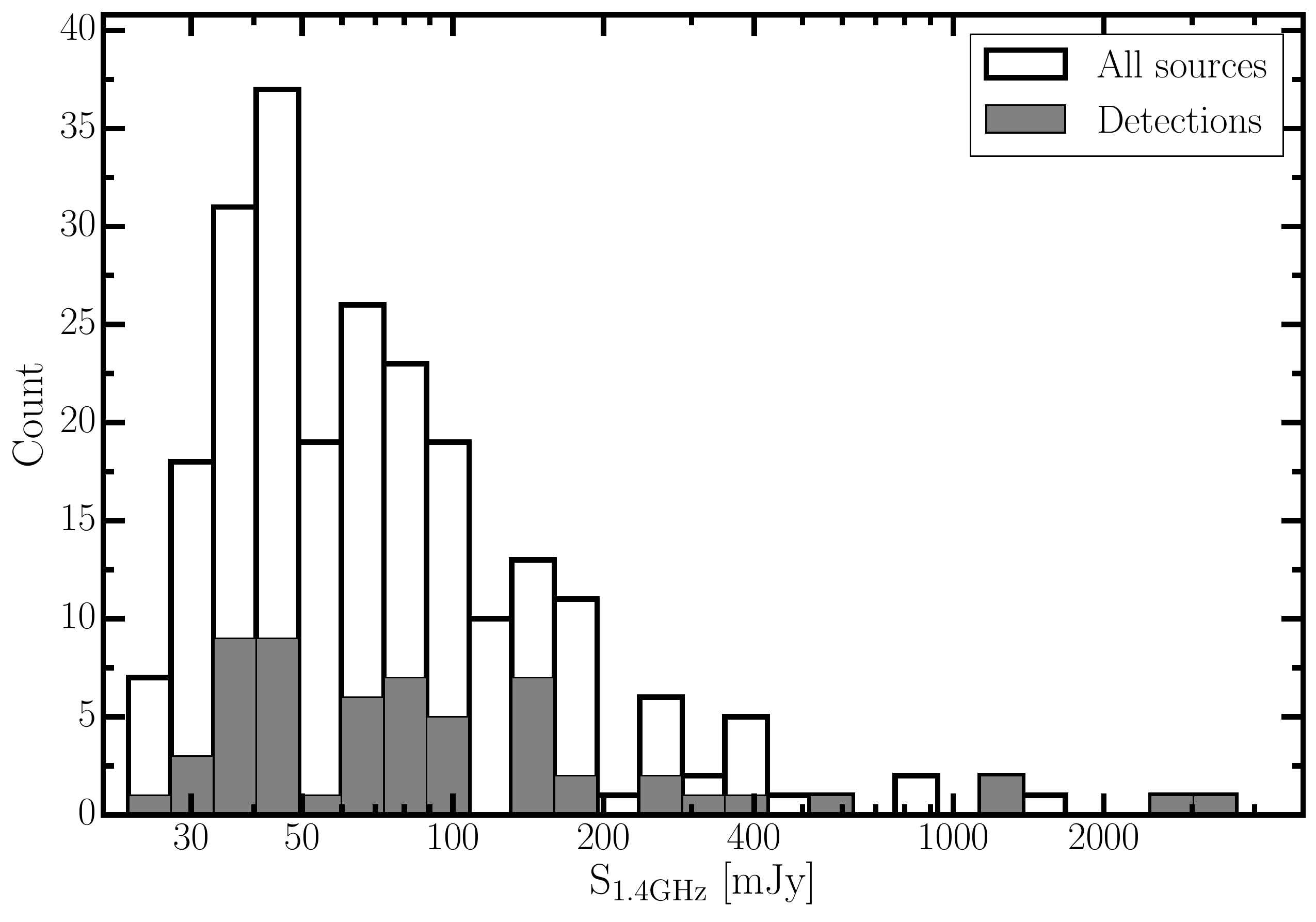}
\includegraphics[trim = 0 0 0 0, clip,width=.495\textwidth]{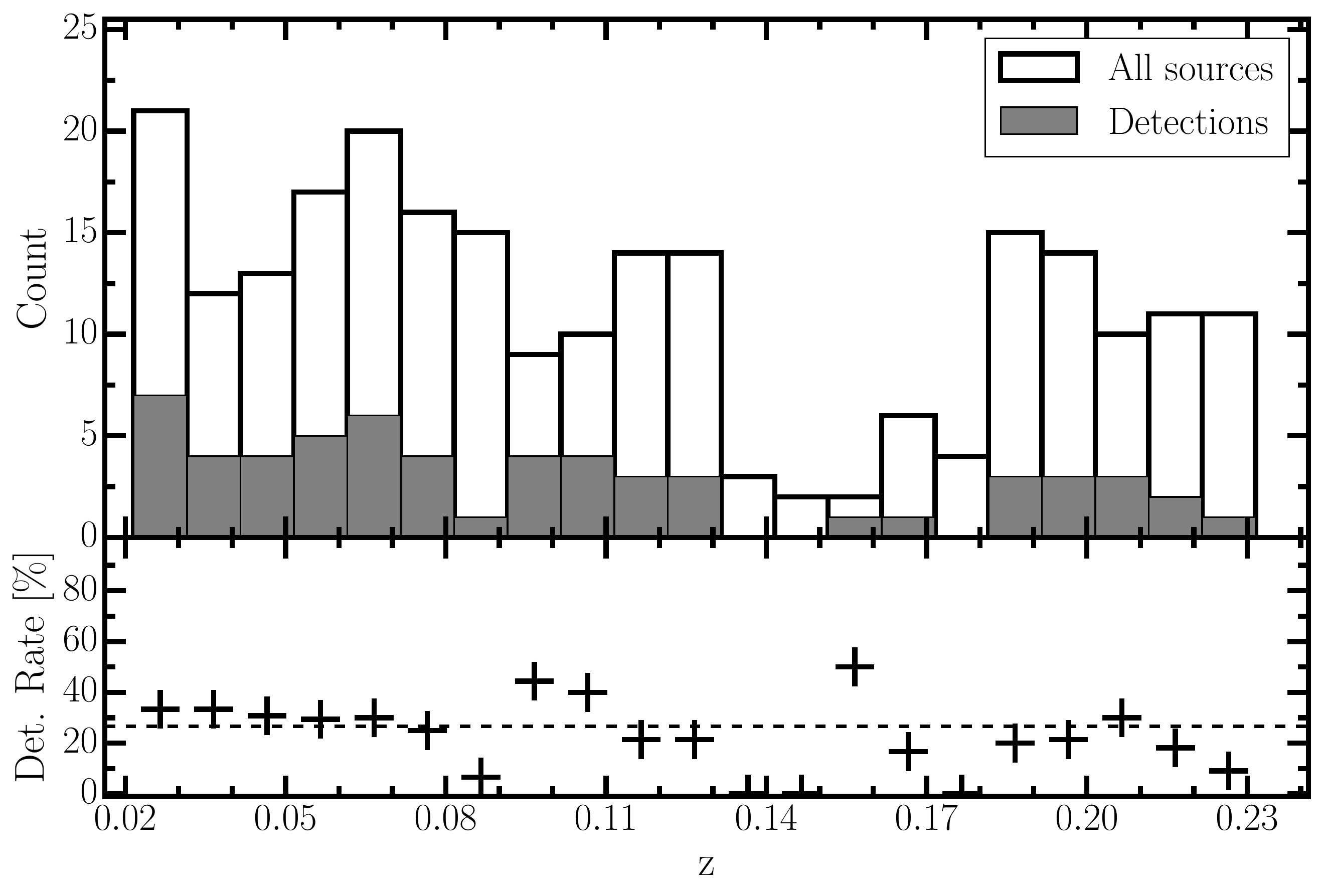}
\includegraphics[trim = 0 0 0 0, clip,width=.48\textwidth]{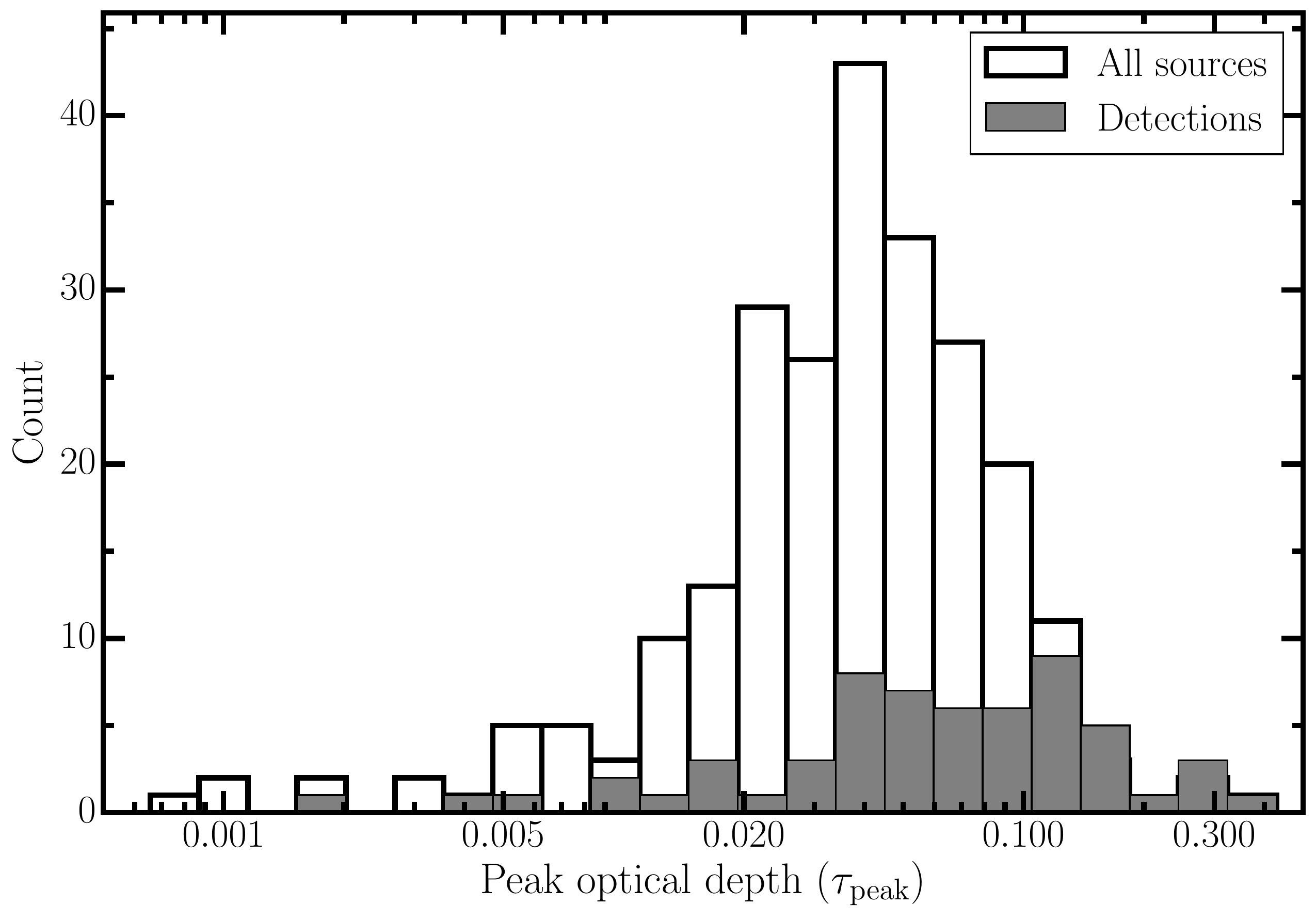}
\caption{(Top left panel) Radio power distribution of the full sample and the detections. The sub-panel below shows the detection rate for each bin of the histogram and the average detection rate (dashed line). (Top right panel)  Radio continuum flux (\flux) distribution of the full sample (open bars) and of the detections (grey bars). (Bottom left panel) Redshift ($z$) distribution of the full sample of $248$ radio sources and of the \HI\ detections. The sub-panel shows the detection rate for each bin of the histogram. (Bottom right panel) Distribution of the optical depth detection limits of the full sample overlaid with the distribution of the optical depth of the peak of the detected absorption lines.}
\label{fig:zf_distribution}
\end{center}
\end{figure*}

\begin{figure*}[tbh]
\begin{center}
\includegraphics[trim = 0 0 0 0, clip,width=.48\textwidth]{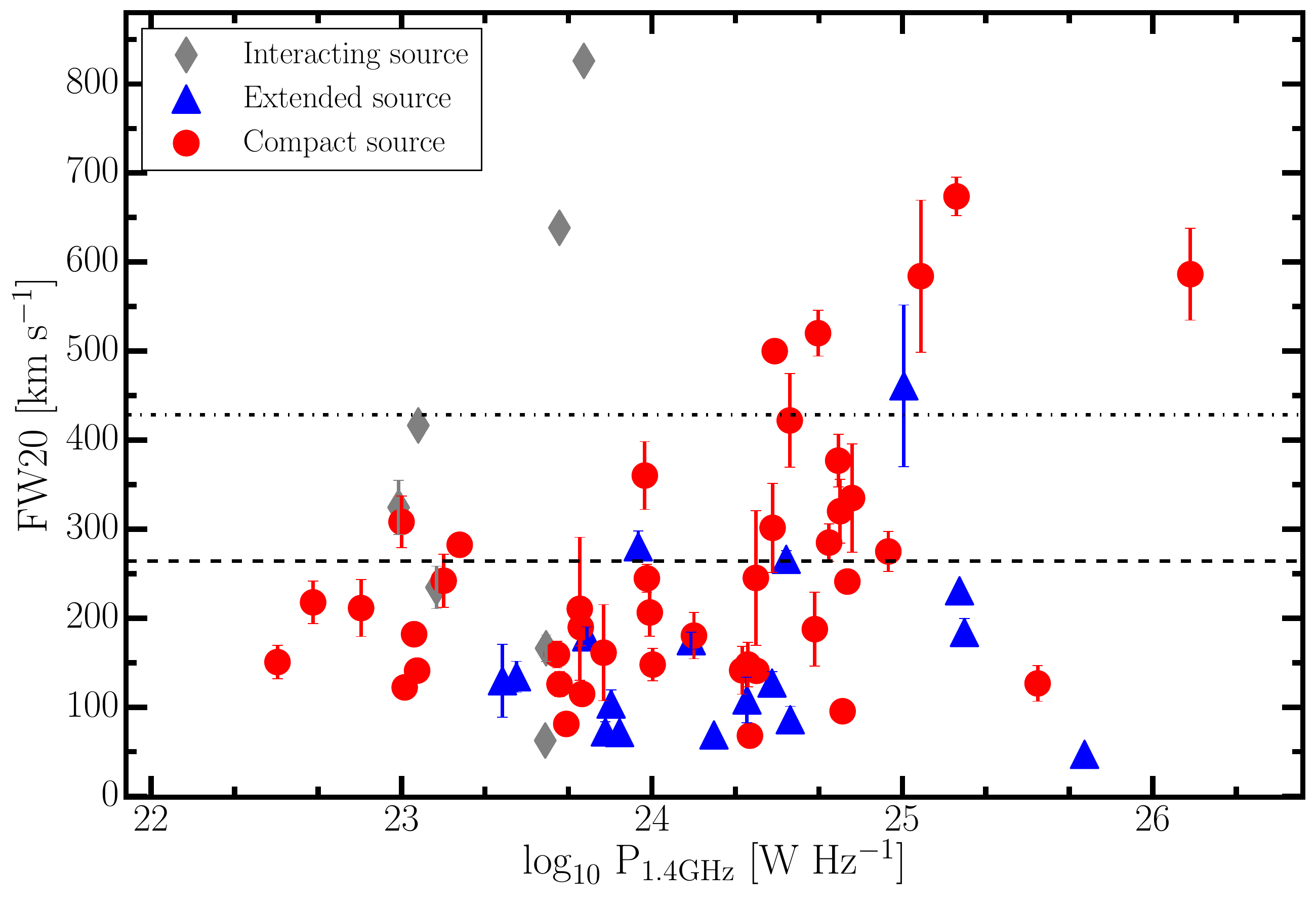}
\includegraphics[trim = 0 0 0 0, clip,width=.48\textwidth]{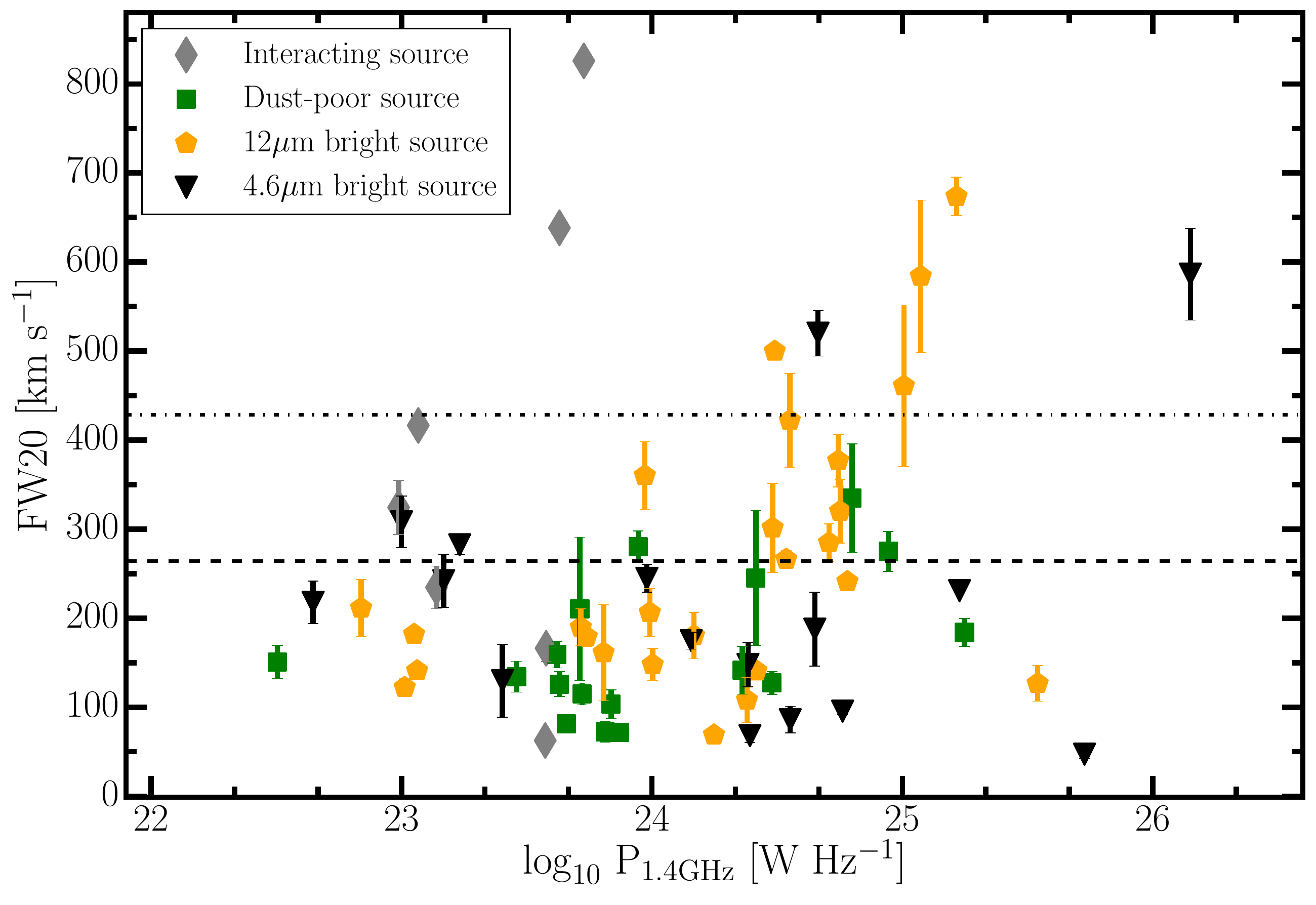}
\caption{(Left panel) Full width at 20$\%$ of the intensity (FW20) of the \HI\ profiles vs. the radio power of the sources.Sources are classified according to the extension of their radio continuum (see Fig.~\ref{fig:ce}). The dashed line indicates the mean of the distribution of rotational velocities of the sources of the sample. The fine dashed line shows the $3$$\sigma$ upper limit of the distribution (see Sect.~\ref{sec:det} for further details). (Right panel) Same as in the left panel with symbols following the WISE colour-colour plot shown in Fig.~\ref{fig:wise}.}
\label{fig:rp_linvar}
\end{center}
\end{figure*}

\section{Results}
\label{sec:res}

From the new observations presented in this paper (objects with radio flux between \flint), we detect \HI\ absorption in 34 galaxies with $3$$\sigma$ significance above the noise level.  Since the observations are relatively shallow, for the weaker radio sources (\flux$\sim30$~\mJy) we are sensitive to absorption lines with peak optical depth of $\tau_{\rm peak}\sim0.08$. 

Following \pz, we use the busy function (BF;~\citealt{westmeier2014}) to measure the main properties of the lines in a uniform way, i.e. centroid, peak optical depth ($\tau_{\rm peak}$), the integrated optical depth ($\int \tau dv$) full width at half-maximum (FWHM), and full width at $20\%$ of the peak flux (FW20). The detected lines and fits produced by the BF are shown in Figs.~\ref{fig:Profiles1},~\ref{fig:Profiles2}, and~\ref{fig:Profiles3}. In Table~\ref{tab:long} we summarise these parameters  for each detection. 

In the following, we discuss the results making use of the full sample by combining the new data and the \ps\ and \pz\ data.

\subsection{Occurrence of \HI\ in radio sources}
\label{sec:full}
Considering the full sample, we detected  \HI\  absorption in $66$ sources out of $248$, leading to a detection rate of $27\%\pm5.5\%$\footnote{We compute the errors on the detection rates as the $95\%$ confidence level of a binomial proportion.}. Compared to \ps\ and \pz, we have now extended the range of the sample to low radio powers (down to \pow$=22.5$~\whz), and it is worth mentioning that, albeit with small number statistics in some of the bins, the detection rate is similar across the range of radio powers covered by the survey, as illustrated in the top left panel of Figure~\ref{fig:zf_distribution}. The top right panel of the figure shows that we detect \HI\ absorption lines, with a peak that is three times above the noise level, in sources throughout the entire range of fluxes. The bottom left panel shows that we detect \HI\ in all redshift intervals. The absorption lines have a broad range in peak optical depths, i.e. approximately between $0.3$ and $0.003$ (bottom right panel), and we detect lines across the full range of optical depths to which we are sensitive. These results confirm and extend what already observed in the subsample of higher flux sources (\flux$>50$~\mJy) of \ps\ and \pz. 

In Table~\ref{tab:stats}, we show the detection rate of \HI\ absorption depending on the classification of their radio continuum emission and of their WISE colours. We detect \HI\ in all different types of galaxies but with different detection rates. In the sources that we classify as compact, \HI\ is detected twice as often as in the extended sources ($32\%\pm7.9\%$ and $16\%\pm6.8\%$, respectively). This behaviour was also observed in \ps\ and in previous works on smaller and less homogeneously selected samples (e.g.~\citealt{emonts2010,curran2011,curran2013,curran2013b}).

\begin{figure*}[tbh]
\begin{center}
\includegraphics[trim = 0 0 0 0, clip,width=.48\textwidth]{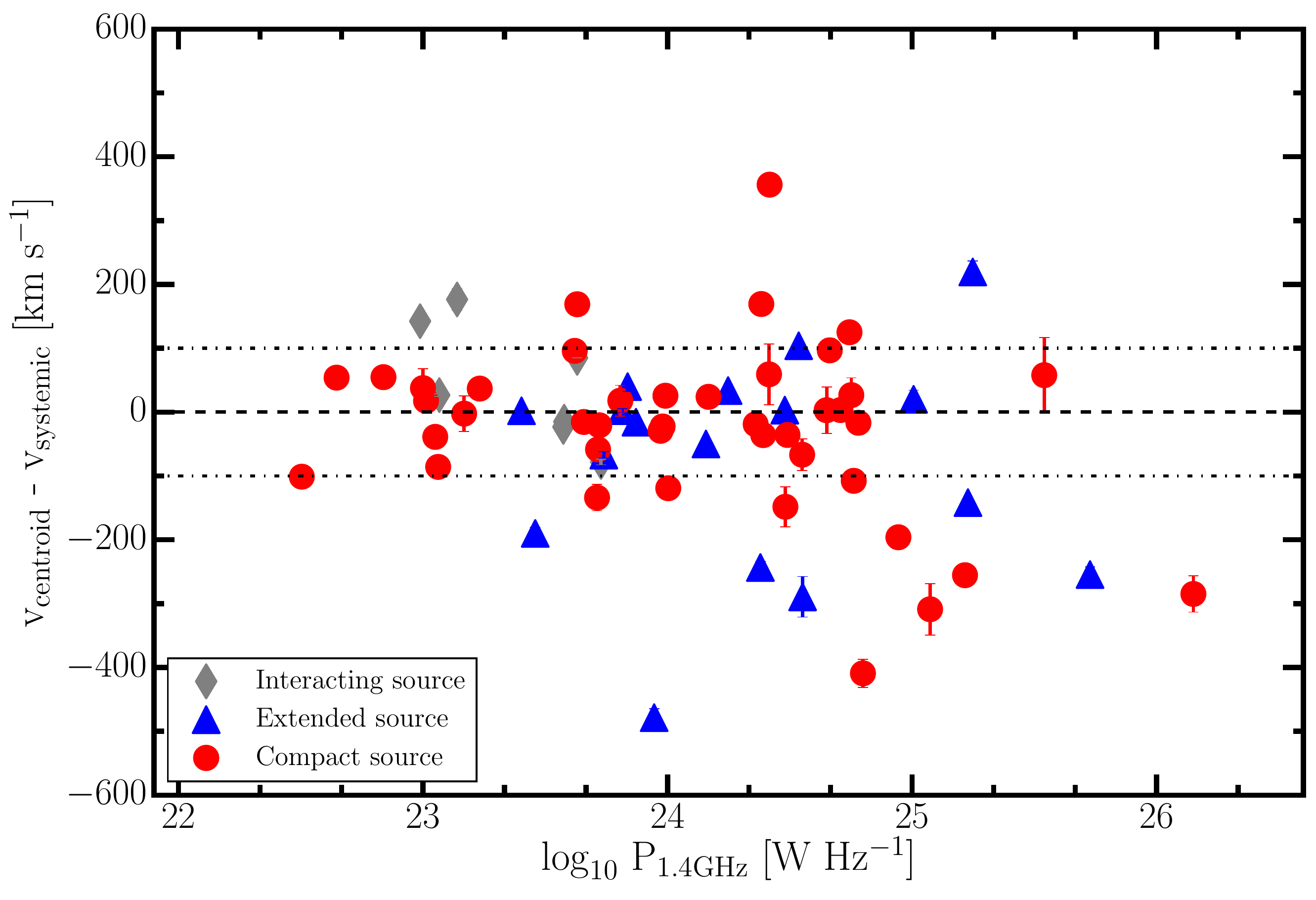}
\includegraphics[trim = 0 0 0 0, clip,width=.48\textwidth]{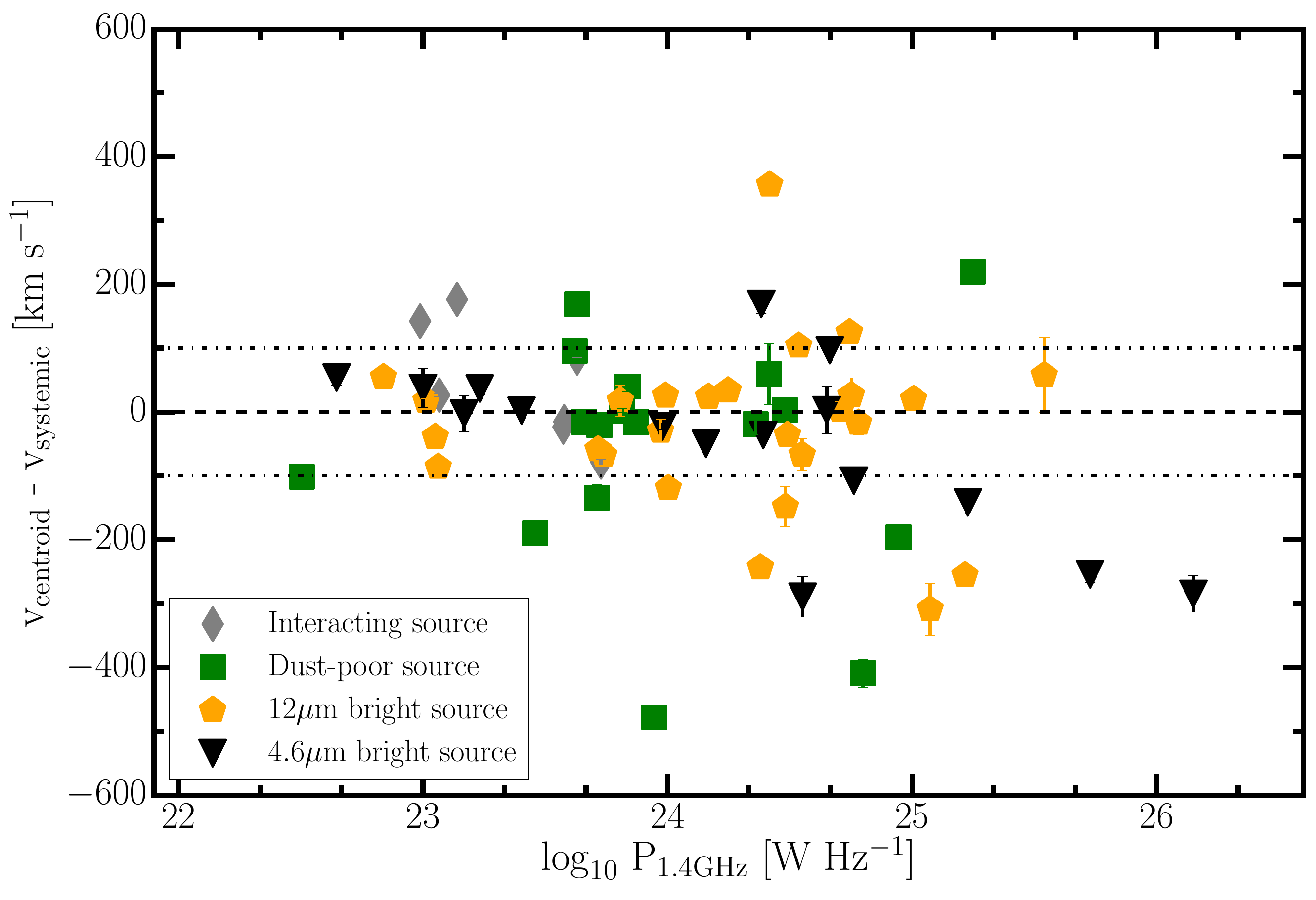}
\caption{(Left panel) Blue shift and red shift of the \HI\ line centroid with respect to the systemic velocity vs. the radio power of the sources. Sources are classified according to the extent of their radio continuum (see Fig.~\ref{fig:ce}). The fine dashed lines indicate the $\pm 100$\kms\ velocities. (Right panel)   Same as in the left panel with symbols following the WISE  colour-colour plot shown in Fig.~\ref{fig:wise}.}
\label{fig:rp_cen}
\end{center}
\end{figure*}

For dust-poor galaxies the detection rate is low ($13\%\pm5.8\%$) compared to $12$~\um\ bright sources ($38\%\pm11\%$) and $4.6$~\um\ bright galaxies ($38\%\pm16\%$). 

\subsection{Kinematics of \HI}
\label{sec:det}

Figure~\ref{fig:rp_linvar} shows the FW20 of the \HI\ absorption lines versus the radio power of the sources. The dashed horizontal line indicates the mean of the distribution of rotational velocities of the sample.\footnote{We determine the rotational velocity of the sources from their K magnitude using the Tully-Fisher relation for red-sequence galaxies~\citep{heijer2015} and correcting for the average inclination of the sources, measured from the axis ratio of the stellar body of the host galaxies.} The fine dashed horizontal line indicates the $3\sigma$ value of the distribution of the rotational velocities. In this study, we consider an \HI\ line broad when it has FW20 above this latter line. When a line is broad, the \HI\ cannot simply rotate within the galaxy, but it must also have a component with non-ordered kinematics. At high radio powers (\rp), $30\%\pm15\%$ of the lines are broad.

Figure~\ref{fig:rp_cen} shows the shift of the centroid of the line with respect to the systemic velocity of the galaxy versus the radio power of the sources. At low radio powers (\rpl), the majority of the lines are centred at the systemic velocity ($\Delta v \pm100$\kms). At \rp, $36\%\pm16\%$ of the lines are offset with respect to the systemic velocity.

As shown in the left panel of Fig.~\ref{fig:rp_linvar}, broad lines are found only in compact radio sources. One exception is \mbox{3C 305}, which is classified as extended according to our classification, but is known to be a compact steep spectrum source of 4 kpc~in size \citep{jackson2003}. The broad \HI\ is known to trace a fast outflow \citep{morganti2005b}, which is consistent with what found in this paper.

The lines that are broad in Fig.~\ref{fig:rp_linvar} also have a shifted centroid. This because, as shown in \pz, these lines have, apart from a main component, a second, shallower component that extends to blue-shifted velocities, which we call wing. Interestingly, the majority of these wings are blue-shifted indicating that there is at least a component of the gas outflowing. 

At low radio powers (\rpl), we only detect relatively narrow lines with the exception of the interacting galaxies. At the sensitivity of our observations, we would have not been able to detect broad and shallow wings with ratio between the wing and peak of the absorption line $< 0.3$, as they would have been hidden in the noise. We know that in some low-power radio sources, for example~\mbox{NGC 1266}~\citep{alatalo2011} and \mbox{IC 5063}~\citep{oosterloo2000,morganti2013b,morganti2015},  \HI\ outflows can be present and are traced by a broad and shallow ($\tau< 0.005$) blue-shifted wing in the absorption line. If these shallow and broad wings were present in the HI lines of the low-power sources, we would not have detected them. Despite this limitation, it is interesting to see that the great majority of the lines in low power sources are centred on the systemic velocity (Fig.~\ref{fig:rp_cen}). This means that the dominant component traced by the \HI\ absorption is associated with settled gas. 
 
In Fig.~\ref{fig:rp_linvar} (right panel), we show FW20 versus the radio power of the sources, classifying sources according to their WISE colours (see Fig.~\ref{fig:wise}). Dust-poor sources do not show broad lines. Mid-infrared bright sources (both at $12$~\um\ bright and at $4.6$~\um) have broad lines if the radio power is \rp.

\begin{figure}[tbh]
\begin{center}
\includegraphics[trim = 0 0 0 0, clip,width=.48\textwidth]{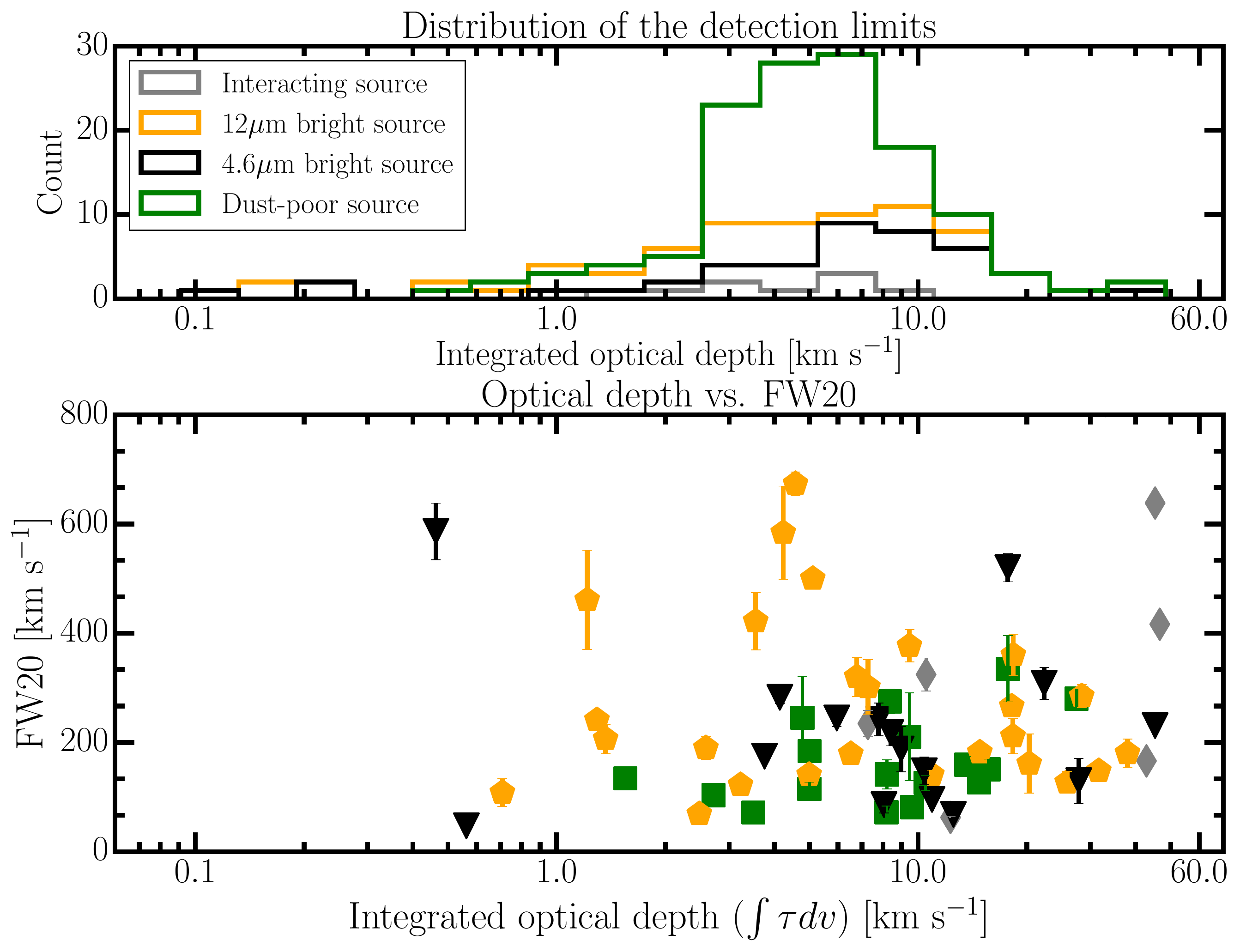}
\caption{ Full-width at $20\%$ of the intensity (FW20) vs. the integrated optical depth of the \HI\ lines ($\int \tau dv$). Symbols follow the WISE colour-colour classification shown in Fig.~\ref{fig:wise}. The histogram in the top panel shows the distribution of the $3\sigma$ detection limit in optical depth for the non-detections of the each subsample.}\label{fig:tau_width}
\end{center}
\end{figure}

Figure~\ref{fig:tau_width} shows the FW20 versus the integrated optical depth of the line, classifying the sources according to their WISE colours. In dust-poor sources, we only detect narrow \HI\ lines (FW20 $< 300$\kms) with integrated optical depth $\gtrsim 1.5$\kms. In $12$~\um\ bright and $4.6$~\um\ bright galaxies we detect both shallow and broad lines. The histogram at the top of the figure shows the estimated upper limit to the integrated optical depth of the non-detections of each subgroup of galaxies. For each source, this is equal to three times the noise of the spectrum in optical depth times the mean width of the detected lines ($145$~\kms). For most dust-poor galaxies the detection limit is $\int \tau dv >1.5$\kms. For ten dust-poor sources the detection limit is lower and should allow us to detect the shallow broad lines that we detected in the other sources, but we detect none in these ten sources. This suggests that in dust-poor sources the \HI\ traced by the broad and shallow wings of the lines is absent.

\subsection{Stacking in search for \HI\ absorption}
\label{sec:stacking}

Sources where we do not directly detect an absorption line may still have \HI\ that  can be uncovered with stacking techniques. In a stacking experiment, the noise of the final stacked spectrum decreases with the square root of the number of stacked spectra. Stacking the non-detections of our sample allows us to explore the statistical presence of \HI\ absorption at low optical depths. 

\begin{figure*}[tbh]
\begin{center}
\includegraphics[trim = 0 0 0 0, clip,width=.33\textwidth]{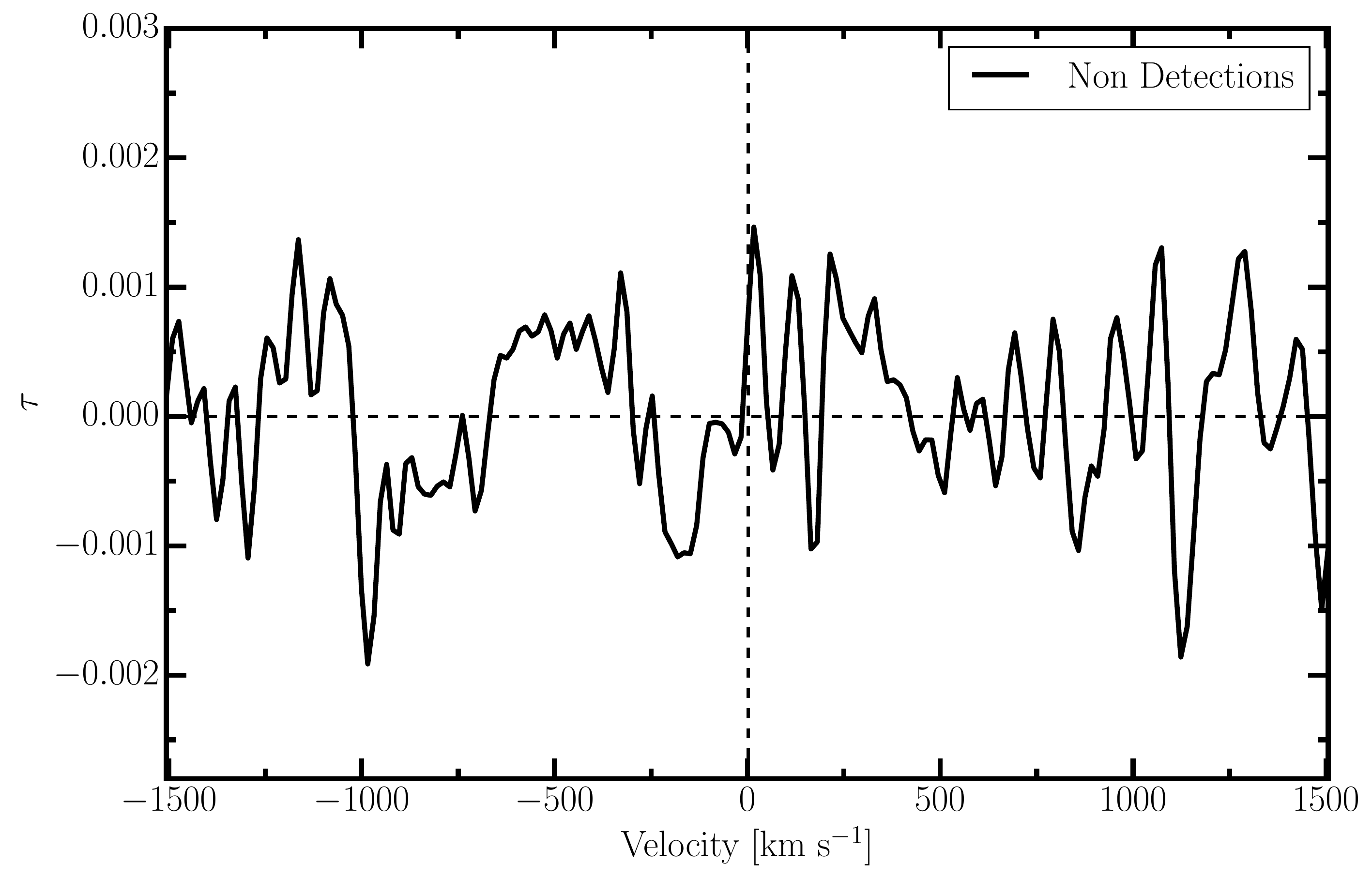}
\includegraphics[trim = 0 0 0 0, clip,width=.33\textwidth]{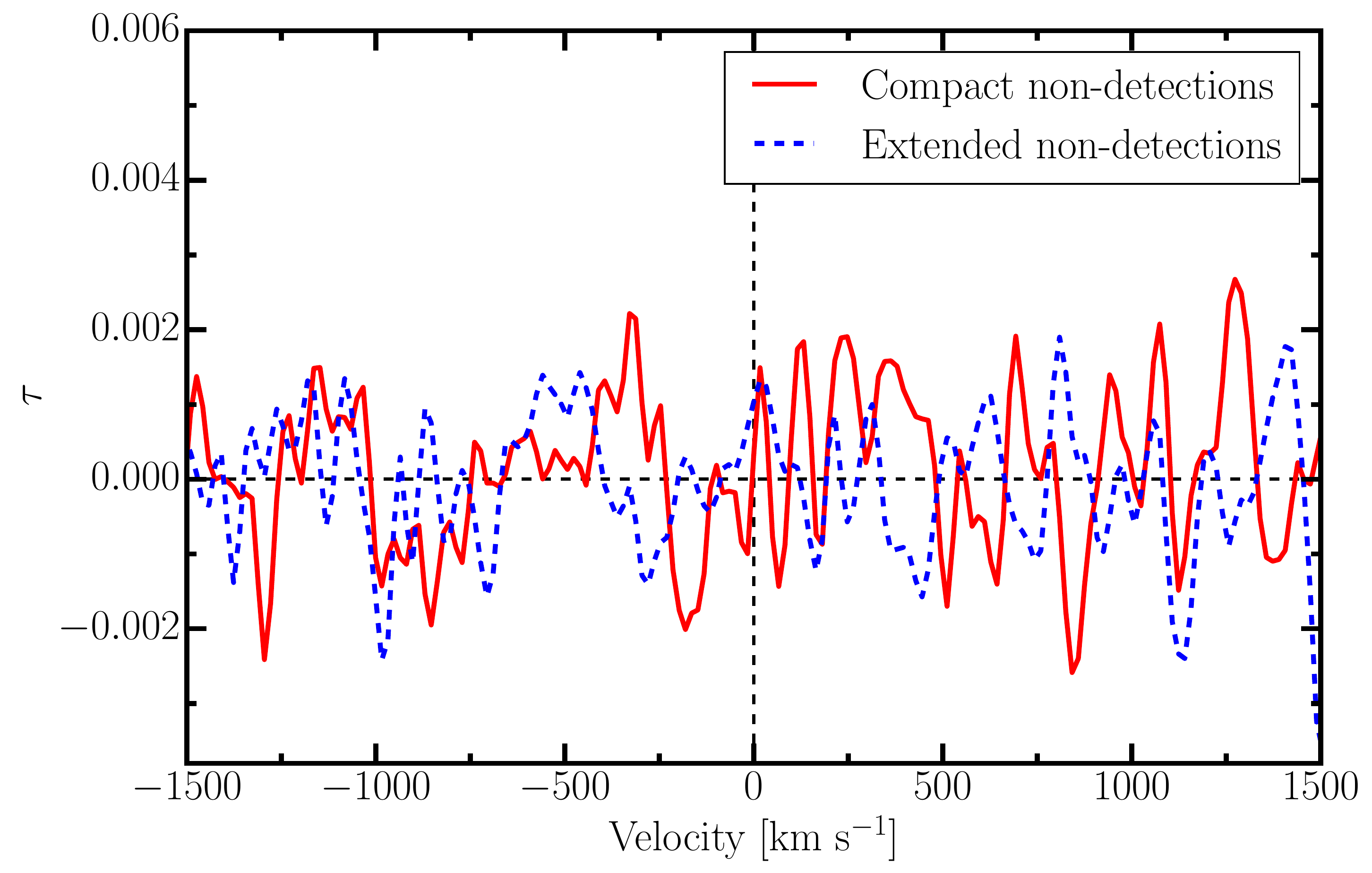}
\includegraphics[trim = 0 0 0 0, clip,width=.33\textwidth]{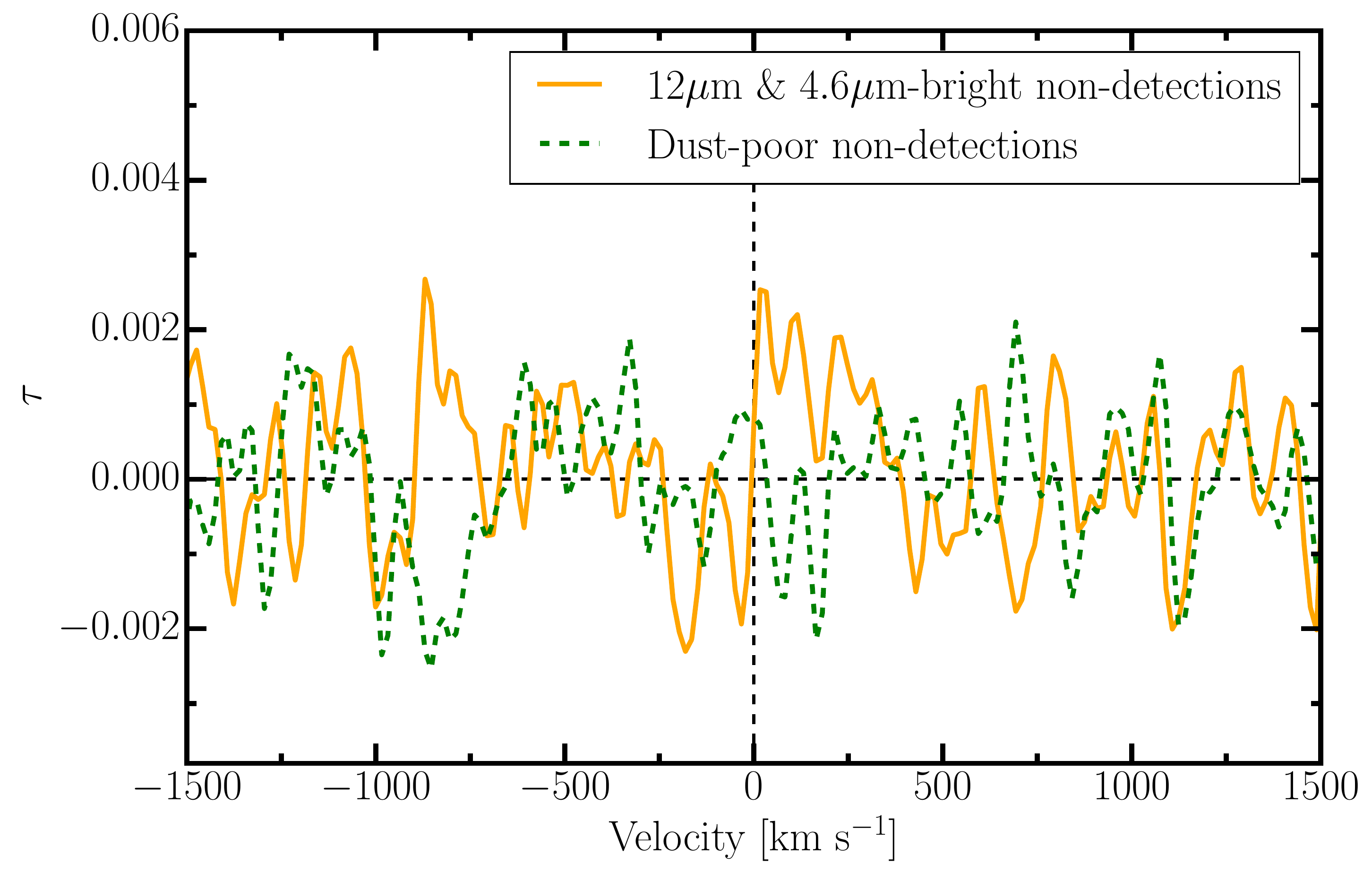}
\caption{(Left panel) Stacked spectrum of $170$ non-detections. (Middle panel) Stacked spectrum of $72$ non-detections with compact radio emission or interacting sources (red) and of $80$ non-detections with extended radio emission (blue). (Right panel) Stacked spectrum of $71$ non-detections classified as $12$~\um\ bright or $4.6$~\um\ bright galaxies (orange), and of $97$ non-detections classified as dust-poor (green), according to the WISE colours.}
\label{fig:stacking}
\end{center}
\end{figure*}

In \ps, we stacked the spectra of $66$ non-detections with \flux$>50$~ mJy, reaching a detection limit of $\tau\sim0.002$ ($3\sigma$) without detecting any absorption. In this study, we stack $170$ non-detections of the sample. We stack the spectra in optical depth aligning them at the SDSS redshift. Figure~\ref{fig:stacking}~(left panel) shows the final co-added spectrum. We reach a detection limit of $\tau=0.0015$ ($3\sigma$),\textbackslash\ without detecting any line. A non-detection at such low optical depth confirms the results of \ps. There, we pointed out that the peak optical depth of the detected \HI\ lines is much higher than the detection limit reached by the stacking experiment. This suggests a dichotomy between \HI\ detections and \HI\ non-detections. In the latter, if \HI\ is present, it must have much lower column densities or higher spin temperature ($T_{\rm spin}$) than in the former.

In Sect.~\ref{sec:full}, we point out that the detection rate of \HI\ in absorption is higher in compact sources than in extended sources. Likewise, it is higher in MIR bright sources than in dust-poor sources. Here, we explore whether this can also be seen by stacking the undetected objects belonging to these groups. 

We do not detect any absorption line either by stacking the spectra of the compact non-detections ($72$ sources) or extended non-detections ($80$ sources); see Fig.~\ref{fig:stacking}~(middle panel).  We do not detect any line either by stacking the dust-poor non-detections ($97$ sources) or the MIR bright non-detections ($71$ sources), see Fig.~\ref{fig:stacking}~(right panel). In Table~\ref{tab:stacking} we show the $3\sigma$ detection limits of the stacked spectra for these subgroups. On average we reach a detection limit of $\tau\sim0.003$. These results suggest that a dichotomy in the presence of \HI\ holds among all kinds of radio sources in agreement with the dichotomy observed in the stacking experiment of all non-detections.

\begin{table*}
\caption{Statistics of the stacking experiment}             
\label{tab:stacking}      
\centering                          
\begin{tabular}{l c c }        
\hline\hline                 
 Sample &  Number of sources & $3\times$ RMS  [optical depth] \\
\hline
Non-detections & 170  & 0.0015 \\

Compact & 72 & 0.0033 \\

Extended non-detections & 80 & 0.0027 \\

$4.6$~\um\ bright and $12$~\um\ bright non-detections & 71 & 0.0033 \\

Dust-poor non-detections & 97 & 0.0027 \\


\hline                           
\end{tabular}
\tablefoot{Results of the stacking experiment for the non-detections and their subsamples based on the radio classification (compact and extended sources) and on the WISE colour-colour plot ($4.6$~\um\ bright, $12$~\um\ bright and dust-poor galaxies). Since no lines are detected in absorption, we provide as upper limits three times the noise level of the final stacked spectrum. }
\end{table*}


\section{Discussion}
\label{sec:disc0}

\subsection{\HI\ morphologies and kinematics in different radio AGN}
\label{sec:disc}

\HI\ absorption has been studied in radio AGN for many years. Previous works (\eg~\citealt{vangorkom1989,morganti2001,vermeulen2003,morganti2005a,gupta2006}) have already suggested the possibility of a range of structures of the absorbing material that can, to the first order, be identified from the shape of the absorption lines. In some cases, this has been verified with follow-up high-resolution observations or by \HI\ emission observations~\citep{beswick2004,struve2010a,struve2010b,mahony2013,maccagni2014}.

Absorption lines detected against the nuclear radio component and close to the systemic velocity with narrow ($< 100$\kms) widths are likely the result of large-scale gas disks (e.g. dust lanes) while broader profiles (a few hundred \kms) can result from a circumnuclear disk,~such as~\mbox{Cygnus A}~\citep{conway1995,struve2010a}and \mbox{Centaurus A}~\citep{vanderhulst1983,struve2010b,struve2012}.
Distinguishing these structures is not straightforward because, as for all absorption studies, the shape and width of the line also depend on the structure of the background radio continuum. 

Absorption lines with centroid offset with respect to the systemic velocity or lines with a broad (\eg\ beyond the expected rotational velocity) red-shifted or blue-shifted wing, can trace gas that is unsettled, either infalling or outflowing, as in,~for example~\mbox{NGC 315}~\citep{morganti2009}, \pks~\citep{maccagni2014},~\mbox{B2 1504 +377}~\citep{kanekar2008},~\mbox{NGC 1266}~\citep{alatalo2011},~\mbox{4C 12.50}~\citep{morganti2013b}, and~\mbox{PKS B1740-517}~\citep{allison2015}.


In this work, we relate the \HI\ morphologies traced by the $66$ detected absorption lines of our sample to the radio power, extent of the radio continuum, and MIR colours of the associated host galaxies. Figure~\ref{fig:rp_linvar} shows that in sources with \rpl, we only detect narrow lines (excluding interacting sources, see Sect.~\ref{sec:sample}), while at \rp\ we also detect lines that are broad because of an asymmetric wing (with wing-to-peak ratio $\gtrsim0.3$). Figure~\ref{fig:rp_cen} shows that at radio powers below $10^{24}$~\whz\ the lines are centred at the systemic velocity, while above this threshold, the broad lines are also offset by more than $\pm 100$\kms. This suggests that, in our sample, in sources with \rpl\ the \HI\ lines trace a large-scale rotating disk, while in more powerful sources the \HI\ is not only rotating but it can also have unsettled kinematics. If in a low power source, a broad line had wing-to-peak ratio $<0.3$, the broad wing would have not been detected at the sensitivity of our observations. Cases of lines with such broad, shallow, and blue-shifted wings are known to be present in low-power sources,~such as~\mbox{NGC 1266}~\citep{alatalo2011} and~\mbox{IC 5063}~\citep{oosterloo2000,morganti2013b,morganti2015}. However, since these wings are very shallow, they do not affect the FW20 of the bulk of the line, where the peak lies. Hence, we can confirm that the deep absorption is on average narrower in low-power sources.

Figures~\ref{fig:rp_linvar} and ~\ref{fig:rp_cen} also show the sources where the \HI\ is unsettled, have compact radio continuum (left panels), and are MIR bright (right panels). Figure~\ref{fig:tau_width} shows that in dust-poor galaxies we do not detect unsettled gas but only narrow and deep \HI\ absorption lines. These results suggest that in dust-poor sources the \HI\ may be rotating in a disk. If, instead, the radio AGN is powerful (\rp) and compact,~\ie\ the jets are embedded within the host galaxy and the host galaxy is MIR bright, then the \HI\ may have unsettled kinematics, which possibly originates from the interplay between the radio activity and surrounding cold gas.

\subsection{Stacking experiment and comparison with the \atlas\ sample}

The absorption lines detected in our sample show that, depending on the radio power, they may trace \HI\ with different kinematics. Nevertheless, the \HI\ is detected with a detection rate of $27\%\pm5.5\%$ that is independent of the radio power of the sources~(see Sect.~\ref{sec:det}). The results of the stacking experiment have shown no detection of \HI\ absorption, even after expanding the number of stacked sources compared to \pz\ (see Sect.~\ref{sec:stacking}). This confirms a dichotomy in the presence and/or properties of \HI\ in radio AGN. For the detections, the peak of the absorption is often found at high optical depths ($\tau\gtrsim0.01$). In the non-detections, if \HI\ is present, then this peak must have a much lower optical depth than in the detections or its detection must be affected by orientation effects.

\begin{figure}[tbh]
\begin{center}
\includegraphics[trim = 0 0 0 0, clip,width=.48\textwidth]{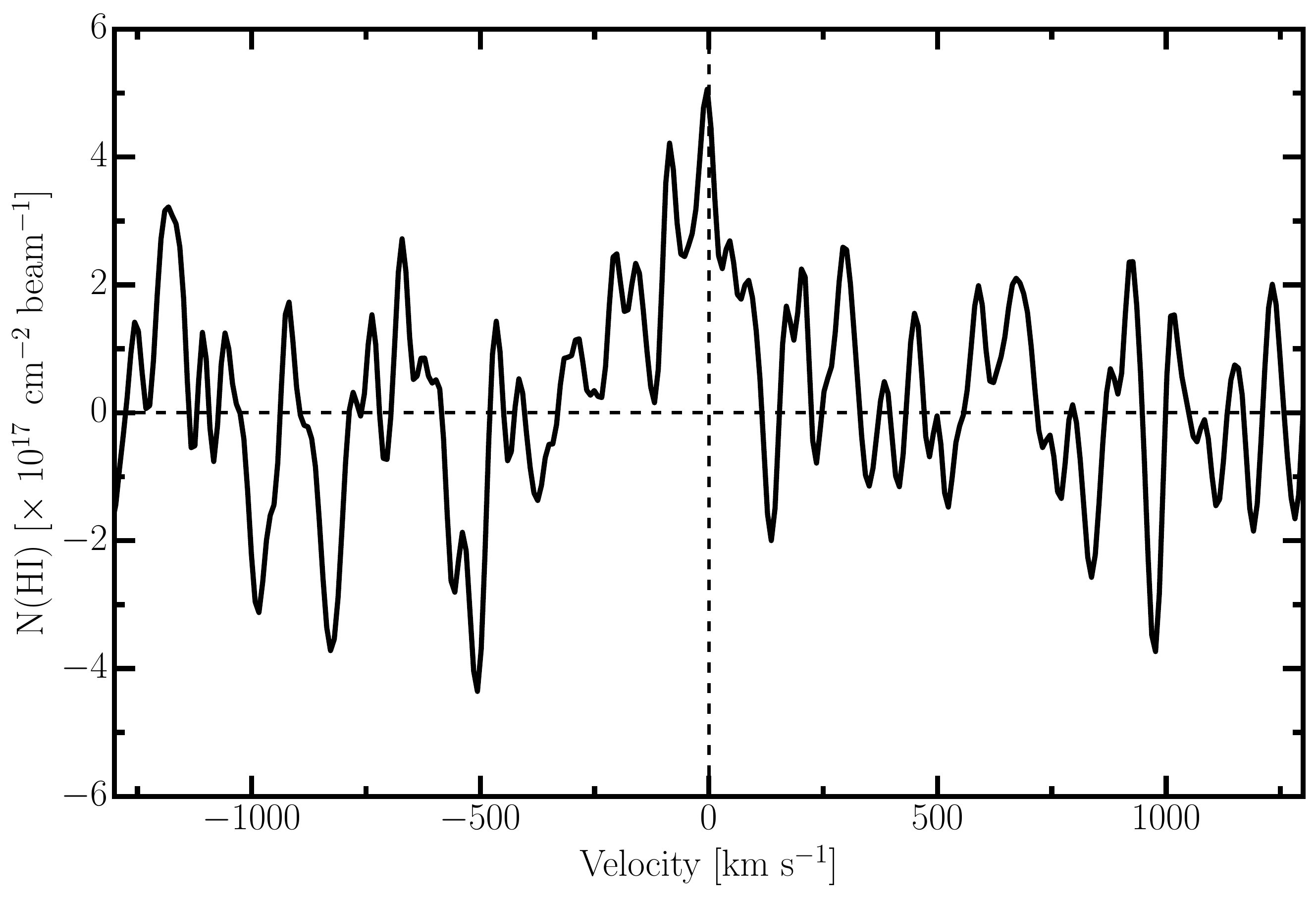}
\caption{Stacking in column density of the spectra of $81$ \atlas\ sources undetected in \HI\ emission.}
\label{fig:stacking_a3d}
\end{center}
\end{figure}

In order to investigate the origin of this dichotomy, we compare our results with what is found for \HI\ emission, \ie\ not affected by orientation effects and not limited by the size of the continuum emission. We use the \atlas\ sample~\citep{cappellari2011} because it represents the only complete volume-limited sample of early-type galaxies of the local Universe (\ie\ objects similar to the host galaxies of our radio sources) with deep \HI\ observations~\citep{oosterloo2010a,serra2012}. The \atlas\ sources represent the low radio-power end of our sample, since they have \pow$<22.5$~\whz~\citep{nyland2016b} and lie at fainter magnitudes in the red sequence than our sample ($-21.5<M_{\rm r}<-19$) . Among the \atlas\ sources, $139$ were observed in \HI. \HI\ has been detected in emission in the centre of $25\%$ of these sources,~\ie\ in the same region where we search for \HI\ in the sources of our sample. In most cases, the \HI\ appears settled in disk and ring morphologies, but there are many exceptions. \HI\ is not detected in the centre of $81$ galaxies. A stacking experiment on these sources allows us to determine the typical \HI\ column density, or its upper limit, in the centre of early-type galaxies.

Before stacking, we convert each spectrum to column density, correcting for the beam ($\theta$, in arcminutes) of the observation,
\begin{equation*}
N_{\rm\HI}\sim3.1\times10^{17}\cdot S\cdot dv/\theta^2,
\end{equation*}
where $S$ is the flux per channel ($dv$).  In Fig.~\ref{fig:stacking_a3d}, we show the result of stacking the \HI\ spectra of the $81$ non-detections. An emission line is detected with $\sim3\sigma$ significance at the systemic velocity. The FWHM of the line is $\sim200\pm50$\kms.

The detection of \HI\ in the early-type galaxies of \atlas\ shows that the majority have a low amount of \HI\ that is  below the typical detection level but that can be recovered by stacking.  We use this detection to derive the corresponding optical depth if this gas would have been observed in absorption. The integrated column density of the emission line is \nhi$\sim2.1\times10^{19}$~\cmsq.  The corresponding optical depth can be estimated as
\begin{equation*}
\tau\sim \frac{N_{\rm\HI}}{1.8216\times10^{18}\times{\rm FWHM}\times T_{\rm spin}/c_f}\sim 0.06\times \frac{c_f}{T_{\rm spin}},
\end{equation*}
where $c_f$ is the covering factor, i.e. the fraction of radio continuum covered by the absorbed gas, and $T_{\rm spin}$ is its spin temperature. The spin temperature of the \HI\ can range between a few $10^2$ K in the extended disks to a few $10^3$ K in the central regions of AGN \citep{maloney1996,kanekar2003,holt2008,kanekar2011,morganti2016}. The detection limit we reach by stacking the spectra of our sample ($\tau=0.0015$, see Table~\ref{tab:stacking}) can be converted to a column density, assuming that the width of the absorption line is equal to the width of the emission line of the stacked \atlas\ sources. This column density is~\nhi$=3.5\times10^{17}\,(T_\mathrm{spin}/c_f)$~\cmsq. For $T_\mathrm{spin}\gtrsim 100$ K and $c_f =1$, this value is higher than the column density of the emission line of the stacked \atlas\ sources. Therefore, the stacking experiment in absorption cannot detect the low-column density gas we detect in the \atlas\ sources. If a similar amount of \HI\ as in the stacked \atlas\ galaxies was present in our radio galaxies, we would expect (assuming $c_f =1$) \HI\ absorption in the stacked profile with optical depth in the range $\tau\sim10^{-4}$ for $T_{\rm spin} \sim 100$~K and $\tau\sim 10^{-5}$ for higher temperatures,~for example\ $T_{\rm spin}\sim 10^3$ K. These values are approximately three times lower than the detection limit we found in the stacking experiment for our sample (see Table~\ref{tab:stacking}). Thus, the stacking of our radio galaxies does not reach yet the sensitivity necessary to detect an amount of \HI\ that is comparable to what was detected for the \atlas. Achieving these limits will only be possible with the larger samples provided by the new surveys (see Sect.~\ref{sec:surveys}).

From the amplitude of the \HI\ detection obtained by stacking the \atlas\ sources, we can infer that the dichotomy between early-type galaxies with detected and undetected \HI\ could be due to a difference in the amount of \HI\ between the two groups. The \HI\ detected in emission in the \atlas\ galaxies (and column-density limited) is likely tracing only \HI\ in large galactic-scale structures. As described above, \HI\ absorption can also trace gas that is distributed in small scale, nuclear structures as its detection depends only on the strength of the background continuum. Thus, in the case of absorption, other effects can affect the non-detection of \HI. These effects can be orientation effects, \HI\ depletion in the nuclear region, or when only the warm component of \HI\ (\ie\ low optical depth for a given column density) is intercepted along the line of sight.

In compact sources the radio emission is embedded within the host galaxy hence the detection of absorption should be less affected by orientation effects of the absorbing structure~\citep{pihlstrom2003,curran2013}. However, the stacking experiment of this class of objects does not reveal an absorption line down to  $\tau\sim0.003$. This may suggest that orientation effects are not the only explanation of why \HI\ is detected (or not) in absorption. Nevertheless, among two subsamples of sources we find hints that orientation effects contribute in detecting \HI\ absorption. In \mbox{Fig.~10} of \ps\ we show that among the high-power sources (\rp) in the most highly inclined galaxies (axial ratio of the stellar disk, measured by SDSS, $b/a < 0.6$), we always detect \HI\ at high column densities (\nhi$\gtrsim10^{19}\cdot c_f/T_{\rm spin}$~\cmsq). Among more face-on galaxies ($b/a < 0.6$), we do not always detect \HI; when we detect  \HI\, the absorption lines can have lower column densities than in edge-on sources. When a source is $4.6$~\um\ bright, the WISE colours (W2--W4) allow us to estimate the orientation of the circumnuclear dust with respect to the AGN~\citep{crenshaw2000,fischer2014,rose2015}. Among the $42$ $4.6$~\um\ bright sources of our sample, $28$ have W2--W4$< 6, $ which suggests an unobscured AGN, \ie\ the circumnuclear dust is face-on with respect to the line of sight, while $14$ have W2--W4$> 6$, which suggests an obscured AGN, \ie\ the circumnuclear dust is edge-on with respect to the line of sight. Among obscured AGN the detection rate of \HI\ is $50\%\pm25\%$, while in unobscured AGN the detection rate is $32\%\pm17\%$. These detection rates are different, but they are consistent within the errors. Collecting a larger sample of $4.6$~\um\ bright sources should allow us to decrease the errors and understand if the difference in detecting \HI\ in absorption among obscured and unobscured AGN holds.

A different spin temperature could also be relevant to explain the lack of \HI\ absorption in the stacking of the sources. In addition to the high $T_{\rm spin}$ characteristic of circumnuclear gas affected by the radiation from the AGN, the typical ISM has a large warm component. A number of studies (\eg~\citealt{maloney1996,kanekar2003,curran2007,kanekar2009,kanekar2011}) have shown that in the typical ISM \nhi$\,\sim2\times10^{20}$~\cmsq\ is the threshold column density for cold \HI\ clouds ($T_{\rm spin} \lesssim 500$ K) and that lower column density \HI\ has higher spin temperature ($T_{\rm spin} \gtrsim 600$ K). 
The undetected galaxies could be dominated by the low column density component, as the stacking of the \atlas\ galaxies suggests, implying that we observe mostly gas at high spin temperature. For a fixed column density, the optical depth decreases with increasing spin temperature. If in central regions of the non-detections most of the \HI\ was warm, we would not have detected it in the stacking experiments of our survey.

\subsection{ Impact of the radio activity on the cold ISM of galaxies}\label{sec:im}

In Fig.~\ref{fig:rp_linvar} and Fig.~\ref{fig:rp_cen} we point out that among high-power sources (\rp) $30\%\pm15\%$ have broad \HI\ lines and that $36\%\pm16\%$ have lines that are shifted with respect to the systemic velocity. In {Fig.~7} of \pz\, we show that among  high-power sources the broadest lines are also more asymmetric and more blue-shifted with respect to the systemic velocity. We find very few cases of red-shifted absorption and they are all associated with relatively narrow lines. Furthermore, symmetric broad lines are detected only in interacting sources. All this seems to favour a scenario in which the unsettled kinematics of the \HI\ we detect in absorption is due to an outflow driven by the nuclear activity, rather than a scenario in which the \HI\ is unsettled prior to the triggering of the radio activity and may be falling into the radio source to contribute to its feeding. Nevertheless, cases of red-shifted narrow lines have been found, possibly in objects similar to~\mbox{NGC 315}~\citep{morganti2009} and \pks~\citep{maccagni2014}.

\begin{figure}[tbh]
\begin{center}
\includegraphics[trim = 0 0 0 0, clip,width=.48\textwidth]{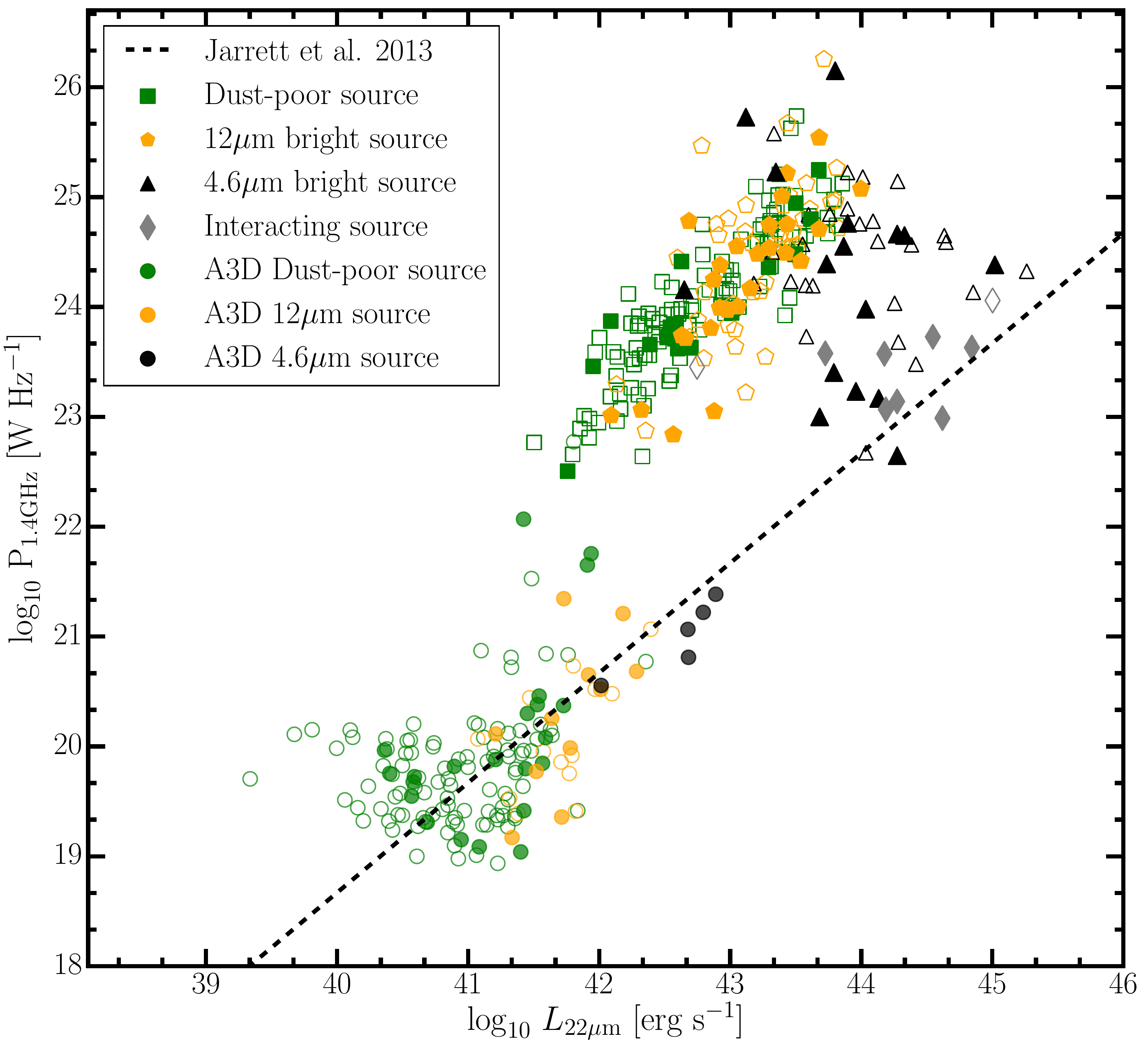}
\caption{Radio power of the sources vs. $22\mu$m luminosity for the sources of our sample and the \atlas\ sample (A3D), which have been observed in \HI. The colour coding follows the WISE colour-colour plot shown in Fig.~\ref{fig:wise}. The dashed line denotes the fit to the MIR-radio relation estimated by \cite{jarrett2013}.} 
\label{fig:mir_radio} 
\end{center}
\end{figure}

In our sample, early-type radio sources rich in heated dust, i.e. MIR bright sources, have a higher detection rate in \HI\ than dust-poor sources. Among high radio power sources, only MIR bright sources show broad lines. Figure~\ref{fig:mir_radio} shows the MIR-radio relation (\pow\ versus $\log_{10}\,{L}_{22 \rm \mu m}$) for all sources in our sample and for the sources of the \atlas\ sample\footnote{We measure ${L}_{22 \rm \mu m}$ from the magnitude found by WISE (W4).}. We also show a linear fit to the relation taken from the literature~\citep{jarrett2013}. The \atlas\ sample fits the relation well in the low star formation end. As expected, most sources in our sample are radio loud with respect to the relation. For the most powerful part of our sample (\rp), the detection of broad, asymmetric lines (see Fig. ~\ref{fig:rp_linvar} and~\ref{fig:rp_cen}) suggests that the energy released by the central AGN through the radio jets can perturb the circumnuclear cold gas. Among the galaxies of our sample that lie on the MIR-radio relation, \HI\ is detected either in narrow lines, most likely tracing a disk, or in interacting sources. In these sources, the origin of the unsettled kinematics and of the star formation may be attributed to the interaction event itself. 


The results of our survey are limited to AGN that were selected according to their radio flux. \cite{mullaney2013} investigated the impact of the radio power of the sources on the circumnuclear ISM in an optically selected sample of AGN. These authors found that only AGN with radio power $\log_{10}\,{\rm P}_{1.4 \rm GHz}>23$~\whz\ show broad [OIII] lines and that compact radio cores play a major role in perturbing this gas, which is in agreement with the results of our study. Hence, it seems that AGN perturb the surrounding ISM mainly via the mechanical power of their radio jets. In our sample, we see the effects of the radio jets expanding through the ISM in the ionised gas as well (Santoro et al., in prep). The major role played by the radio nuclear activity in perturbing the ISM is also suggested when looking at the \HI\ in compact and extended radio sources. The former, where the radio jets have sub-galactic scales and are likely carving their way through its ISM, show broader and more offset lines than the latter, where the radio jets have already exited the galaxy; see Fig.~\ref{fig:rp_linvar} and \ref{fig:rp_cen}~(left panels). 

\section{\HI\ absorption detections in Apertif, ASKAP, and MeerKAT}
\label{sec:surveys}

The main results of our survey show the importance of \HI\ absorption studies to trace the presence of cold gas in the central regions of early-type radio galaxies. Such studies allow us to understand how  \HI\ relates to the other components of the ISM, such as\ the warm dust, and how  \HI\ interacts with the nuclear radio emission. In particular, our survey shows that the detection rate of \HI\ in absorption does not change in the range of redshifts and radio powers of the sample (\zint, \pow$=22.5$~\whz~$\lesssim$\pow$\lesssim 26.2$~\whz ). Hence, \HI\ absorption studies are the best tool to investigate the occurrence of \HI\ in sources at all redshifts and radio powers. 

The upcoming blind \HI\ surveys of the SKA pathfinders and precursors, i.e. MeerKAT~\citep{jonas2009}, the Australian SKA Pathfinder,  (ASKAP,~\citealt{johnston2008}), and Apertif~\citep{oosterloo2010b}, will play a fundamental role in detecting \HI\ in absorption, shed new light on its structure and its interplay with the radio nuclear activity, and they will be able to determine the occurrence of \HI\ and its optical depth distribution down to low flux radio sources and intermediate redshift ($z\sim1$).

In this section, we use the results of our survey to explore the possibilities for expanding this work in the upcoming \HI\ surveys and to explore how the different surveys will cover the parameter space of radio sources and whether they will be complementary in describing the presence and properties of the \HI.

A number of large area, wide bandwidth absorption line surveys are planned with next generation radio facilities. In Table~\ref{tab:surveys}~\citep{gupta2017}, we show the frequencies, redshift ranges, sky coverage, and flux limits of four of these surveys, which we consider in this section. In particular, we consider the survey from the new Apertif system on the WSRT, the
Search for \HI\ with Apertif (SHARP), the MeerKAT Absorption Line Survey (MALS) with the South African SKA precursor MeerKAT, and two surveys with the ASKAP~\citep{johnston2008} facility, i.e. the Wide-field ASKAP L-band Legacy All-sky Blind surveY (WALLABY) and the First Large Absorption Survey in \HI\ (FLASH).  A dedicated, intermediate redshift survey, FLASH is searching for associated and intervening absorption at $0.5<z<1.0$. The WALLABY is a large sky coverage survey to detect \HI\ in emission, which will also be sensitive to \HI\ absorption at low redshifts. As with the survey presented in this paper, we focus in this section on the search for \HI\ associated with the radio sources, although within the surveys, intervening absorption systems will also be investigated. 

\begin{figure}[tbh]
\begin{center}
\includegraphics[trim = 0 0 0 0, clip,width=.5\textwidth]{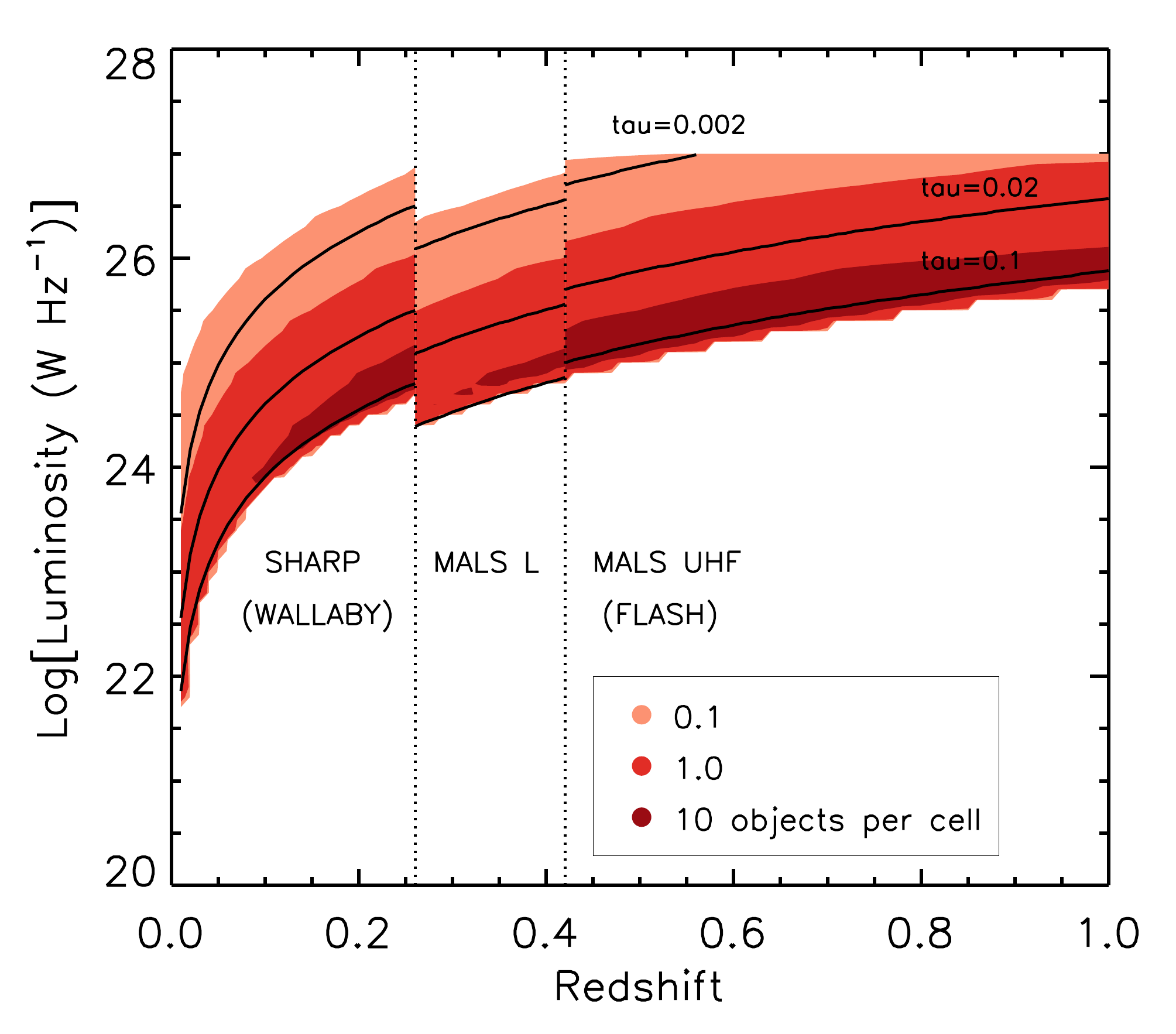}
\caption{Coverage of the luminosity--redshift plane by upcoming absorption surveys. The parameter space is divided into cells of $dz=0.01$ and $dL=0.1$, and the colour coding indicates 0.1 (light), 1 (medium), and 10 (dark shading) objects per cell. As expected, the majority of objects lie close to the flux limits of the surveys. At $z\leq0.26$, SHARP in the northern hemisphere and WALLABY in the south cover very similar parameter space. The full MALS L-band survey covers $0<z<0.58$, but is the only survey probing intermediate redshifts, $0.26<z<0.42$. MALS UHF targets intermediate redshifts, $0.42<z<1$. A similar redshift range is covered by FLASH, but with brighter flux limits. For clarity, we show the coverage of the plane out to $z=1$, even though both surveys extend to $z=1.44$ with the same behaviour. The black lines denote the minimum optical depth visible for sources of a given flux with the channel r.m.s. values given in Table~\ref{tab:surveys}.}
\label{fig:apertif1}
\end{center}
\end{figure}

First, we simulate the continuum source population against which absorption will be seen. We combine the parameters of area, continuum sensitivity, and channel sensitivity of the survey  (listed in Table~\ref{tab:surveys}) with a luminosity function from~\cite{mauch2007} and source number counts extracted from the simulations of~\cite{wilman2008}. From this, we can determine the luminosity--redshift parameter space covered by the various planned surveys, shown in Fig.~\ref{fig:apertif1}. The figure shows SHARP, MALS L band and MALS UHF, and the redshift ranges they cover. The WALLABY covers similar parameter space as SHARP, while FLASH and MALS UHF are both at higher redshift. The flux limits and channel sensitivities of the three surveys are well matched with each other, resulting in continuous coverage of the parameter space with large dynamic range at each redshift.

In Fig.~\ref{fig:apertif1}, at low redshifts, SHARP in the northern hemisphere and WALLABY in the south cover very similar parameter space with WALLABY's larger volume containing more of the rare, high luminosity, low redshift objects. These are the only surveys probing low luminosity, $L<10^{24}$\whz, sources.

At moderate redshifts, the MALS L-band survey is the only survey probing $0.26<z<0.42$, and will provide critical overlap with the intermediate redshift end of the MeerKAT deep HI emission surveys.

At intermediate redshifts ($z\sim1$), MALS UHF and FLASH probe only the high luminosity sources, so care must be taken when comparing the absorption population at intermediate and low redshifts, as the continuum source populations are disparate and, as we have seen here, the \HI\ can have different properties in different types of radio sources. The relatively bright flux limit of FLASH restricts the dynamic range in continuum sources, but the large area will result in excellent statistics of both intervening and associated absorbers, enabling evolution in the absorber population to be probed.

\begin{table*}
\caption{Summary of various upcoming \HI\ 21 cm absorption line surveys}
\centering
\label{tab:surveys}
\begin{tabularx}{\textwidth}{X c c c c c c c}  
\hline\hline                                                         
 Survey &  Redshift & Time per pointing & Spectral r.m.s. & Sky coverage  & Total time & Number \\
    & [\HI\ 21 cm] & [hrs] &[\mJy]  & [deg$^2$] &  [hours] &of lines of sight\\
\hline
Apertif -- SHARP  & 0--0.26 & 12 & 1.3 & 4000 & 6000 & 25000  \\
                                                   &             &      &       &          &          & ($>30$ mJy)  \\

ASKAP -- FLASH  & 0.4--1.0 & 2 & 3.8 & 25000 & 1600 & 65000  \\
                                                   &             &      &       &          &          & ($>90$ mJy)  \\

ASKAP -- Wallaby & 0--0.26 & 8 & 1.6 & 30000 & 8000 & 132000  \\
                                                   &             &      &       &          &          & ($>40$ mJy)  \\

MeerKAT -- MALS  & 0--0.57 & 1.4 & 0.5 & 1300 & 1333 & 16000  \\
        \hspace{16mm}           (L -band)                          &             &      &       &          &          & ($>15$ mJy)  \\

MeerKAT -- MALS  & 0.40--1.44 & 1.7--2.8 & 0.5--0.7& 2000 & 2125 & 33000  \\
  \hspace{16mm}         (UHF-band)                                 &             &      &       &          &          & ($>15$ mJy)  \\

\hline                           
\end{tabularx} 
\tablefoot{The two-part MALS project, MALS L-band, and MALS UHF are targeted surveys focussing on relatively bright, high redshift background sources to search the line of sight for intervening absorption. However, with the more than $1\times1$~deg$^2$ field of view of MeerKAT, a substantial volume for each pointing is blindly, and commensally, probed both for associated and intervening absorbers. SHARP and WALLABY are both commensal HI emission and absorption surveys, primarily investigating associated absorption. FLASH is a blind survey of the southern hemisphere to detect \HI\ absorption in intervening and associated systems at intermediate redshifts ($z\sim1$).}
\end{table*}

After determining the distribution of continuum sources available for each survey, we add the absorption population. We assume that one out of three sources will have an associated absorber, regardless of the source flux or range of observable absorber depths. Figure~\ref{fig:apertif1} shows that for each source and survey channel sensitivity, there is a minimum peak optical depth value that is observable. In the figure, we denote three minimum observable peak optical depths with black lines ($\tau\sim0.1$, $\tau\sim0.02$, and $\tau\sim0.002$). 

Figure~\ref{fig:apertif1} shows that the different surveys complement each other. In high-power radio sources (\pow$\gtrsim10^{25}$~\whz), these surveys will allow us to detect HI at $ 0.002<\tau<0.1$ to $z\lesssim0.5$, and $\tau>0.02$ to $z\lesssim 1$. Hence, in powerful radio sources we will be able to trace the evolution of the properties of \HI\ with redshift. However, the bulk of the population of radio sources has low radio power and its \HI\ absorption content will be explored only by the low redshift surveys, for example the extension of the research presented in this paper.

\begin{figure}[tbh]
\begin{center}
\includegraphics[trim = 0 0 0 0, clip,width=.5\textwidth]{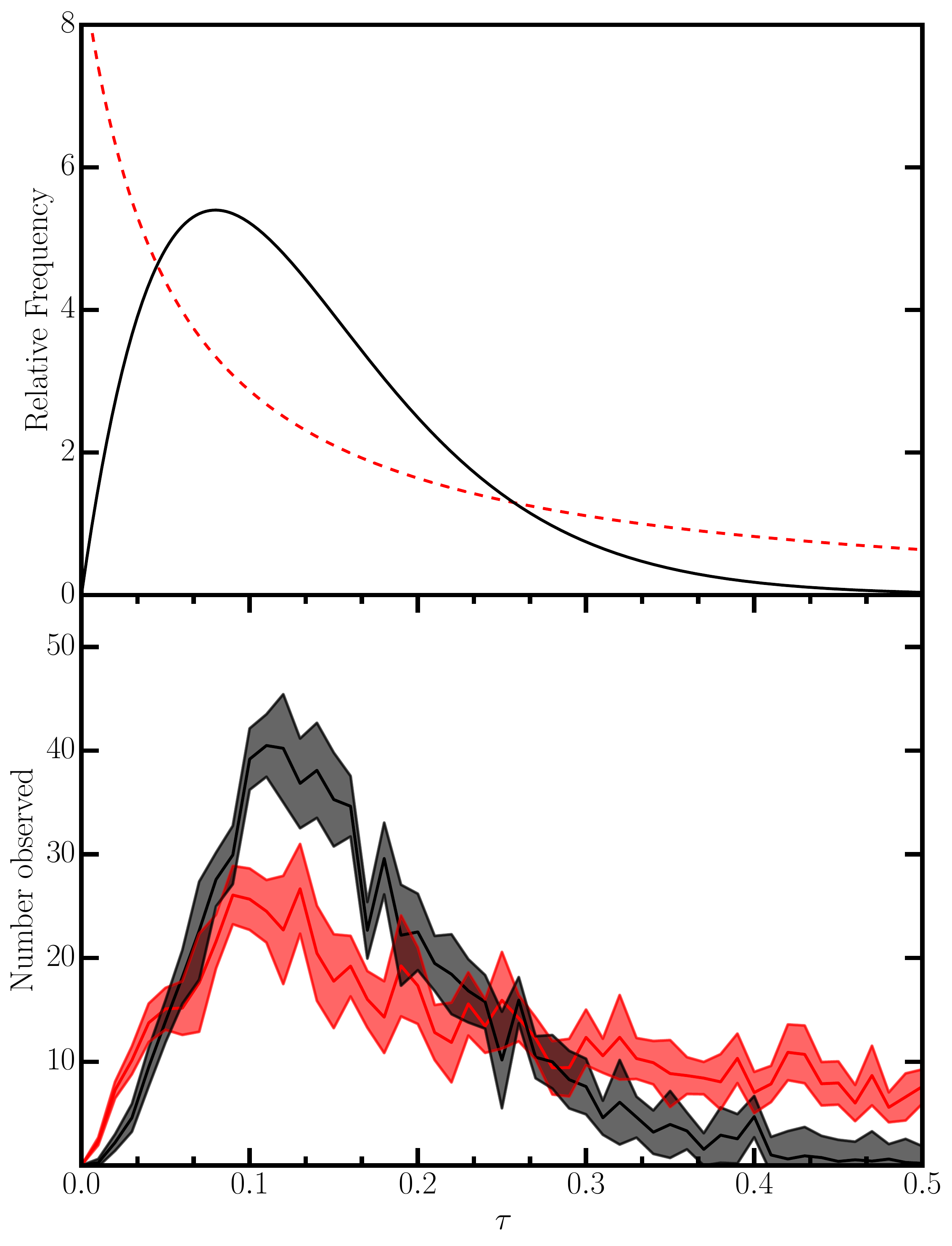}
\caption{(Top panel) Two suggested intrinsic distributions for the frequency of associated absorbers of a given optical depth. The primary difference is at very low optical depth values, where the exponential-shaped distribution is dominated by low values, whereas the other peaks at $\tau\sim0.08$. (Bottom panel) The resulting observed optical depth distributions for the SHARP survey, assuming absorbers are drawn from the two distributions in the top panel. The solid lines indicate the average observed distributions, while the shaded regions their dispersion.}
\label{fig:apertif2}
\end{center}
\end{figure}

In the survey presented in this paper, we find a dichotomy in the detection of \HI, suggesting the gas has high optical depth in the detections and very low optical depth in the non-detections. From this, we assume two different distributions to model the (unknown) intrinsic absorption population (see the top panel of Fig.~\ref{fig:apertif2}) and to extract the distribution we expect to observe with the SHARP survey. The exponential shape might naively be expected, since the most common, low column density \HI\ clouds correspond to very low optical depth absorption. On the contrary, the second shape, which is modelled on a black-body curve, has very few low optical depth absorbers while maintaining large number of absorbers with moderate optical depth. For one out of three sources, an absorber of a given optical depth is chosen randomly from the visible $\tau$ values, separately for the two intrinsic distributions. No redshift evolution in the absorber population is included.

The bottom panel of Fig.~\ref{fig:apertif2} shows the results of this exercise for the SHARP survey. The resulting observed distributions are sufficiently different and upcoming surveys should be able to constrain the intrinsic distribution from which the absorbers are drawn. The observed optical depth distribution is a convolution of the galaxy luminosity function and the intrinsic optical depth distribution. Its shape depends from the fact that most of the observed objects are assigned values of $\tau\ge 0.1$ and are close to the flux limit of the observations. Thus, aside from the normalisation, within this simplistic simulation the shapes of the observed distributions from survey to survey are similar. The targeted aspect of MALS, with deep integrations of bright sources, will help fill in the very low $\tau$ values.

Some evolution of the absorption population is expected and this should be seen in the optical depth distribution observed in the intermediate redshift sources. Also, the occurrence of absorption as a function of optical depth and continuum flux will be better determined from uniform, blind surveys, which provide information about the distribution of the absorbing gas in galaxies.

Interestingly, the lack of absorption seen so far in the stacking exercises indicates that very low optical depth values are not common and this disfavours the exponential underlying optical depth distribution. Larger samples of uniformly selected and observed objects will strengthen this restriction.

\section{Concluding remarks}
\label{sec:conc}
In this paper, we presented new WSRT observations of $147$ radio sources (\flint, $0.02<z<0.25$) observed in search for \HI\ absorption. We detected \HI\ in $34$ sources (see Appendix~\ref{app:a}). We combined this sample with the sample of $101$ sources with \flux$>50$~\mJy, observed in a similar way and presented in \ps\ and \pz. We classified the $248$ sources of the sample according to their radio continuum emission (compact sources and extended sources) and to their WISE colours (dust-poor, $12$~\um\ bright, and $4.6$~\um\ bright sources; see Sect.~\ref{sec:sample}). The vast majority of  the sample lies above the MIR-radio relation, suggesting the bulk of their radio emission comes from the central AGN (see Sect.~\ref{sec:im}). We analysed the occurrence, distribution, and kinematics of the \HI\ with the following results:


\begin{itemize}
\item Twenty-seven percent of radio-galaxies have \HI\ detected in absorption associated with the source. The detection rate does not vary across the range of redshifts and radio powers of the sample (see Sect.~\ref{sec:full}).
\item AGN with radio-power \rpl\ only show narrow \HI\ absorption lines. Broad lines, which can trace a significant component of  \HI\ unsettled by the radio activity, are found only in powerful radio AGN (\rp).
\item Compact sources show broad lines, tracing \HI\ with unsettled kinematics. Compact sources also show a higher detection rate of \HI\ than extended sources.
\item Dust-poor galaxies, at all radio powers, show only narrow and deep \HI\ absorption lines (see Fig.~\ref{fig:tau_width}), mostly centred at the systemic velocity, suggesting that in these galaxies most of the \HI\ is settled in a rotating disk.
\item In MIR bright sources we detect \HI\ more often than in dust-poor sources. Above \rp, these sources often show broad lines, suggesting the \HI\ has unsettled kinematics.
\item Three of the most powerful MIR bright sources show broad (FWHM$>300$\kms) \HI\ lines with a blue-shifted wing. These lines are likely to trace an outflow of neutral hydrogen pushed out of the galaxy by the radio jets. 
\item The broad and asymmetric \HI\ lines we detect all have a blue-shifted wing, while broad lines with a red-shifted component are not found. This may favour a scenario in which the  kinematics of the \HI\ are unsettled by the expansion of the nuclear activity,~\ie\ an outflow, rather than a scenario in which the \HI\ is unsettled prior to the triggering of the radio activity and may be falling into the radio source.
\item Stacking experiments on the non-detections of our sample do not reveal a detection of \HI\ absorption. In stacking the subgroups of sources where the detection rate of \HI\ is higher, i.e. compact sources and MIR bright sources, we still do not detect any absorption line (see Sect.~\ref{sec:stacking}). 
\item As reference, we stack the non-detections of the \atlas\ sample~\citep{serra2012}; we detect an \HI\ emission line with $\sim 3\sigma$ significance, tracing \HI\ with very low column density
\nhi$\sim2.1\times10^{19}$~cm$^{-2}$. If all this \HI\ was cold ($T_{\rm spin}\lesssim100$ K), the corresponding optical depth would be three times lower than
the detection limit of our stacking experiment in absorption (see Table~\ref{tab:stacking}). If the \HI\ was warmer, the optical depth would be even lower. Given that the \HI\ has much higher optical depths when is directly detected, this suggests a bi-modality in the occurrence of \HI\ in early-type galaxies. Of these galaxies, $27\%\pm5.5\%$ have \HI, detectable in absorption with short targeted observations (4--6 hours). The other galaxies, if they are not completely depleted of it, have \HI\ at very low column densities or higher spin temperatures. 
\item Orientation effects do not seem to be the only explanation as to why \HI\ is detected (or not) in absorption. There are suggestions that orientation effects may be important in particular subsamples of sources, such as powerful radio sources where the host galaxy is edge-on and $4.6$~\um\ bright sources. However, these results are affected by the small number of sources of these subsamples. 
\item The upcoming \HI\ absorption surveys of the SKA pathfinders (SHARP with Apertif, MALS with MeerKAT, and FLASH with ASKAP) will allow us to probe the \HI\ optical depth distribution for radio sources out to redshift $z\sim1$. The three surveys are complementary in the redshift intervals, and only the combination of all three will allow us to investigate the \HI\ content in radio sources of all radio powers and at all redshifts. In particular, SHARP and WALLABY are the only surveys which, at low redshifts ($z\lesssim0.26$), will allow us to probe the entire range of radio powers, \mbox{$22$~\whz\ $< \log_{10}\,{\rm P}_{1.4 \rm GHz}<26$~\whz}.
\end{itemize}

\begin{acknowledgements}
The WSRT is operated by the ASTRON (Netherlands Foundation for Research in Astronomy) with support from the Netherlands Foundation for Scientific Research (NWO). This research makes use of the SDSS Archive, funding for the creation and distribution of which was provided by the Alfred P. Sloan Foundation, the Participating Institutions, the National Aeronautics and Space Administration, the National Science Foundation, the US Department of Energy, the Japanese Monbukagakusho, and the Max Planck Society. This publication makes use of data products from the Wide-field Infrared Survey Explorer, which is a joint project of the University of California, Los Angeles, and the Jet Propulsion Laboratory/California Institute of Technology, funded by the National Aeronautics and Space Administration. For this research we made extensive use of the software Karma~\citep{gooch1996} and TOPCAT~\citep{taylor2005}. The research leading to these results has received funding from the European Research Council under the European Union's Seventh Framework Programme (FP/2007-2013) / ERC Advanced Grant RADIOLIFE-320745. The authors wish to thank Dr. Robert Schulz and Dr. Bradley Frank for useful discussions and suggestions.
\end{acknowledgements}


\bibliographystyle{aa}
\bibliography{safari1}

\begin{onecolumn} 
\begin{appendix} 
\section{Ancillary information on the sample}
\label{app:a}

Here, we show the  $34$ \HI\ absorption lines detected with the WSRT from observations between December 2013 and February 2015 in radio sources with \flint. In Table~\ref{tab:long} we show the main properties of the sources and the parameters of the detected lines measured using the busy function.

\begin{figure*}[!hb]
\begin{center}
                \includegraphics[trim = 0 0 0 0, clip,width=.33\textwidth]{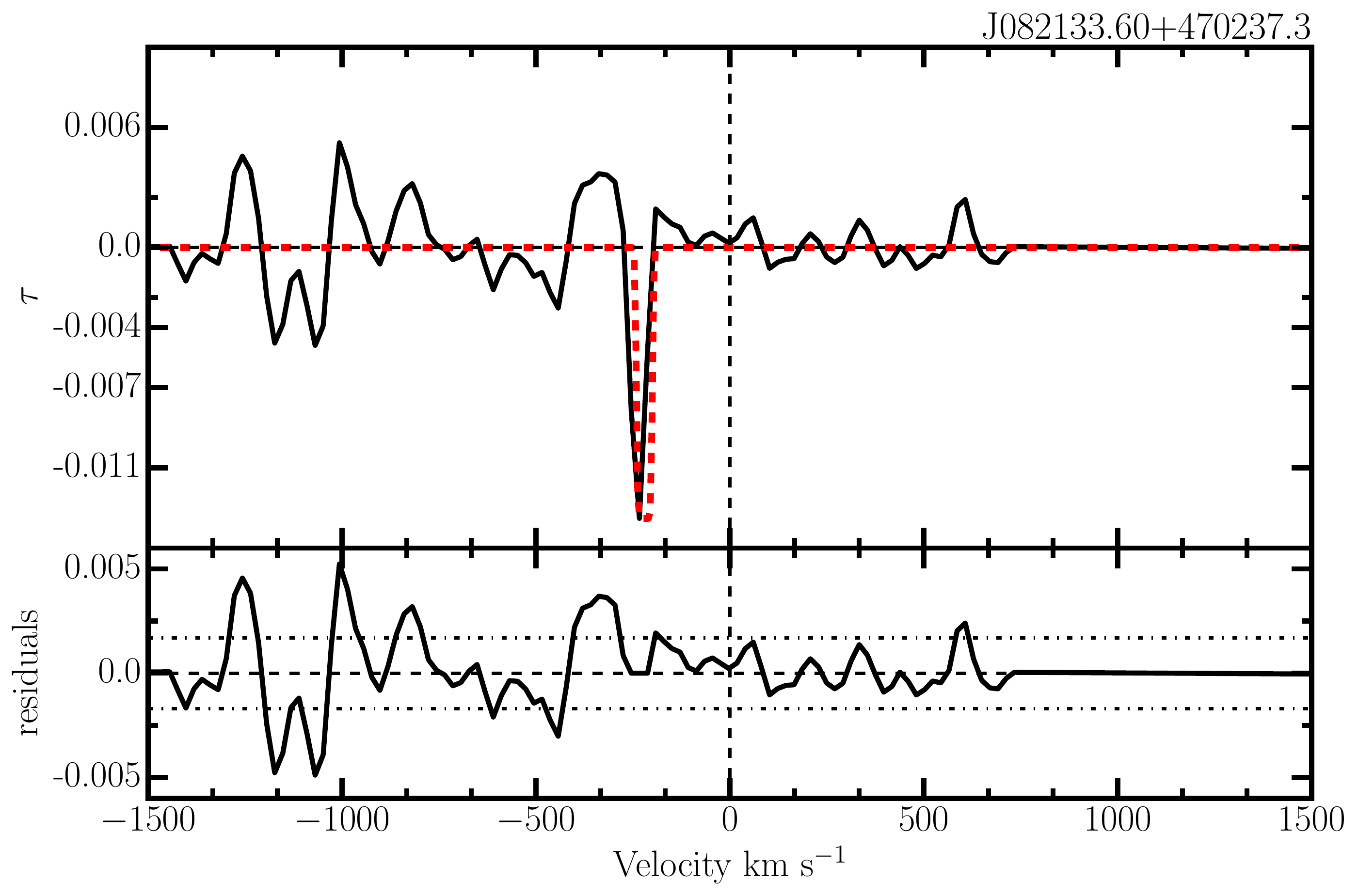}
                \includegraphics[trim = 0 0 0 0, clip,width=.33\textwidth]{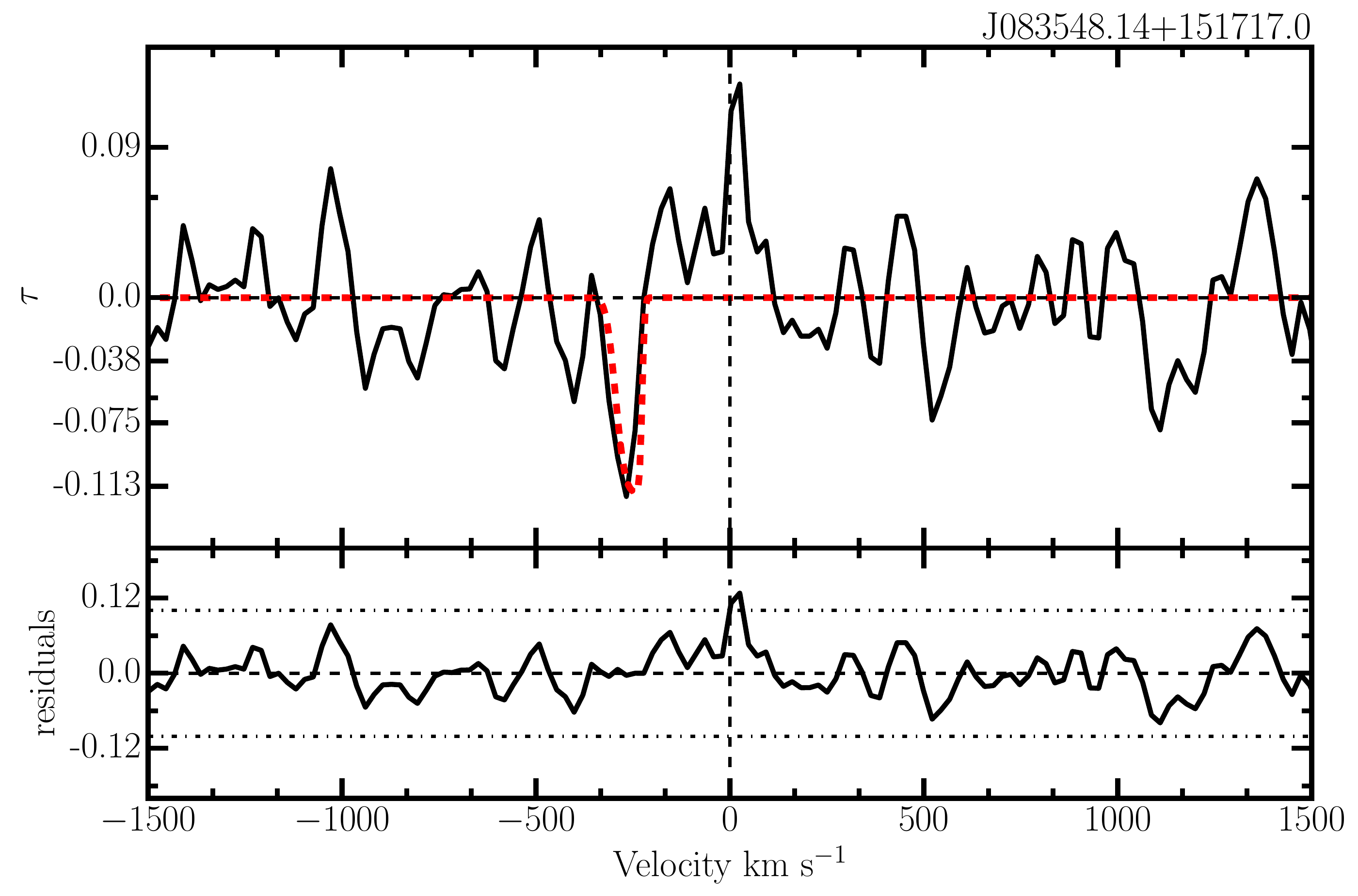}
                \includegraphics[trim = 0 0 0 0, clip,width=.33\textwidth]{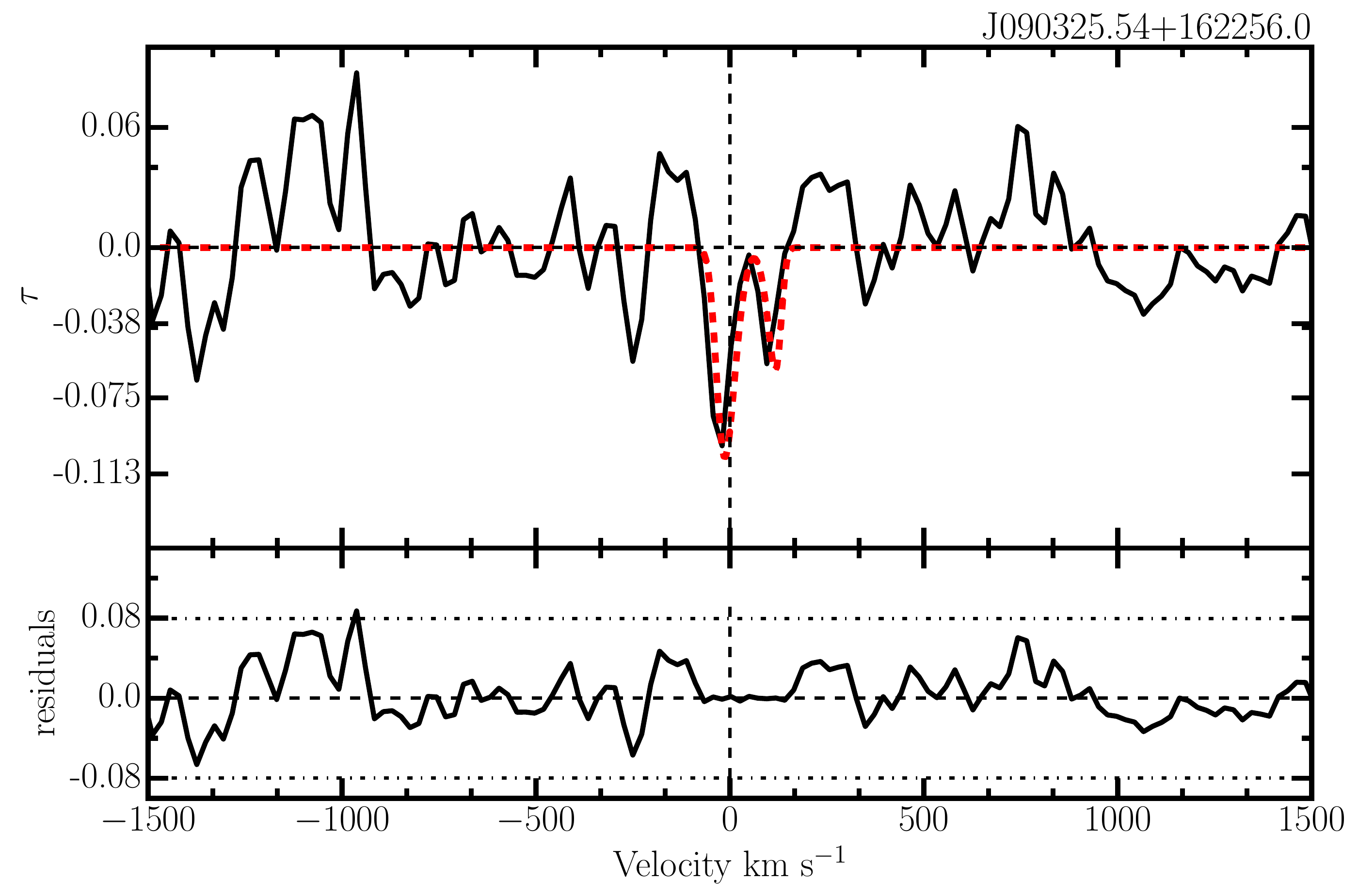}
                \includegraphics[trim = 0 0 0 0, clip,width=.33\textwidth]{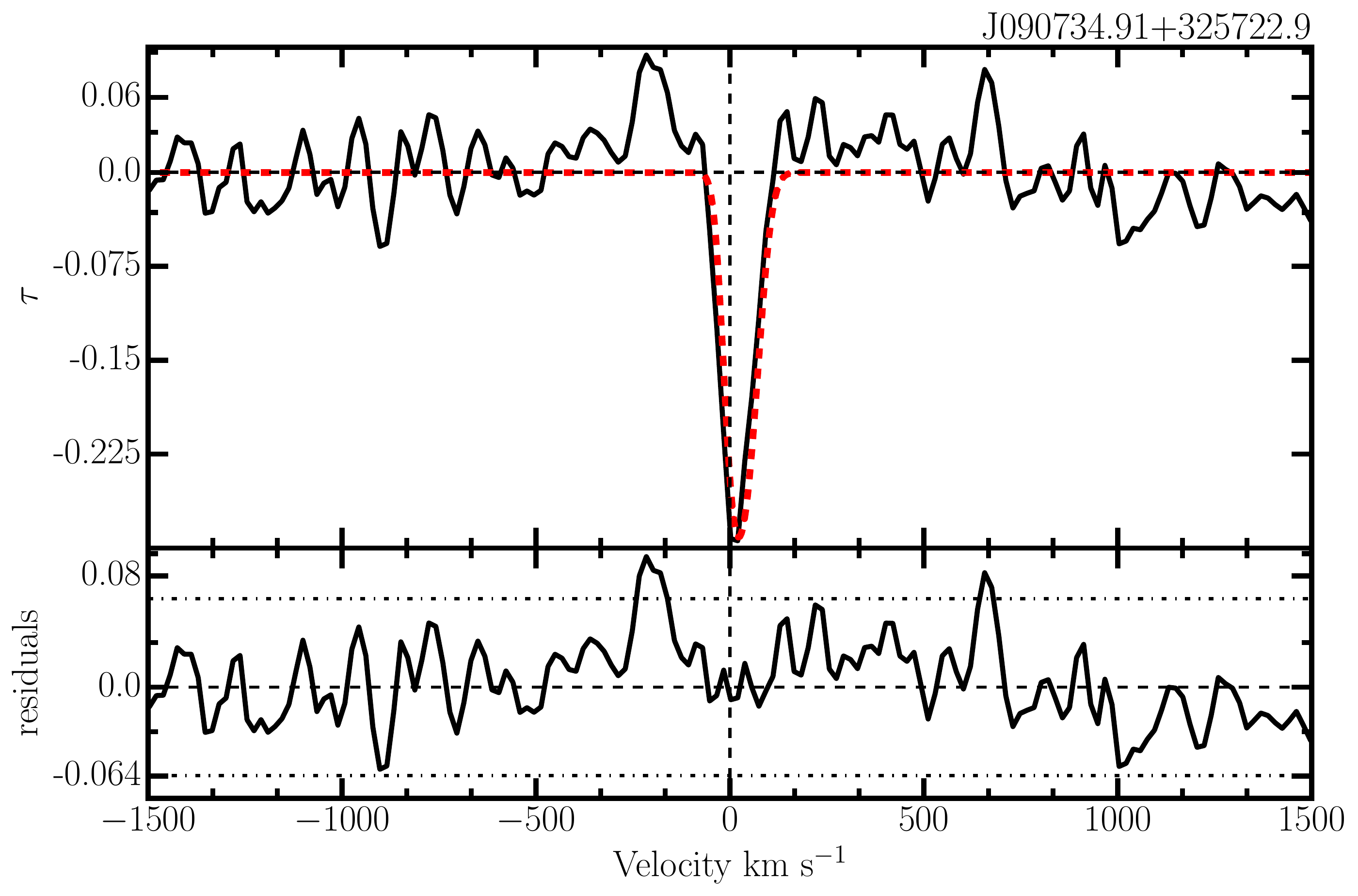}
                \includegraphics[trim = 0 0 0 0, clip,width=.33\textwidth]{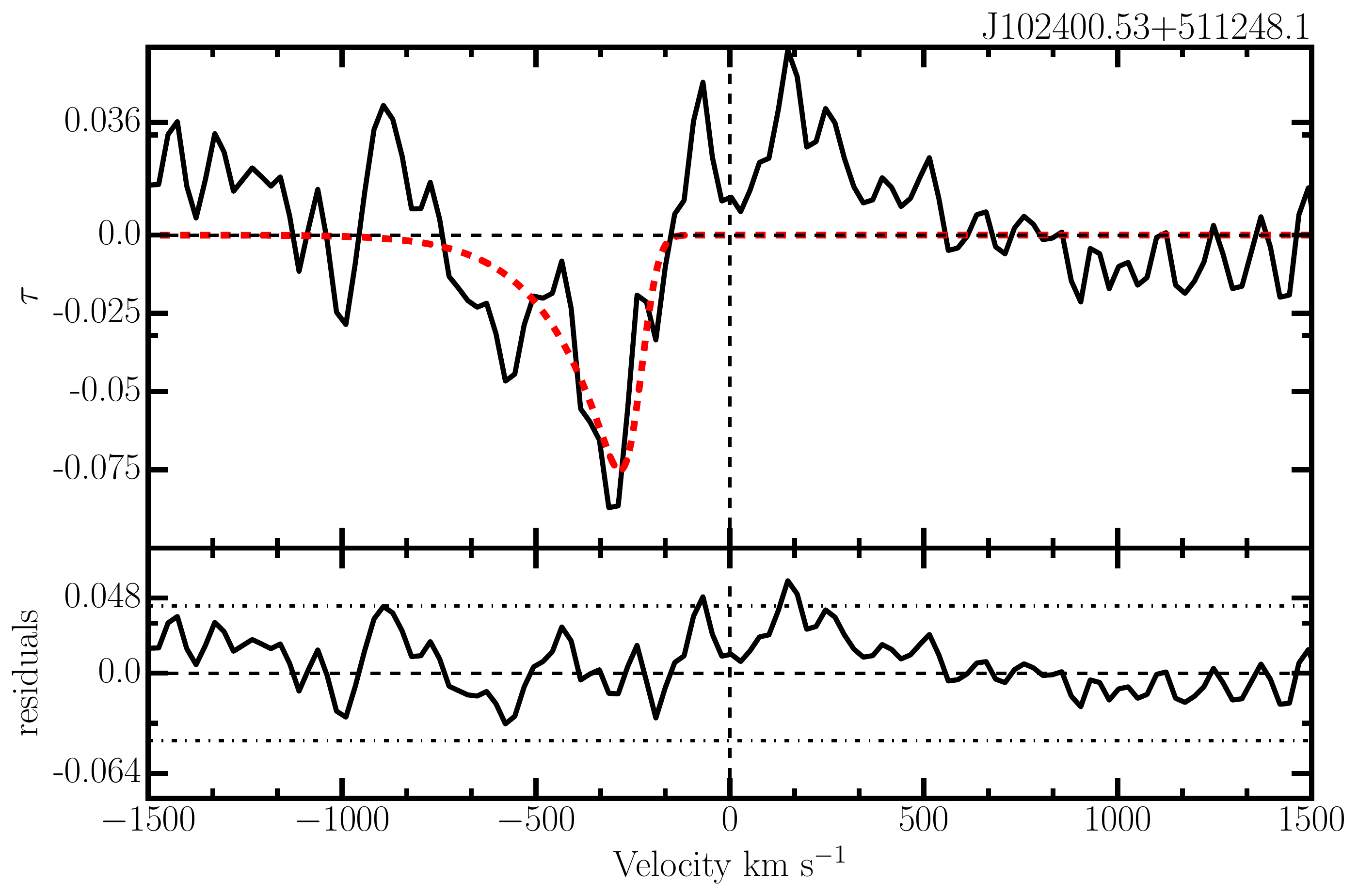}
                \includegraphics[trim = 0 0 0 0, clip,width=.33\textwidth]{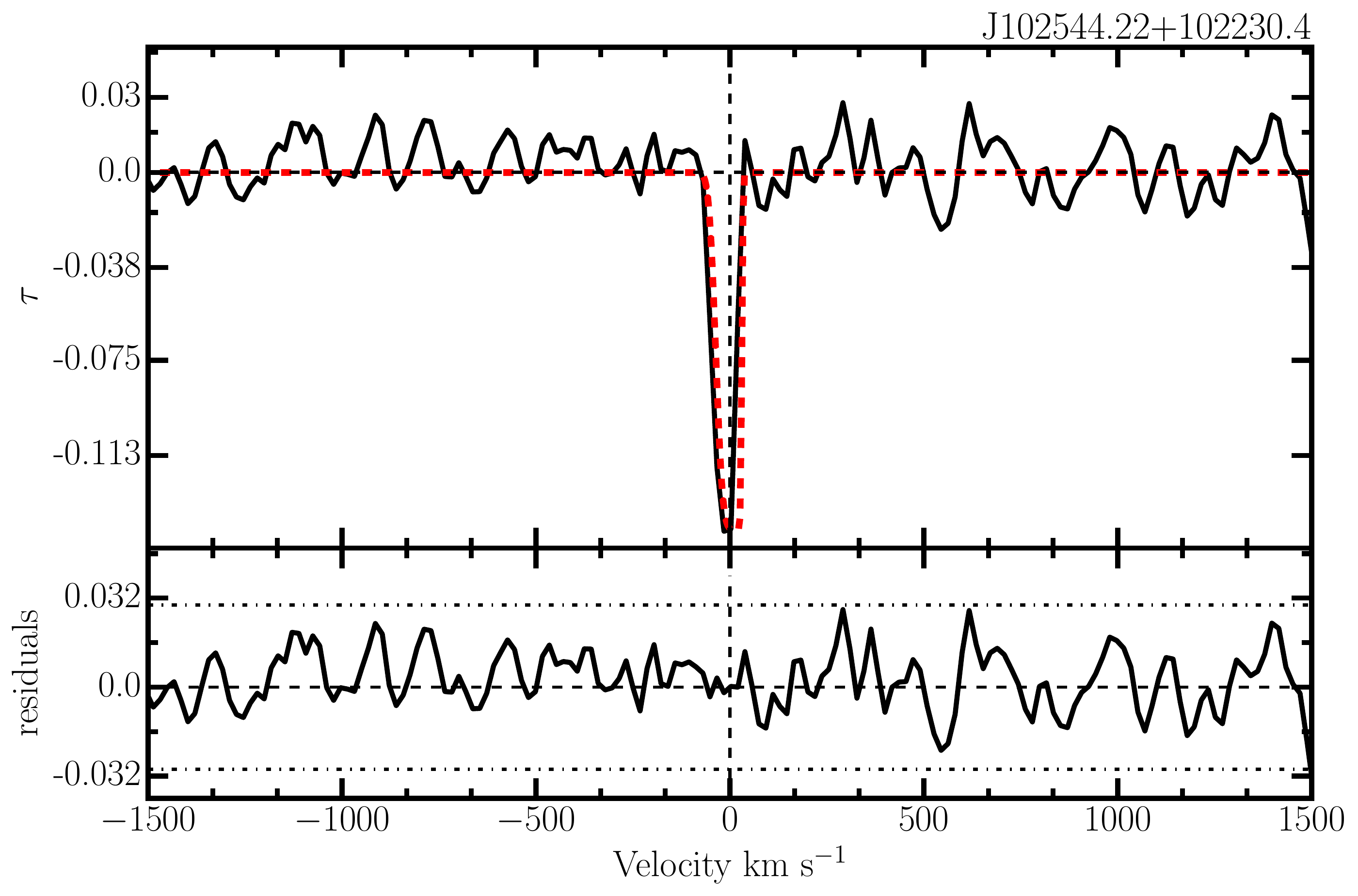}
                \includegraphics[trim = 0 0 0 0, clip,width=.33\textwidth]{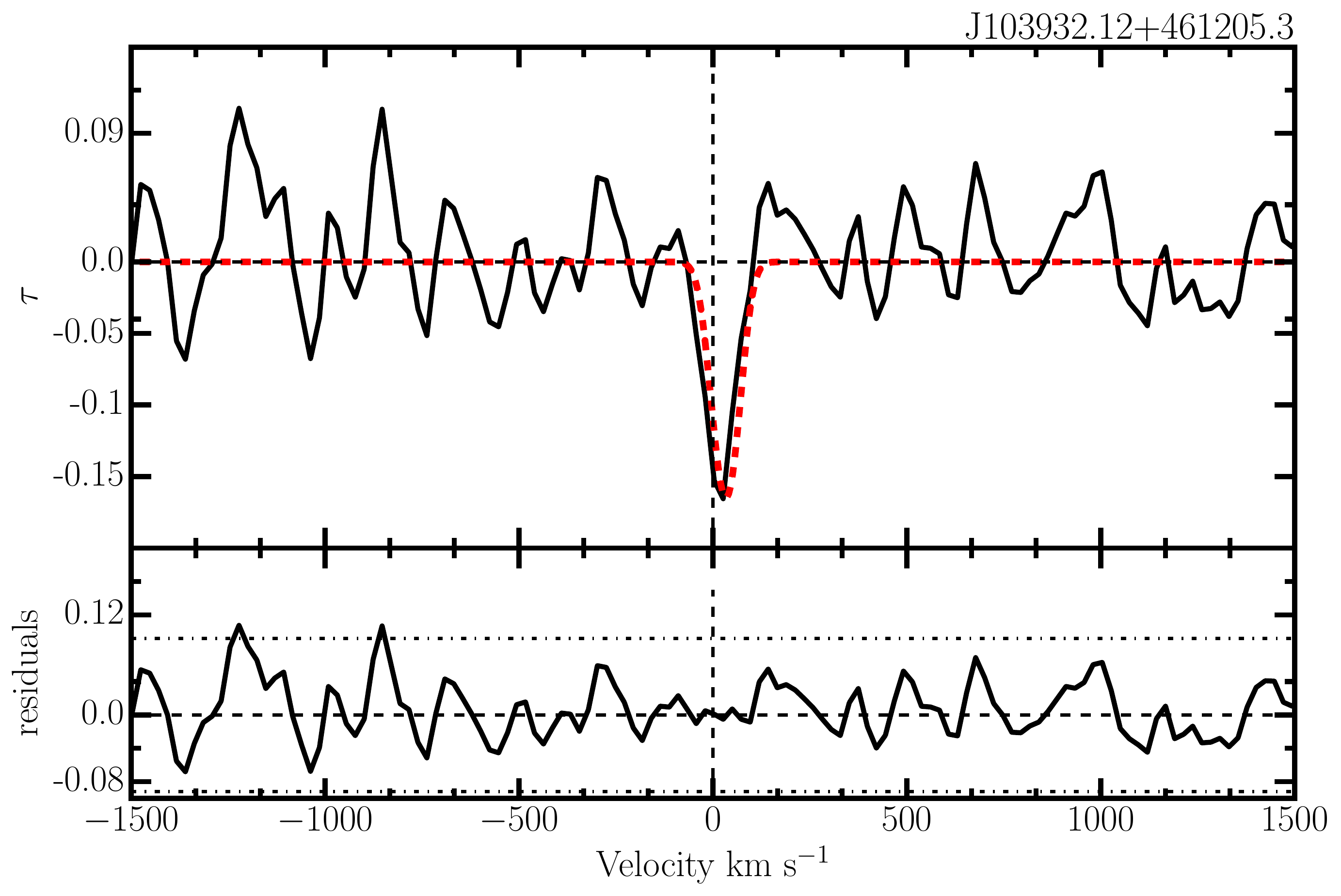}
                \includegraphics[trim = 0 0 0 0, clip,width=.33\textwidth]{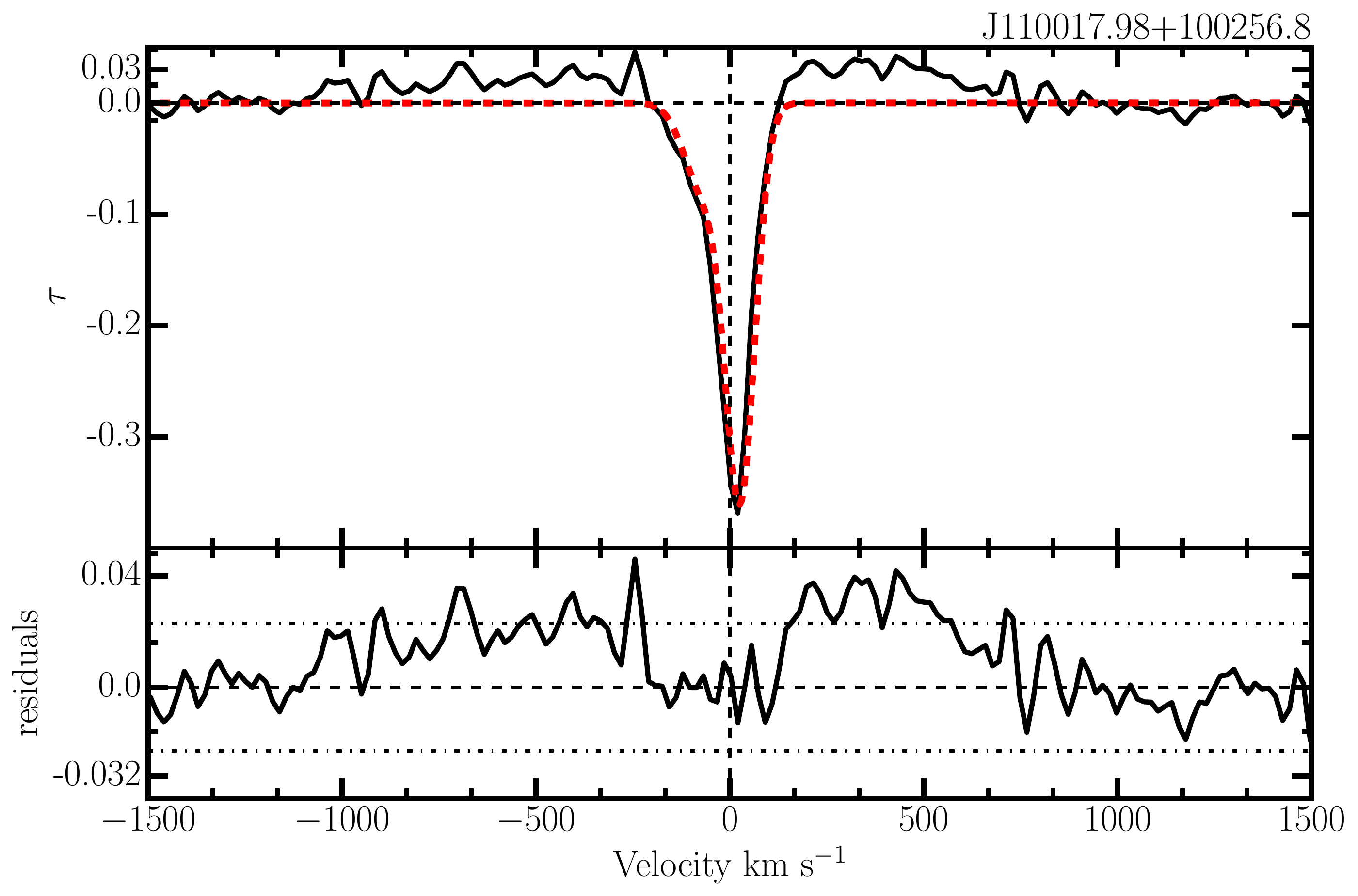}
                \includegraphics[trim = 0 0 0 0, clip,width=.33\textwidth]{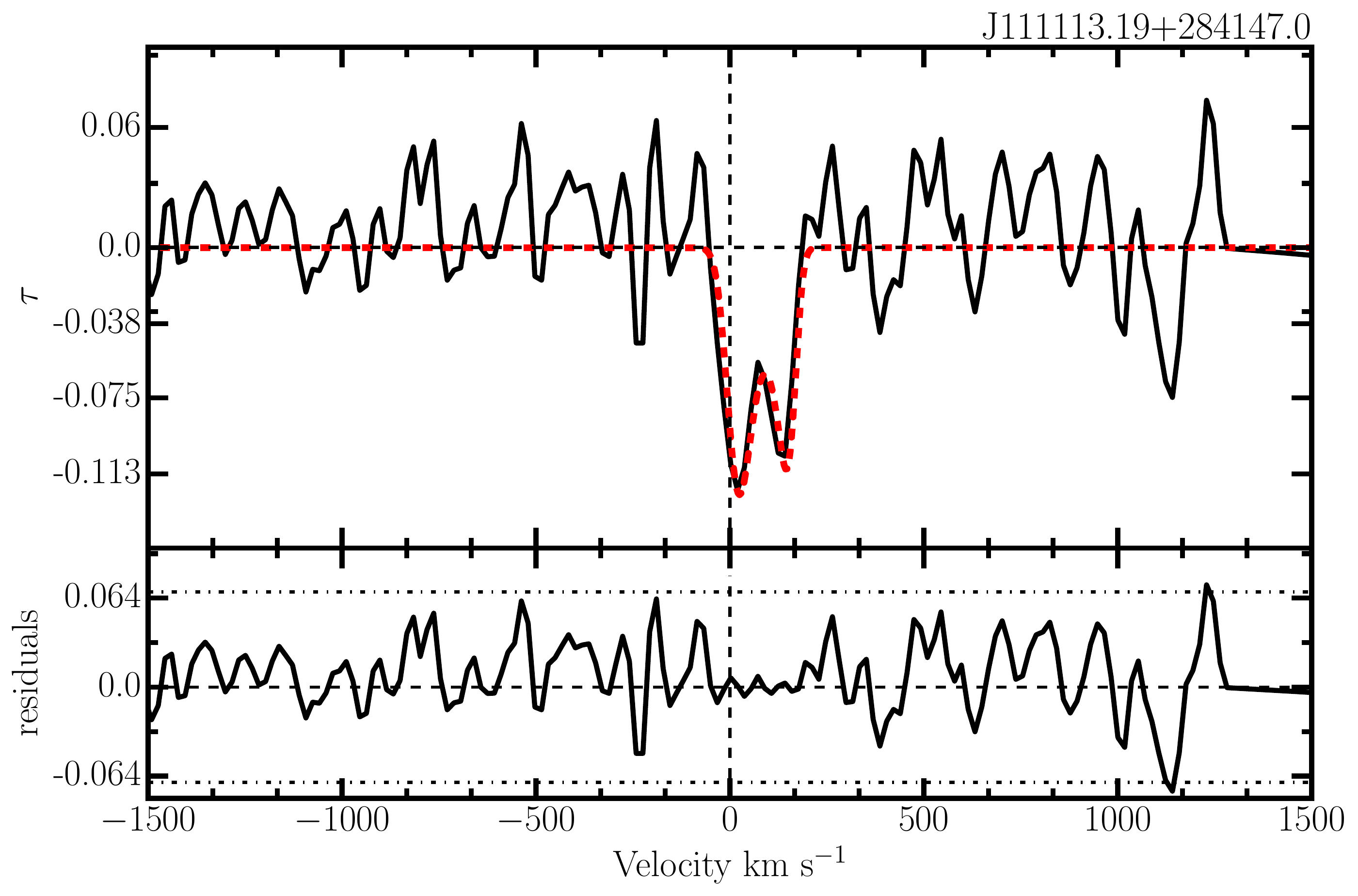}
                \includegraphics[trim = 0 0 0 0, clip,width=.33\textwidth]{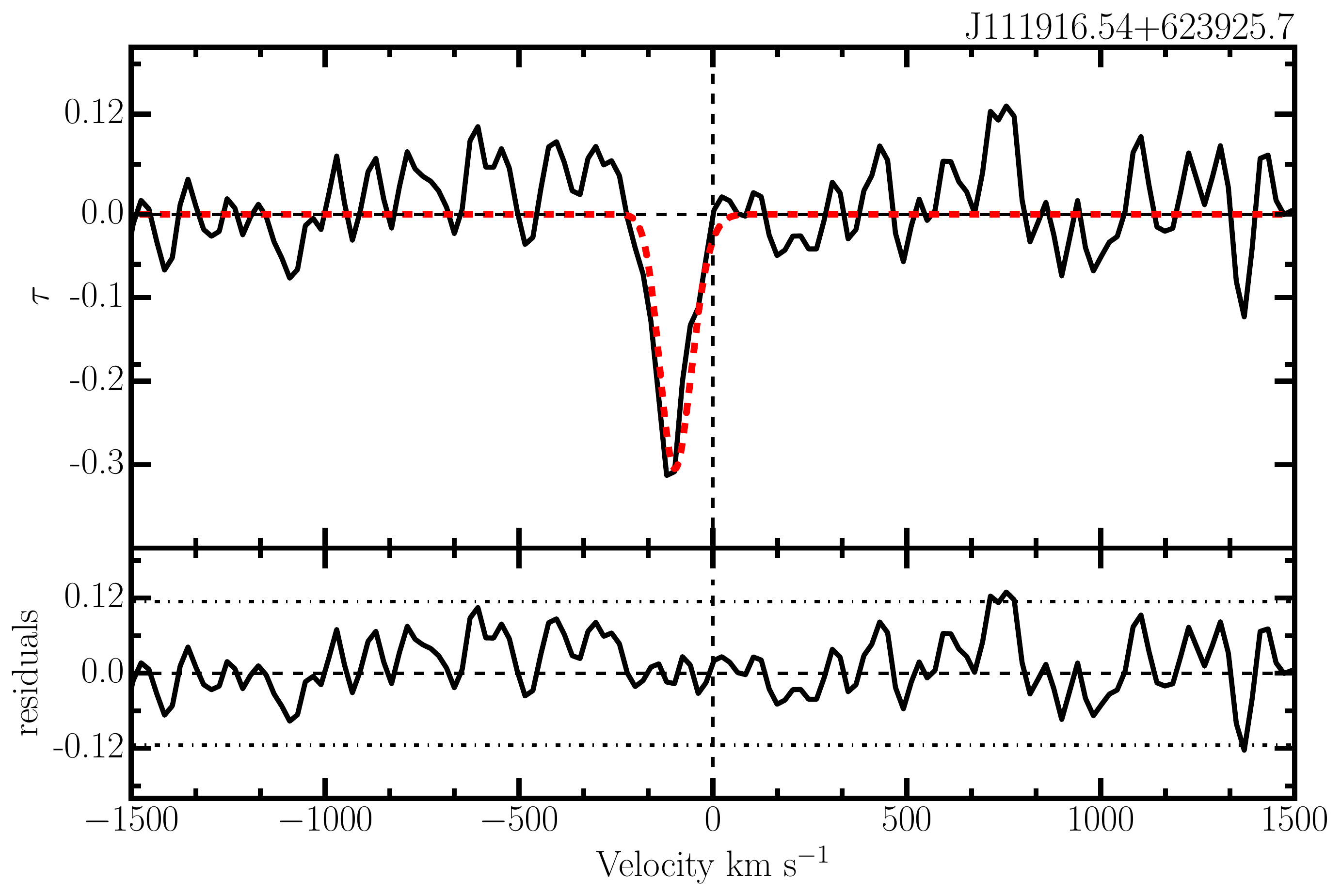}
                \includegraphics[trim = 0 0 0 0, clip,width=.33\textwidth]{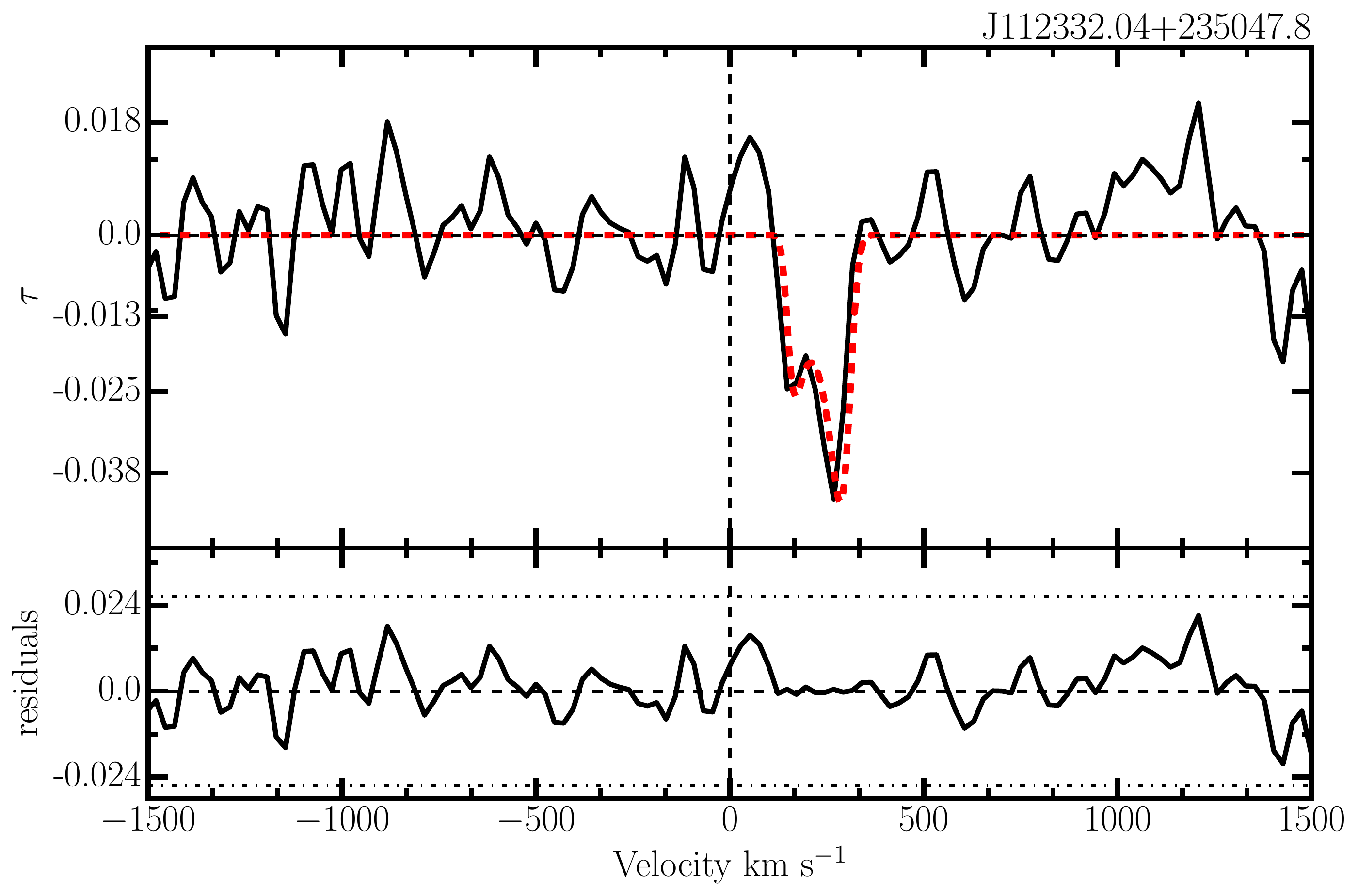}
                \includegraphics[trim = 0 0 0 0, clip,width=.33\textwidth]{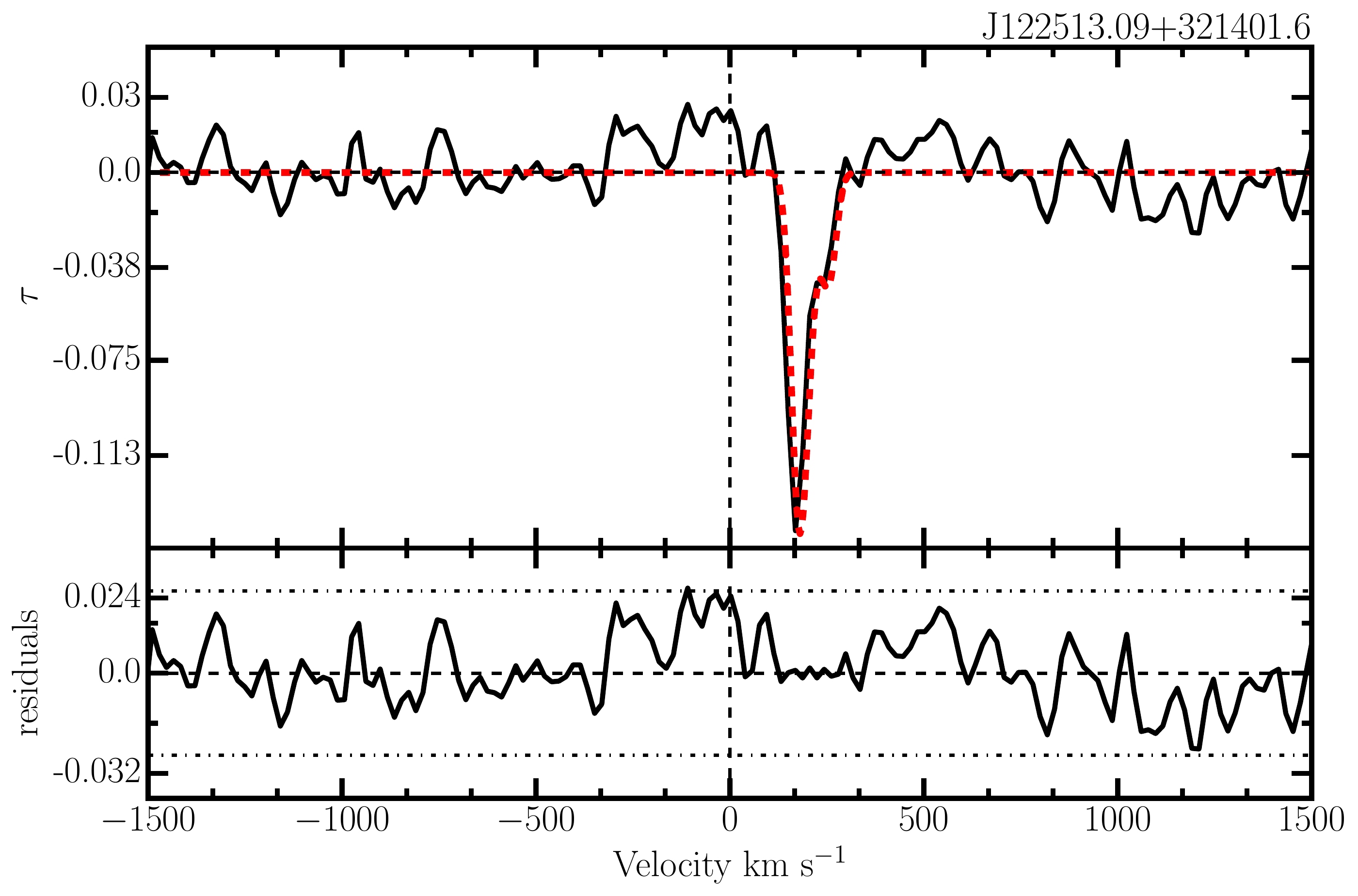}

                \caption{\HI\ absorption detections. The data are shown in black, the BF fit is shown in red. The residuals of the fit are plotted in the bottom panels along with the $\pm1$$\sigma$ noise level (horizontal dotted lines).}
\label{fig:Profiles1}
\end{center}
\end{figure*}

\begin{figure*}
\begin{center}
\includegraphics[trim = 0 0 0 0, clip,width=.33\textwidth]{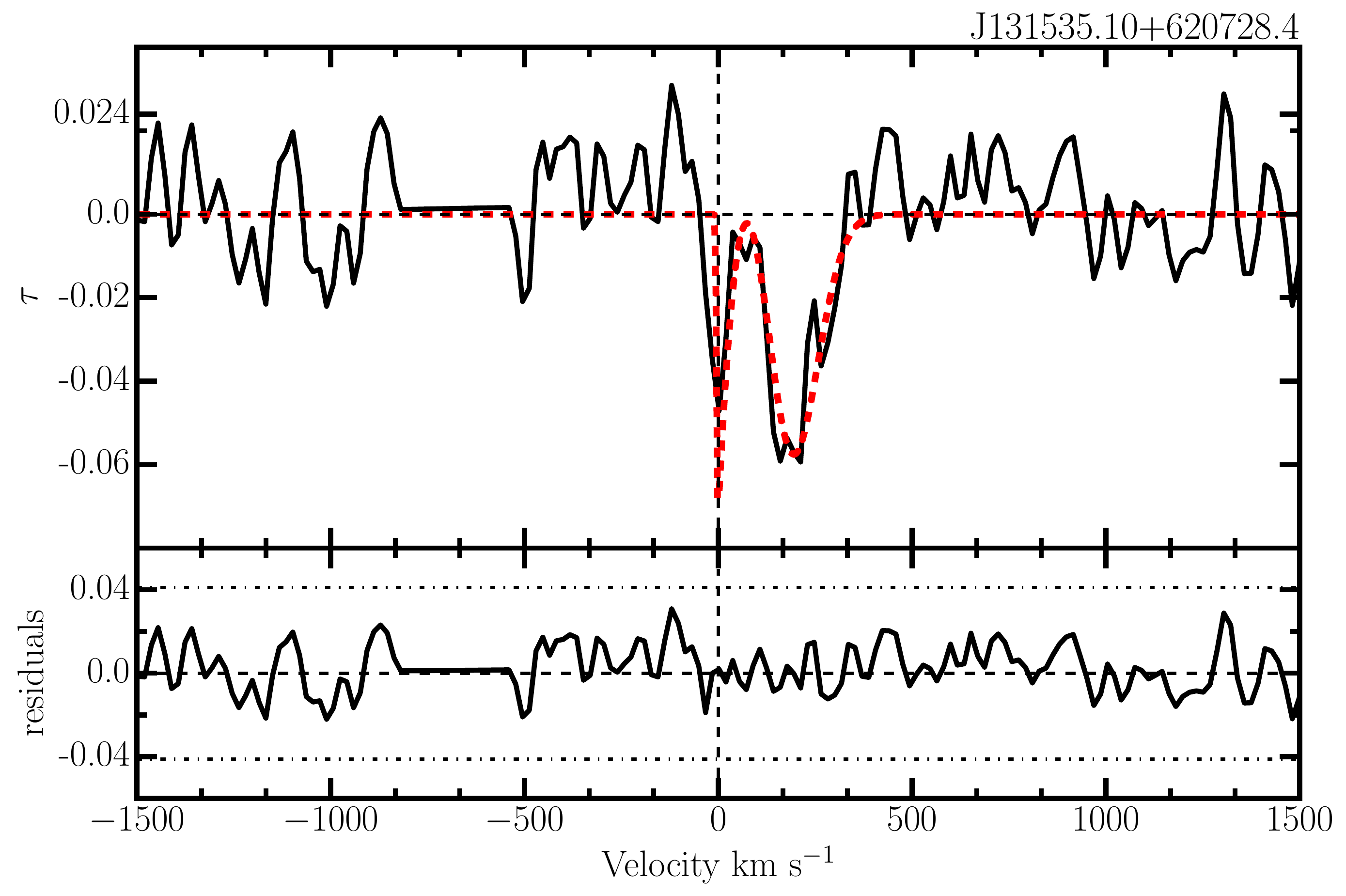}
                \includegraphics[trim = 0 0 0 0, clip,width=.33\textwidth]{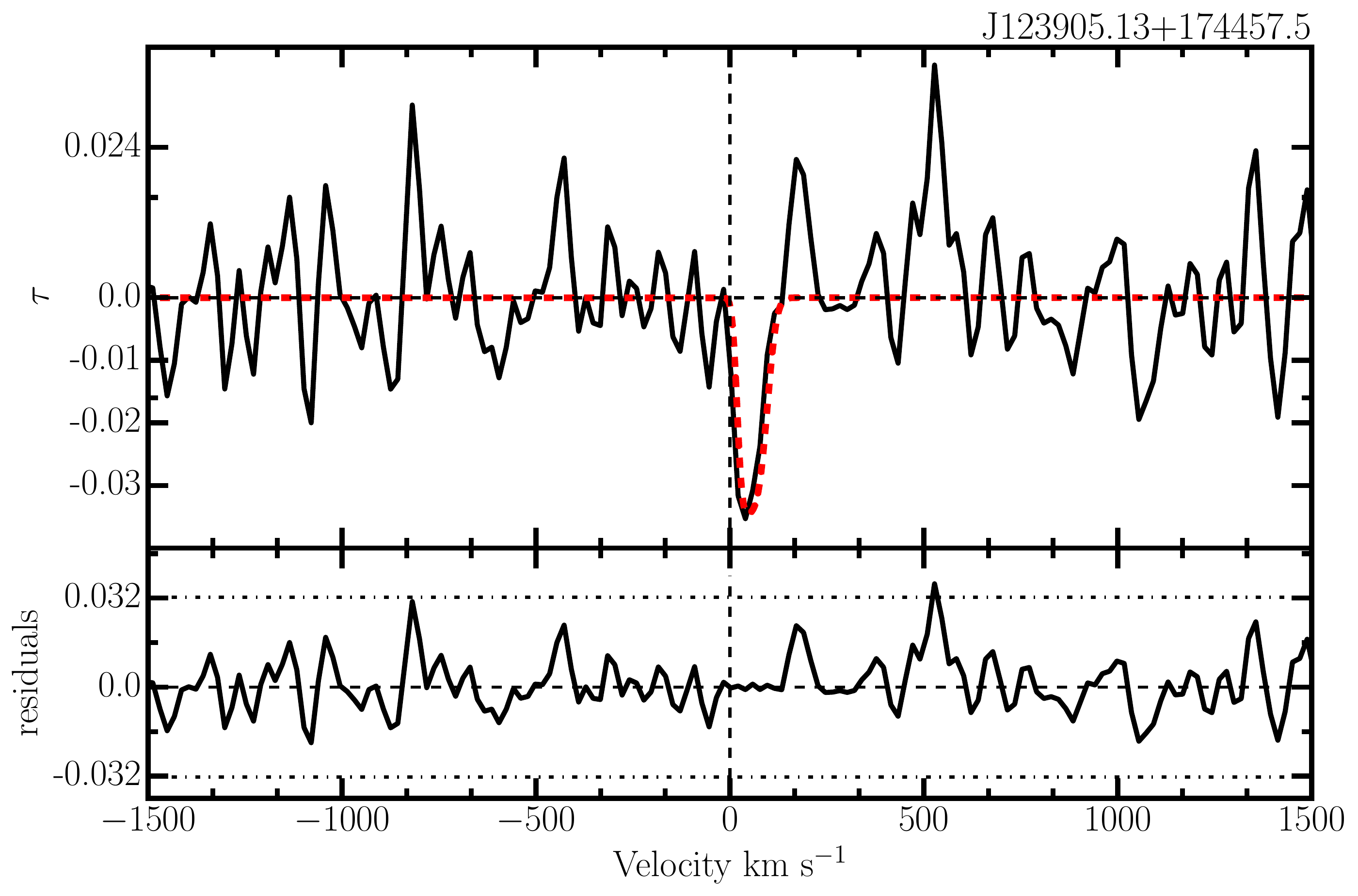}
                \includegraphics[trim = 0 0 0 0, clip,width=.33\textwidth]{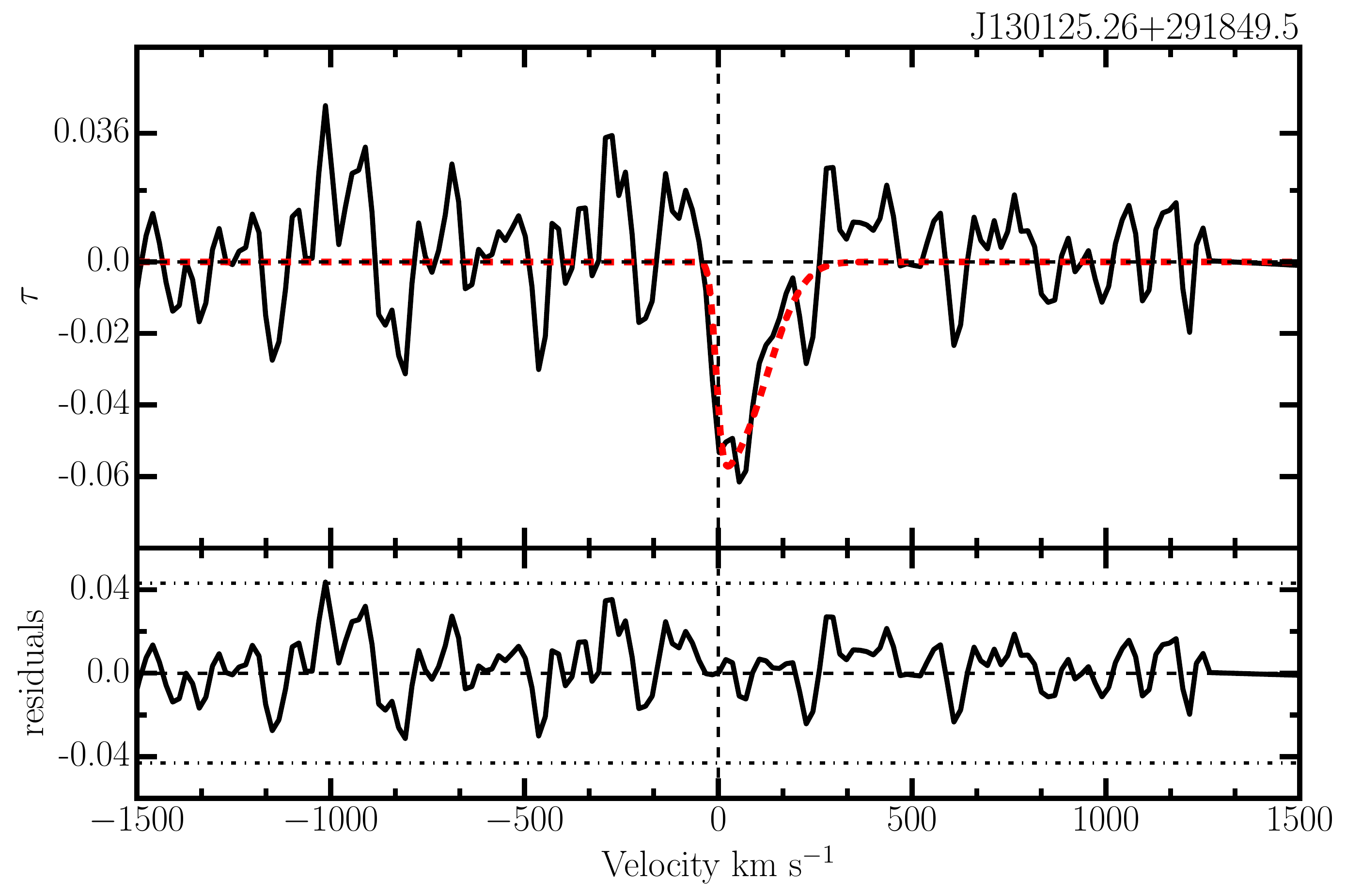}
                \includegraphics[trim = 0 0 0 0, clip,width=.33\textwidth]{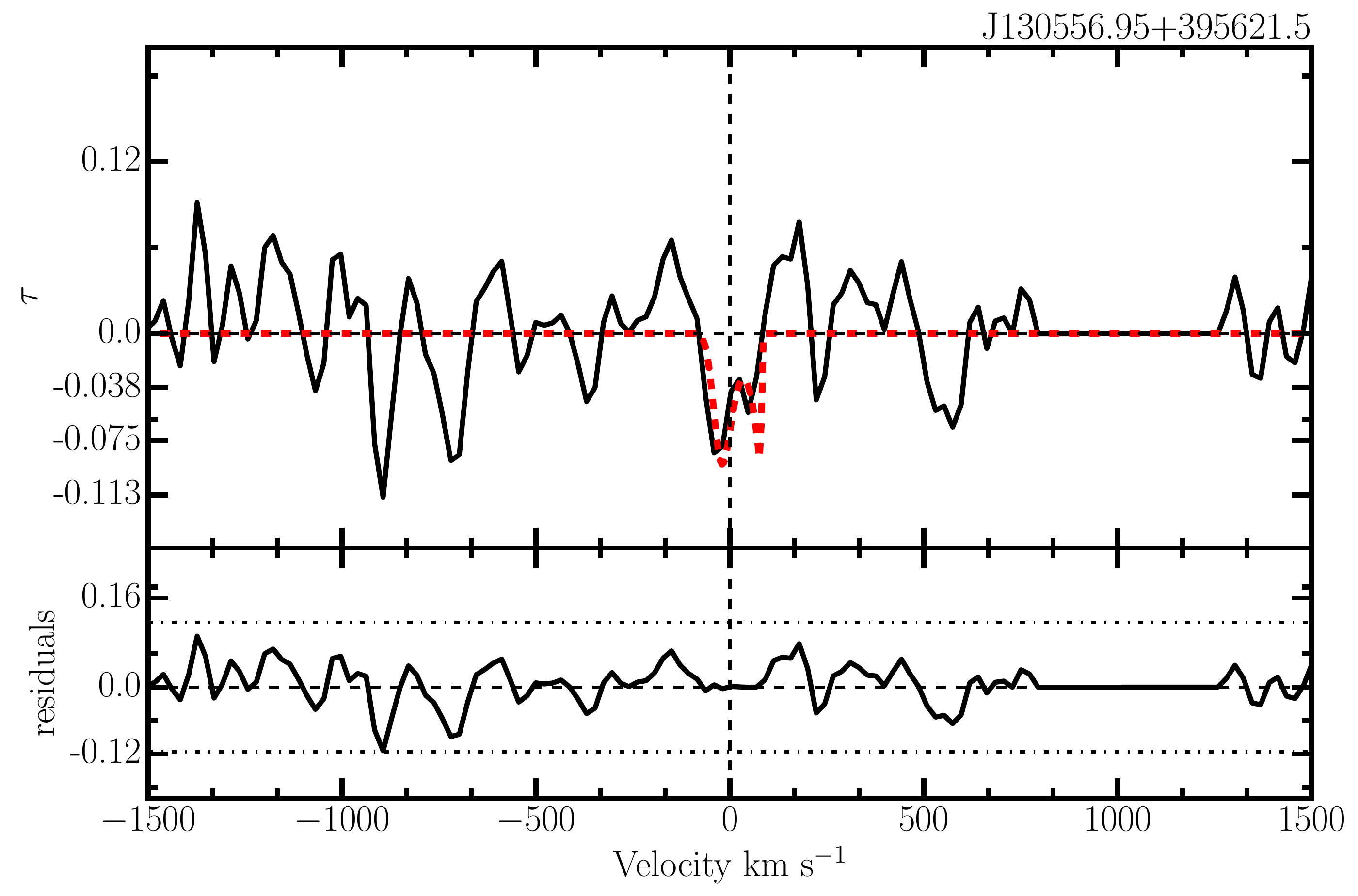}
                \includegraphics[trim = 0 0 0 0, clip,width=.33\textwidth]{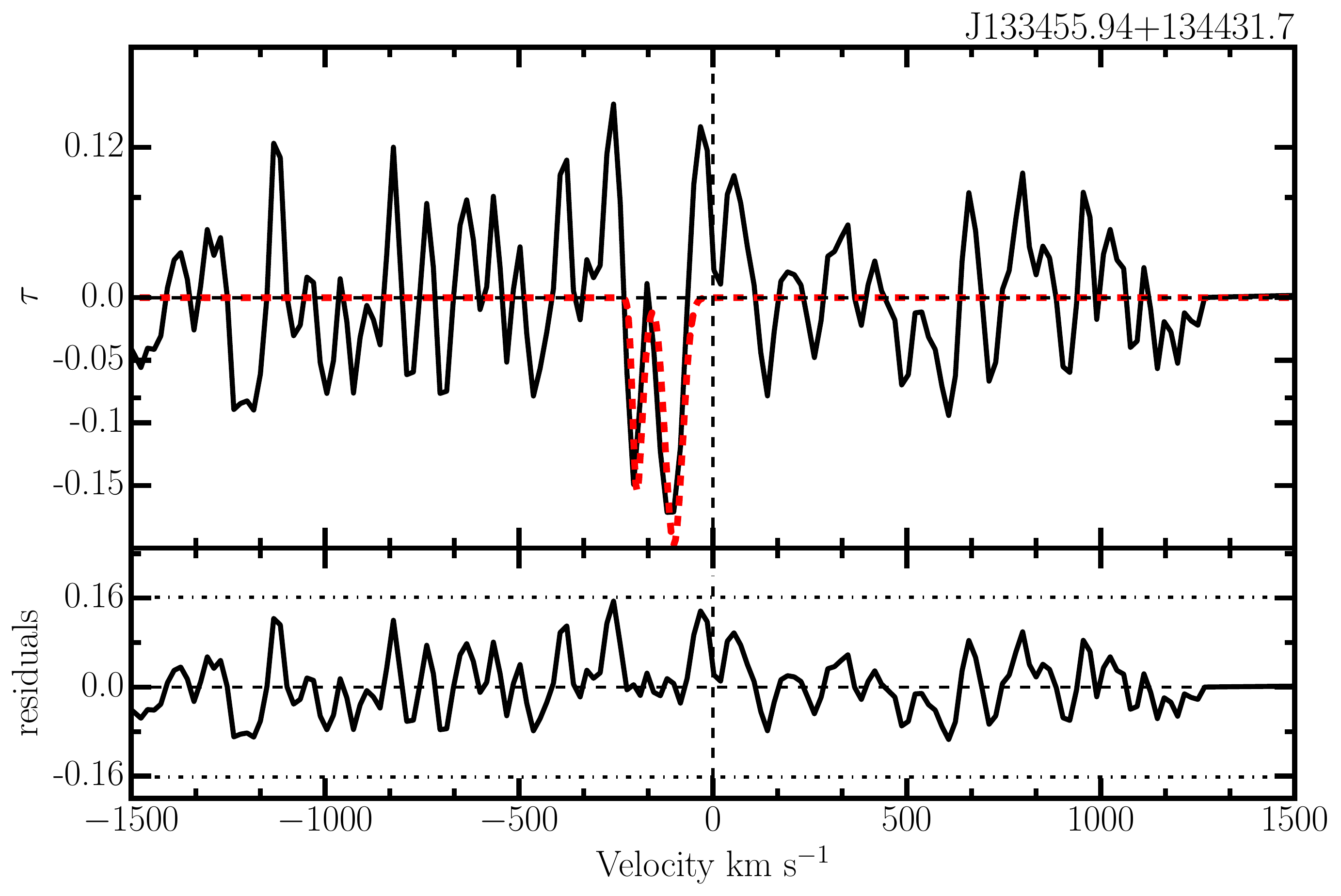}
                \includegraphics[trim = 0 0 0 0, clip,width=.33\textwidth]{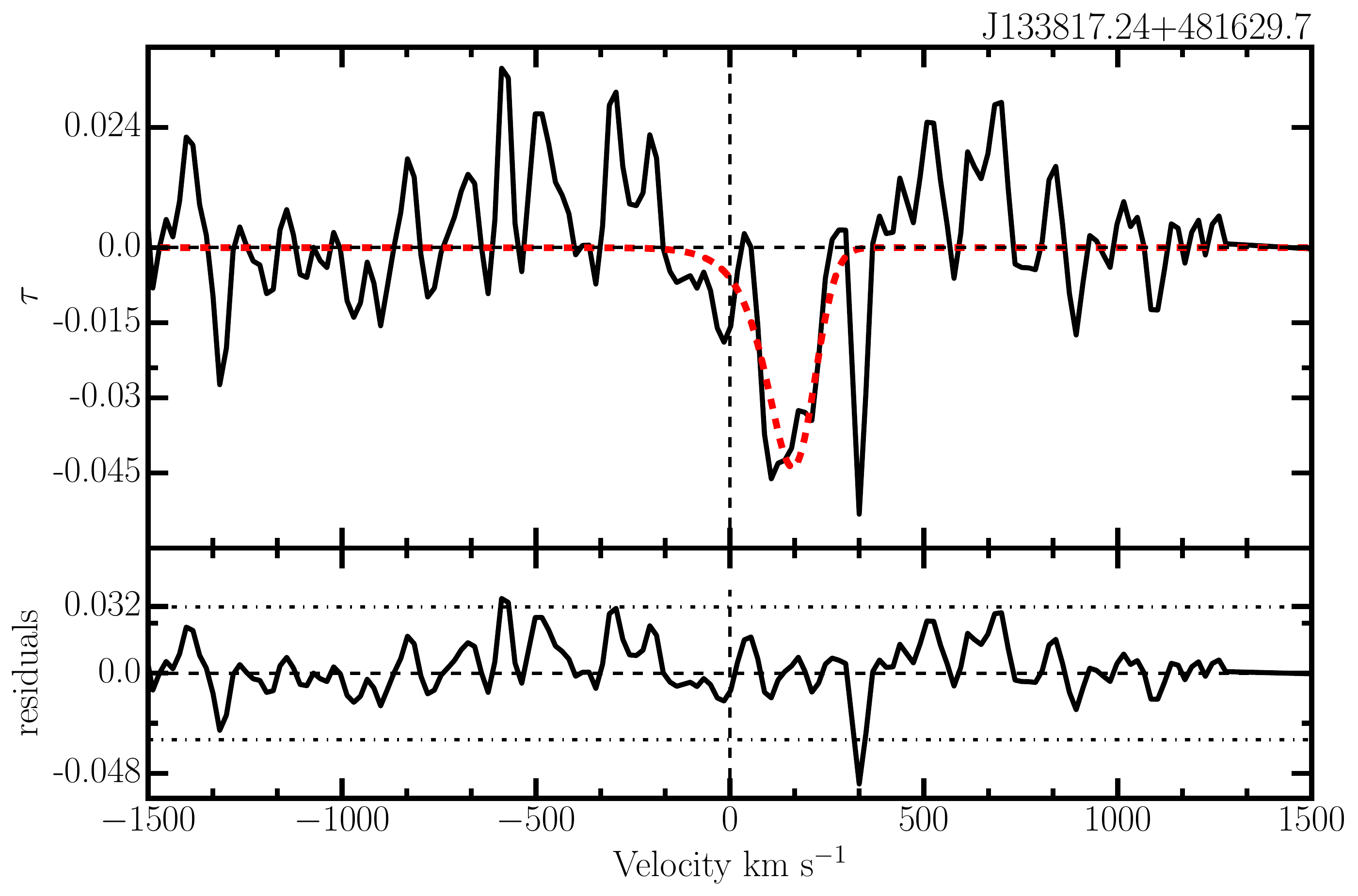}
                \includegraphics[trim = 0 0 0 0, clip,width=.33\textwidth]{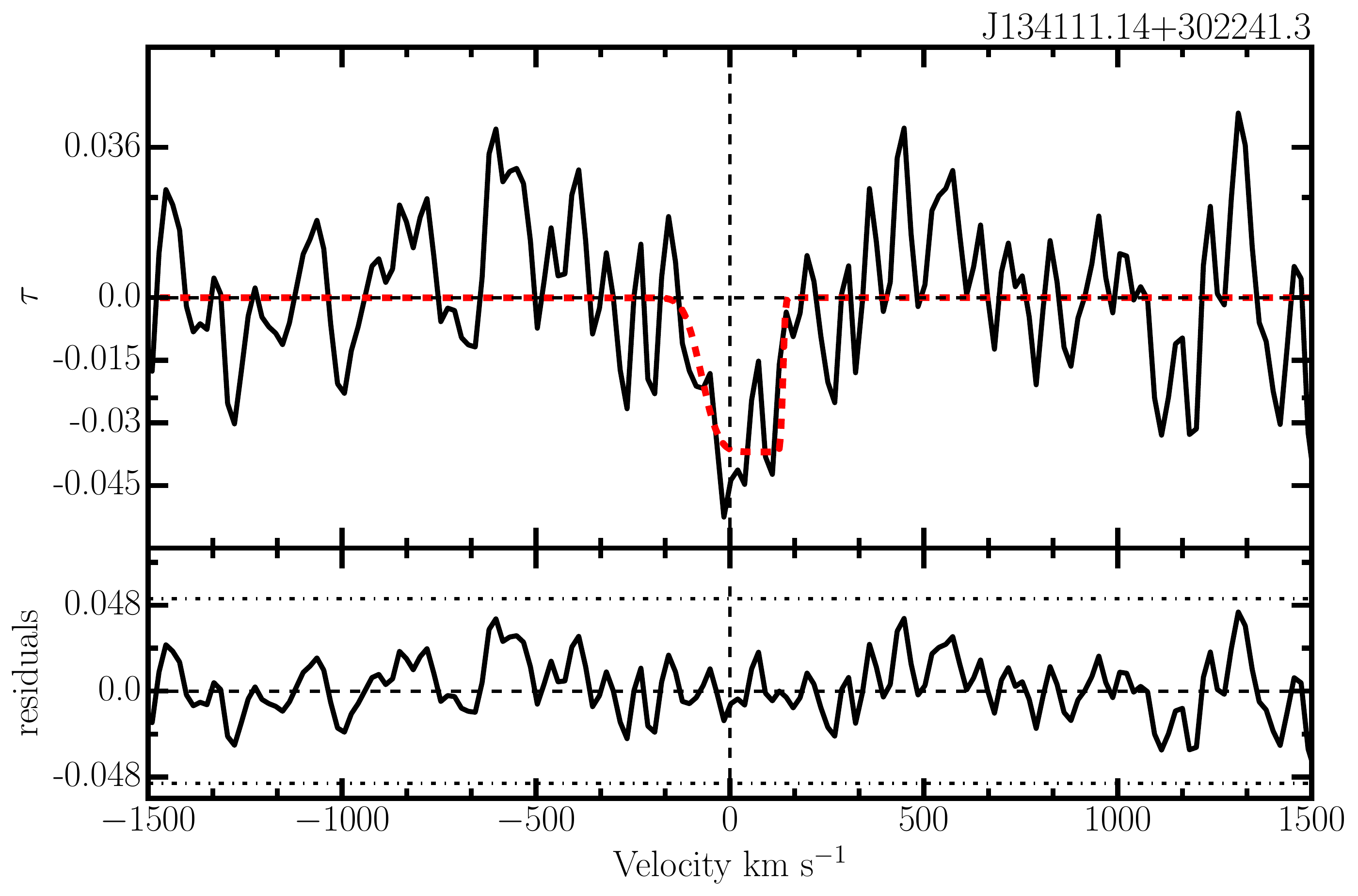}
                \includegraphics[trim = 0 0 0 0, clip,width=.33\textwidth]{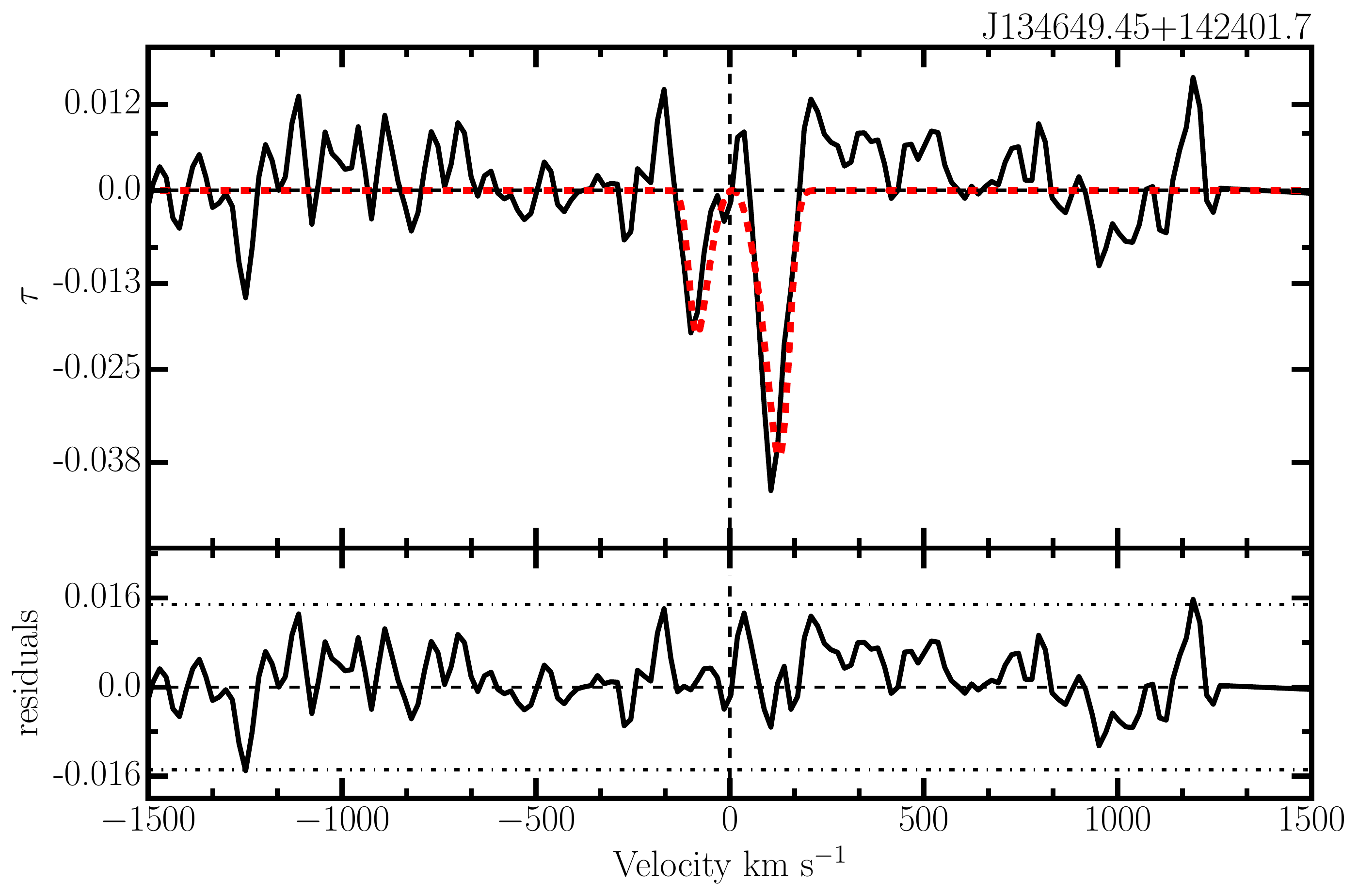}
                \includegraphics[trim = 0 0 0 0, clip,width=.33\textwidth]{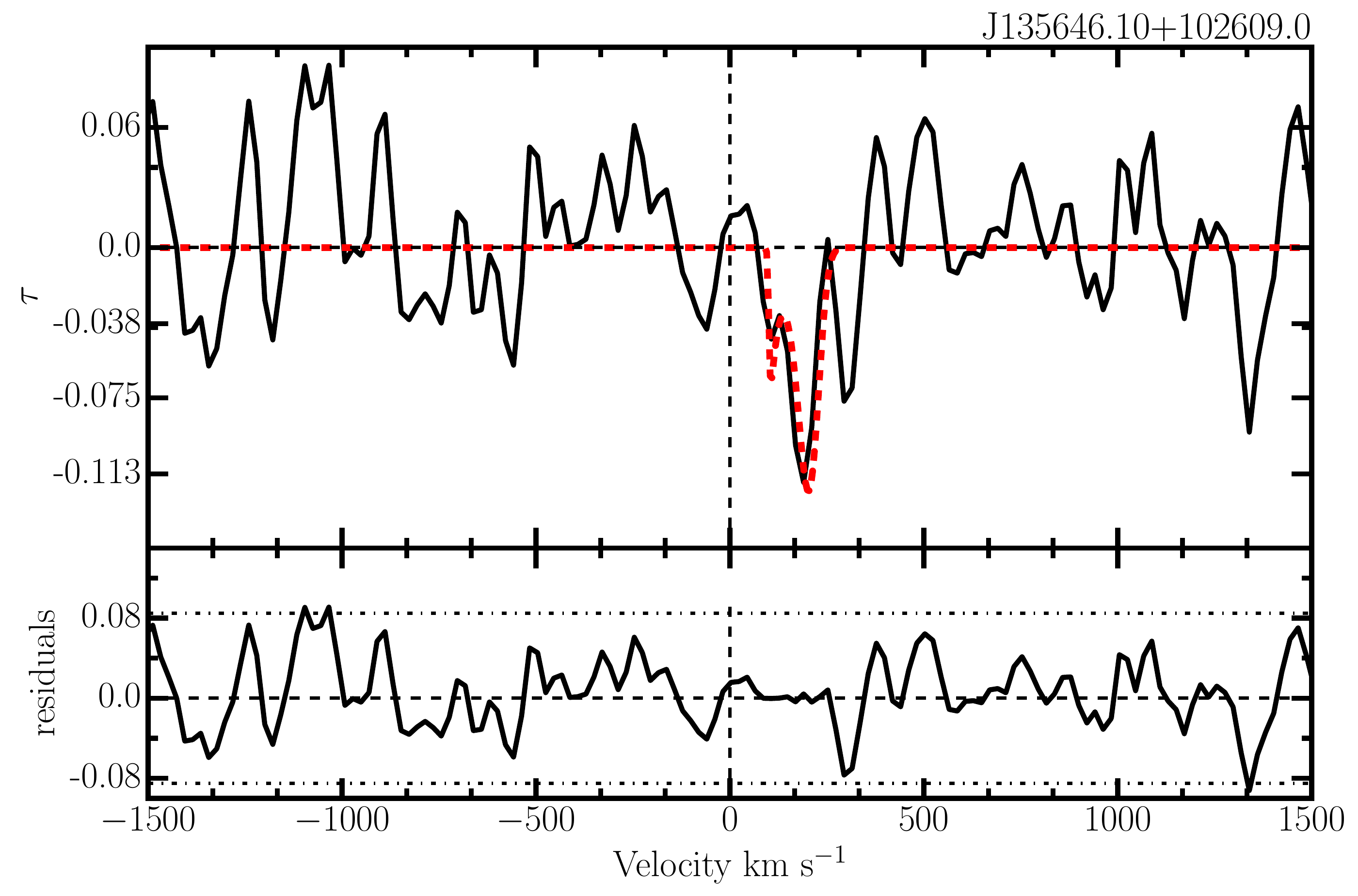}
                \includegraphics[trim = 0 0 0 0, clip,width=.33\textwidth]{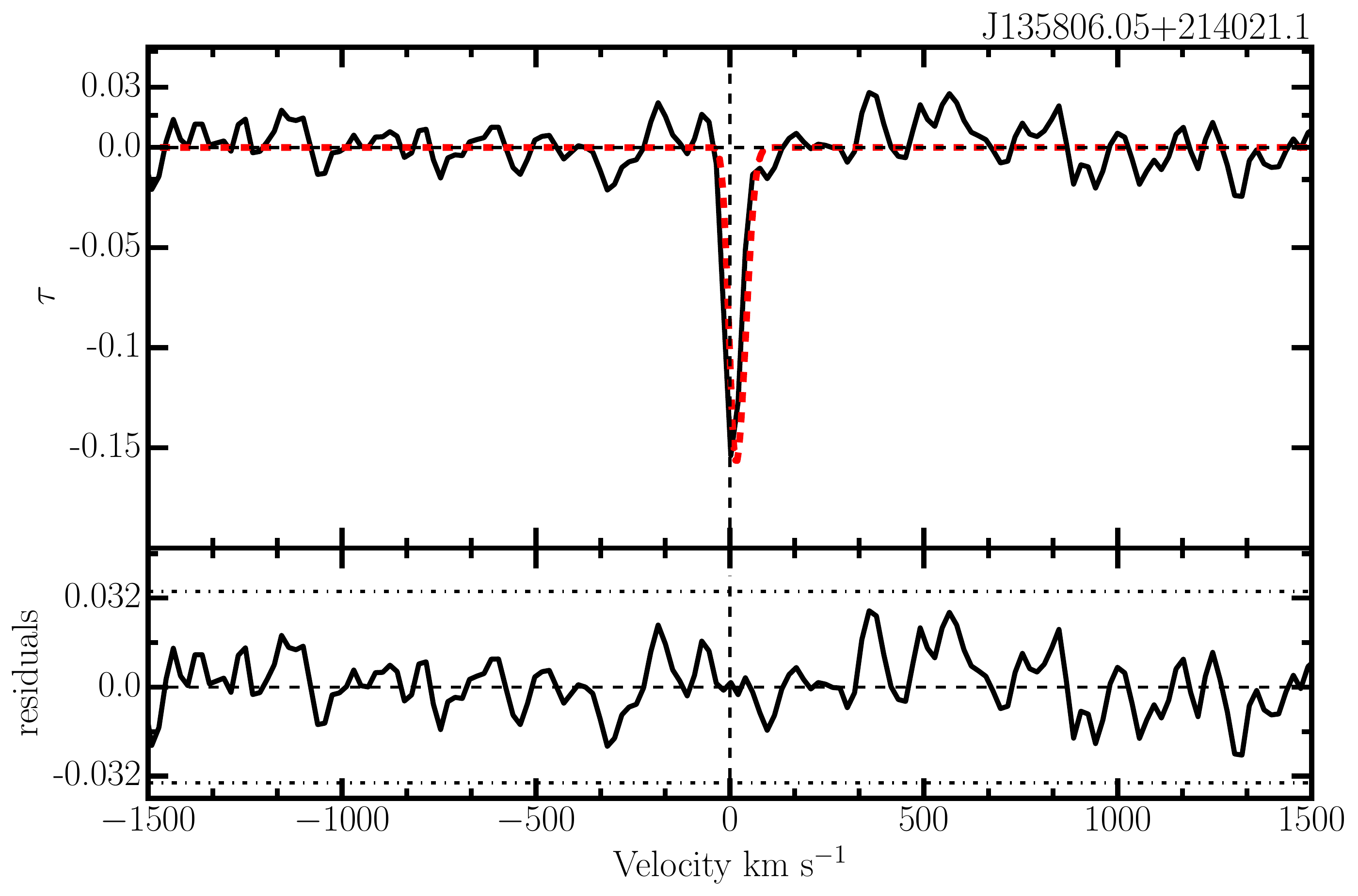}
                \includegraphics[trim = 0 0 0 0, clip,width=.33\textwidth]{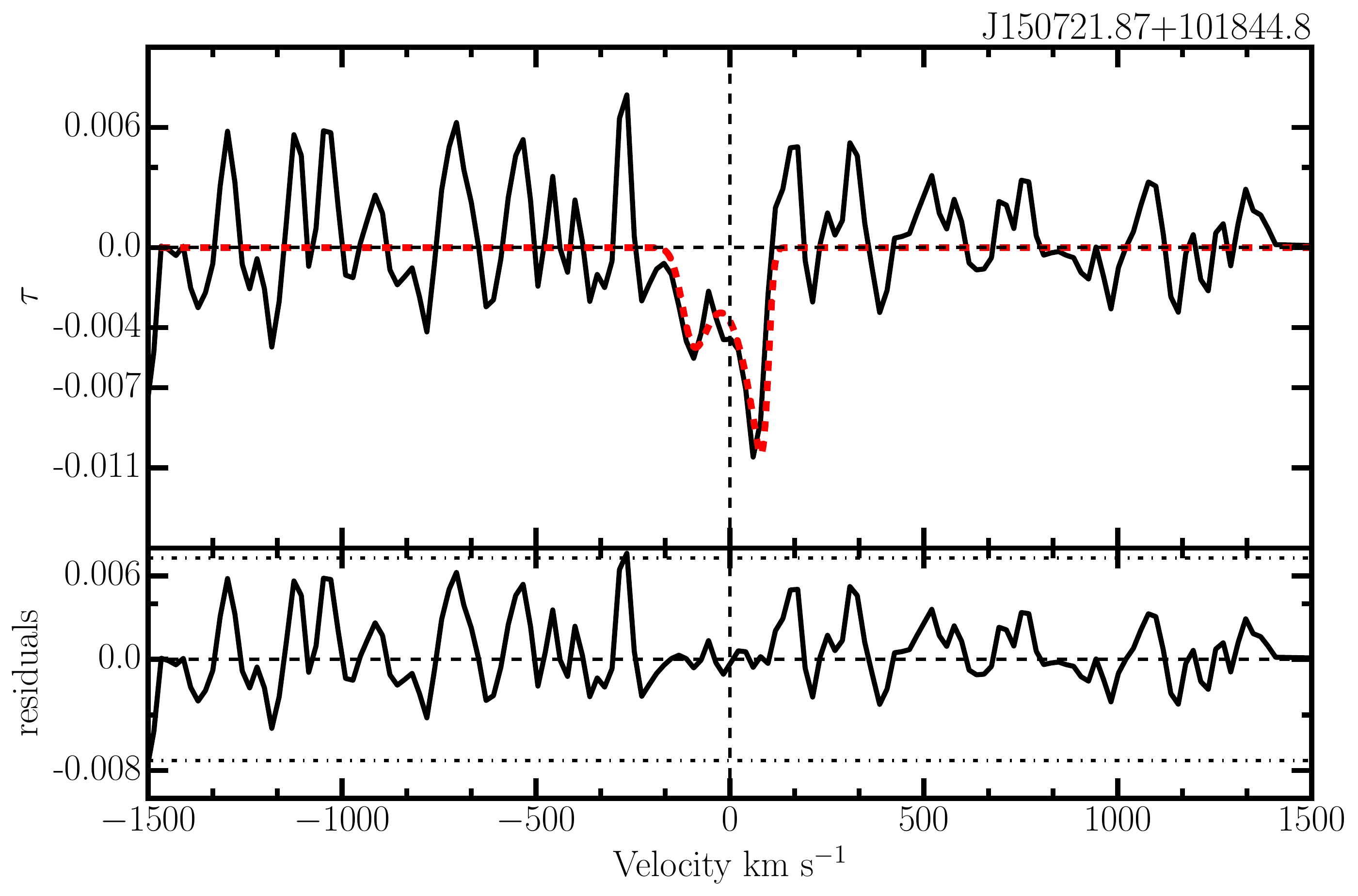}
                \includegraphics[trim = 0 0 0 0, clip,width=.33\textwidth]{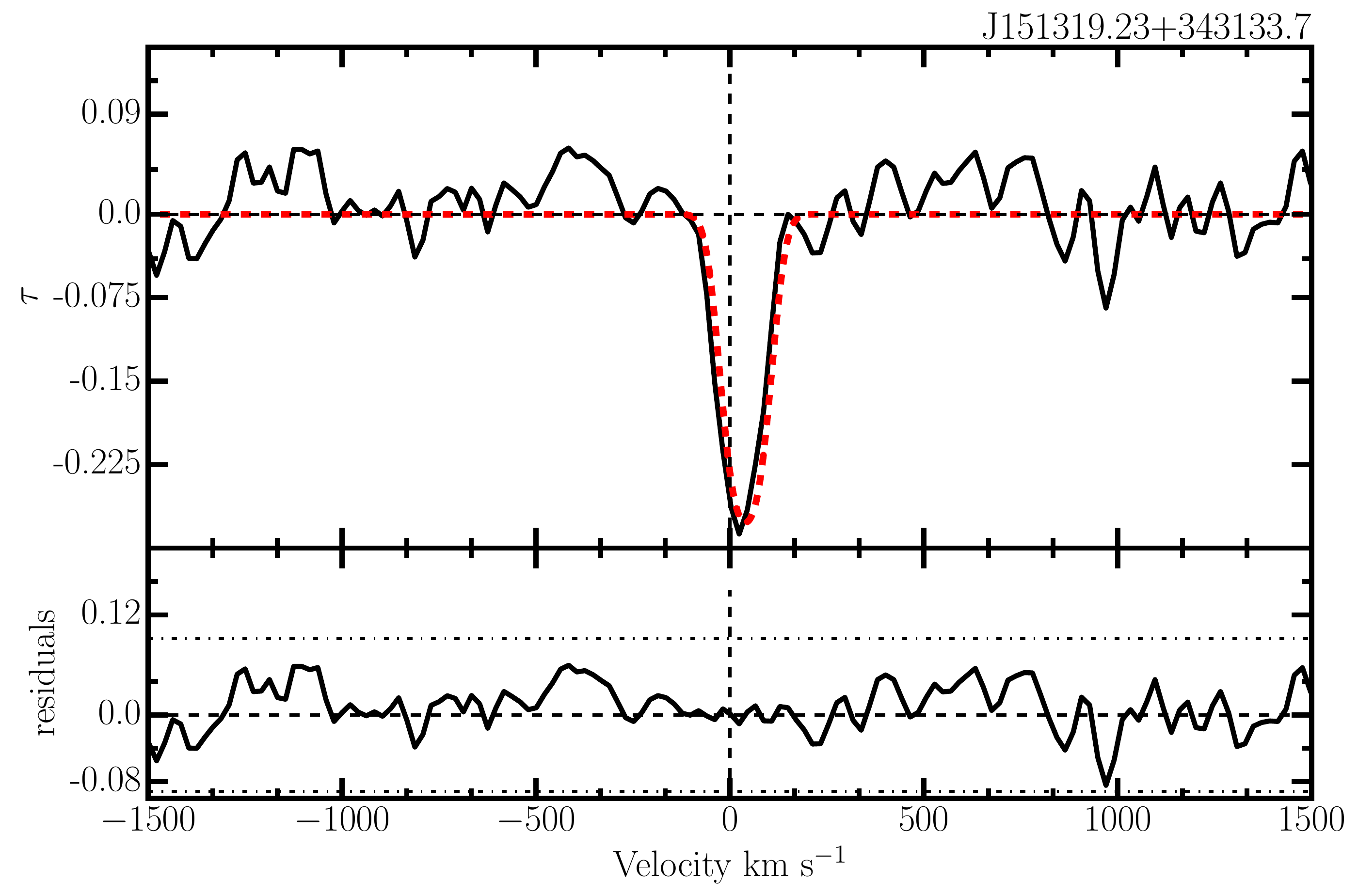}
                \includegraphics[trim = 0 0 0 0, clip,width=.33\textwidth]{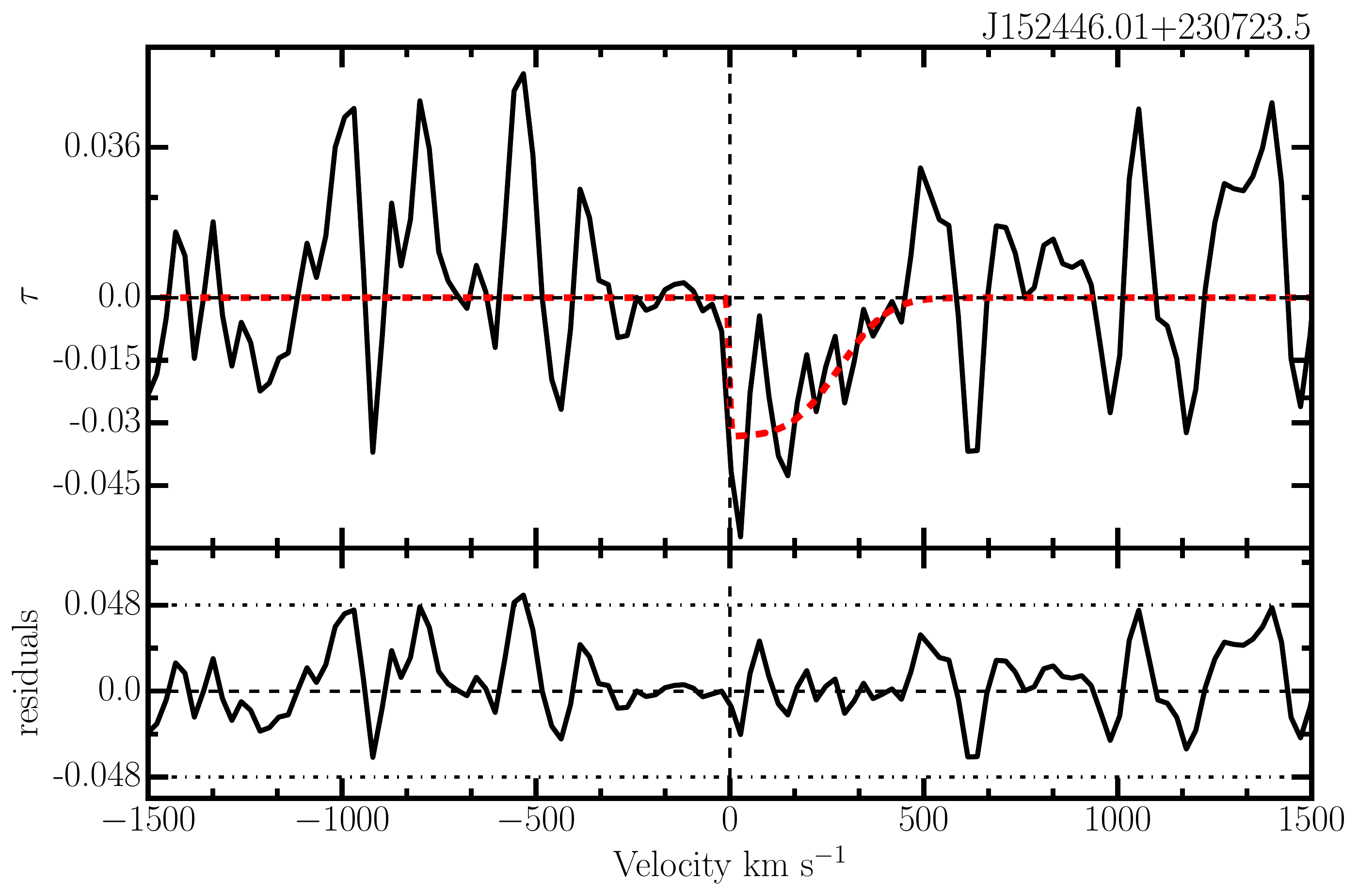}
                \includegraphics[trim = 0 0 0 0, clip,width=.33\textwidth]{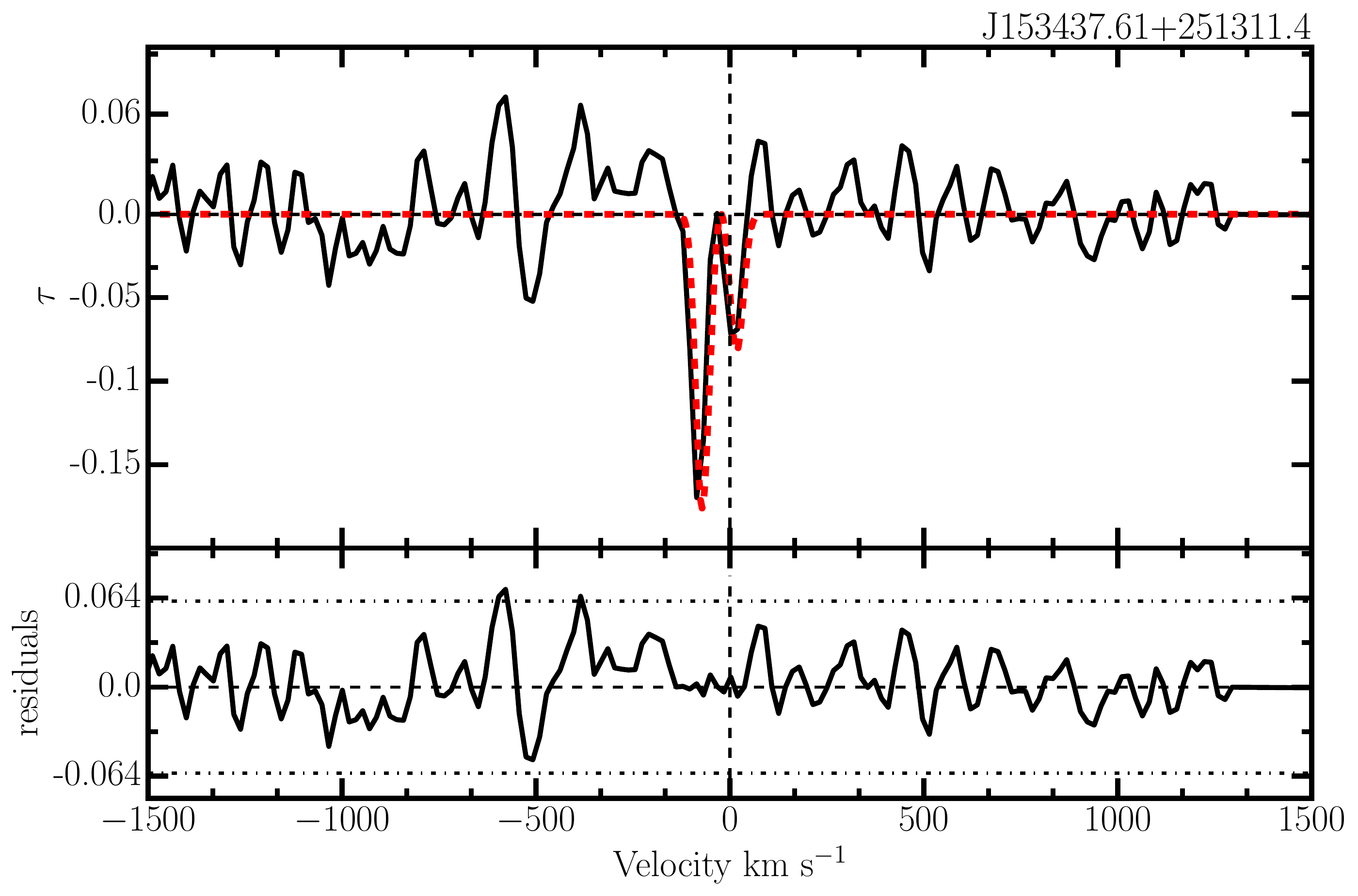}

                \caption{same as Fig.~\ref{fig:Profiles1}.}

\label{fig:Profiles2}
\end{center}
\end{figure*}

\begin{figure*}
\begin{center}
                \includegraphics[trim = 0 0 0 0, clip,width=.33\textwidth]{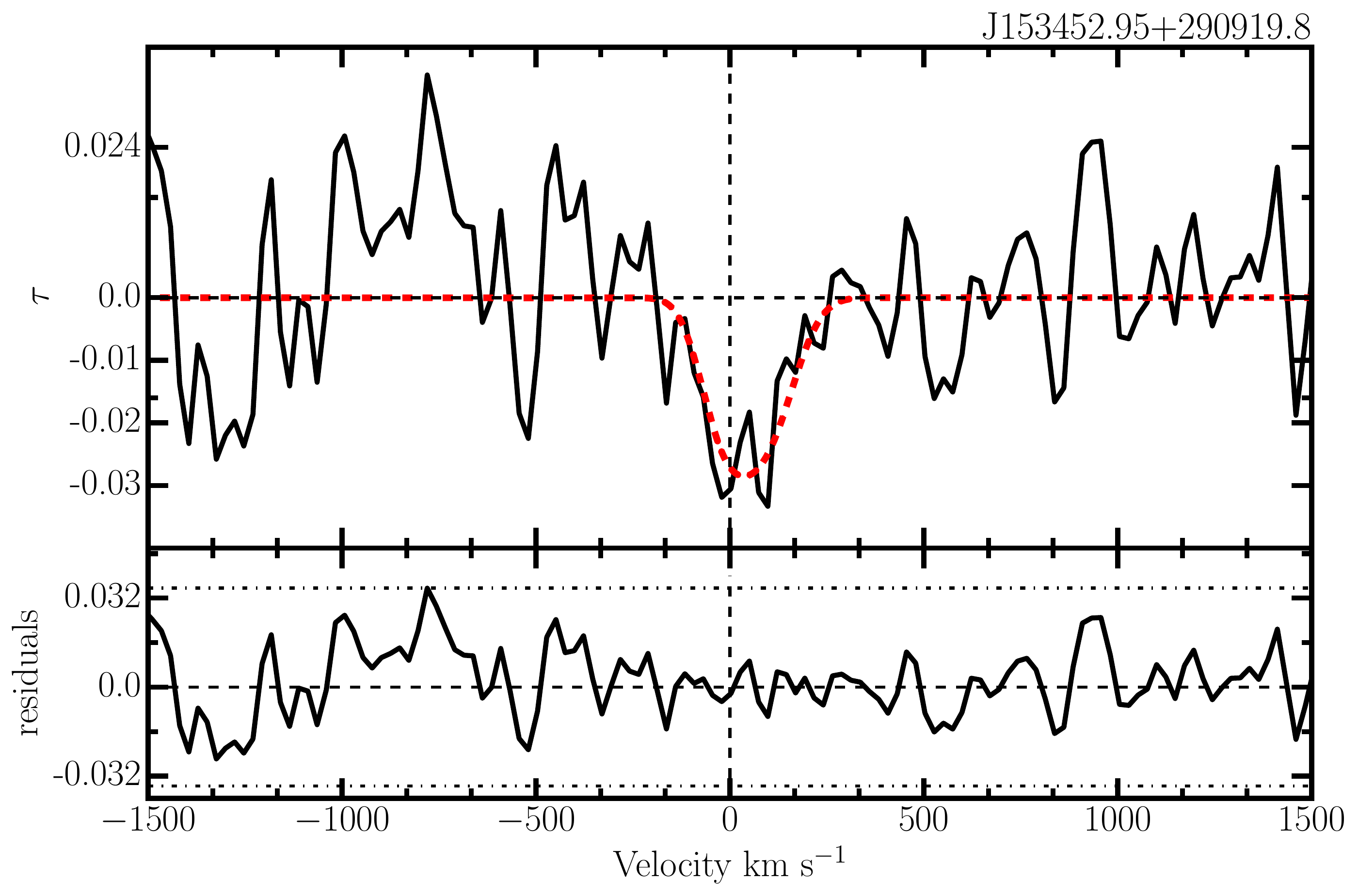}
                \includegraphics[trim = 0 0 0 0, clip,width=.33\textwidth]{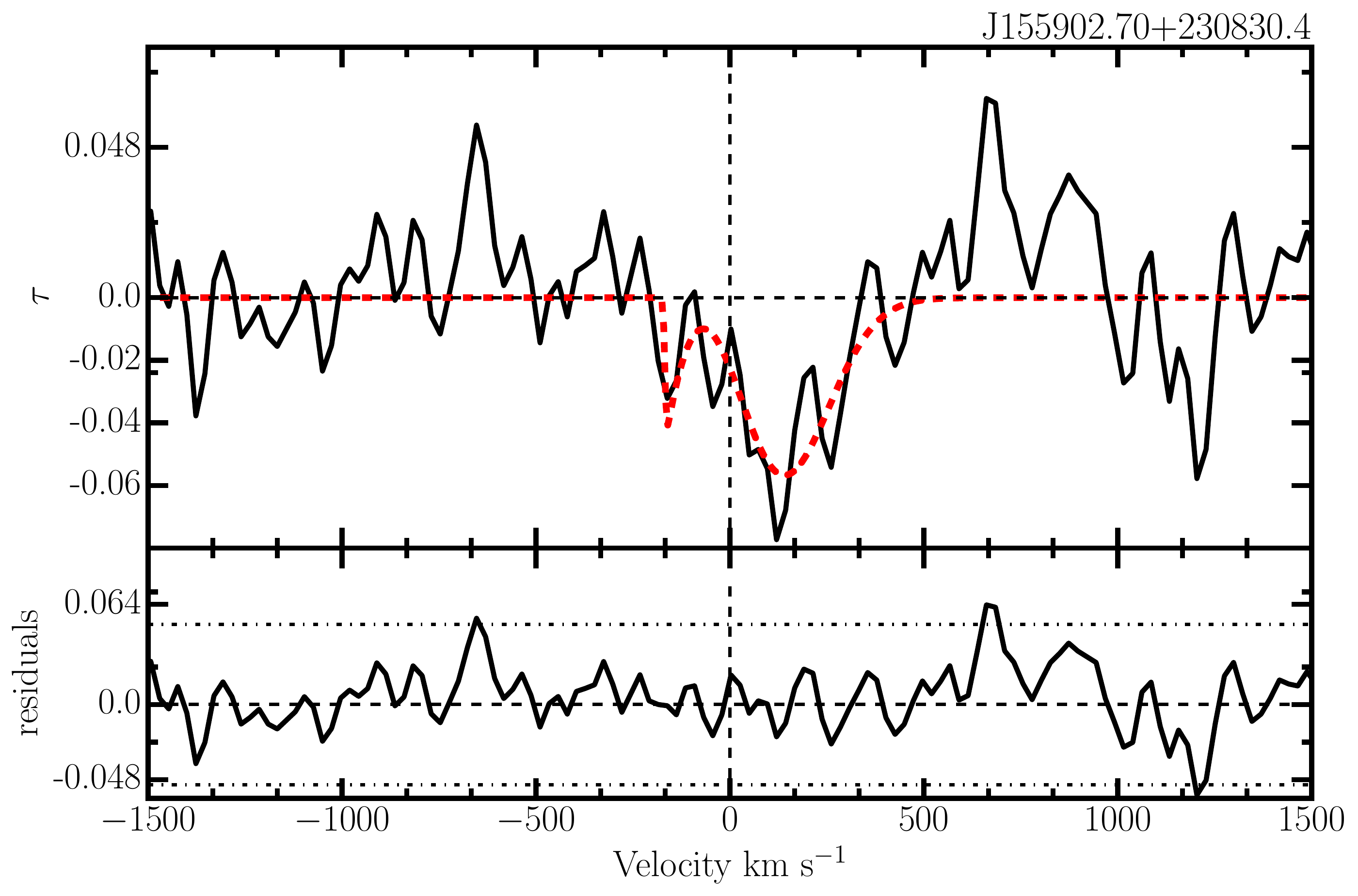}
                \includegraphics[trim = 0 0 0 0, clip,width=.33\textwidth]{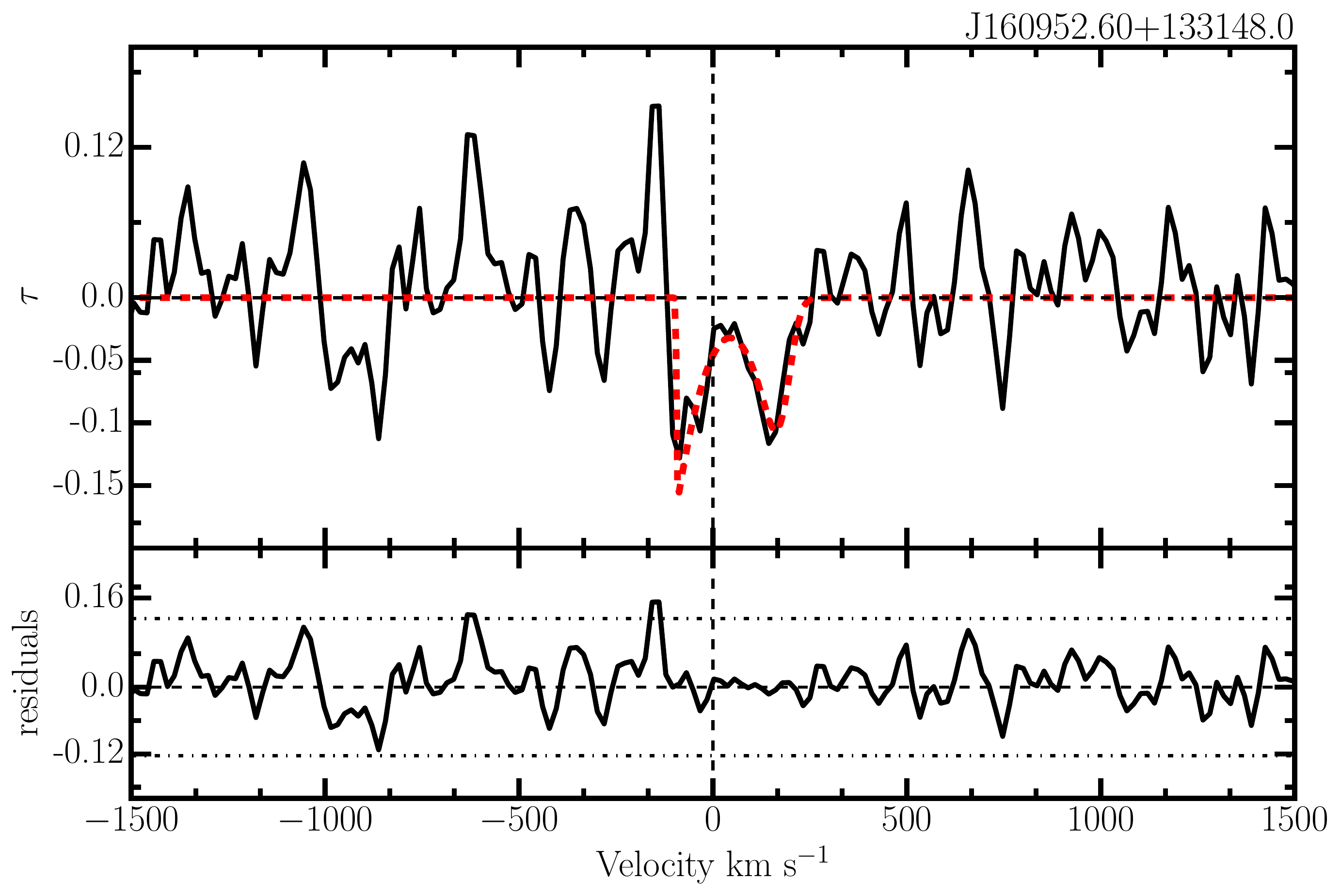}
                \includegraphics[trim = 0 0 0 0, clip,width=.33\textwidth]{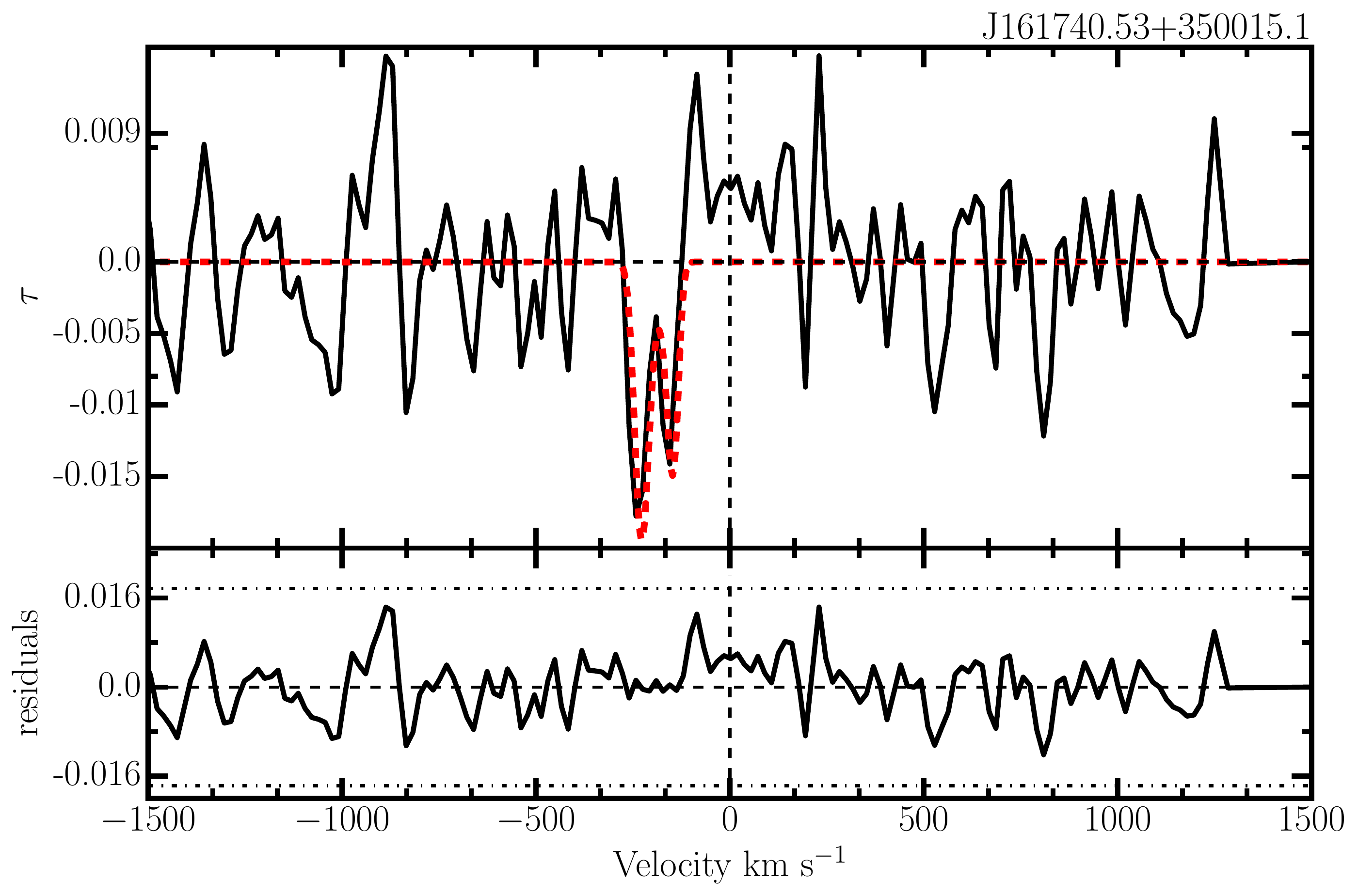}
                \includegraphics[trim = 0 0 0 0, clip,width=.33\textwidth]{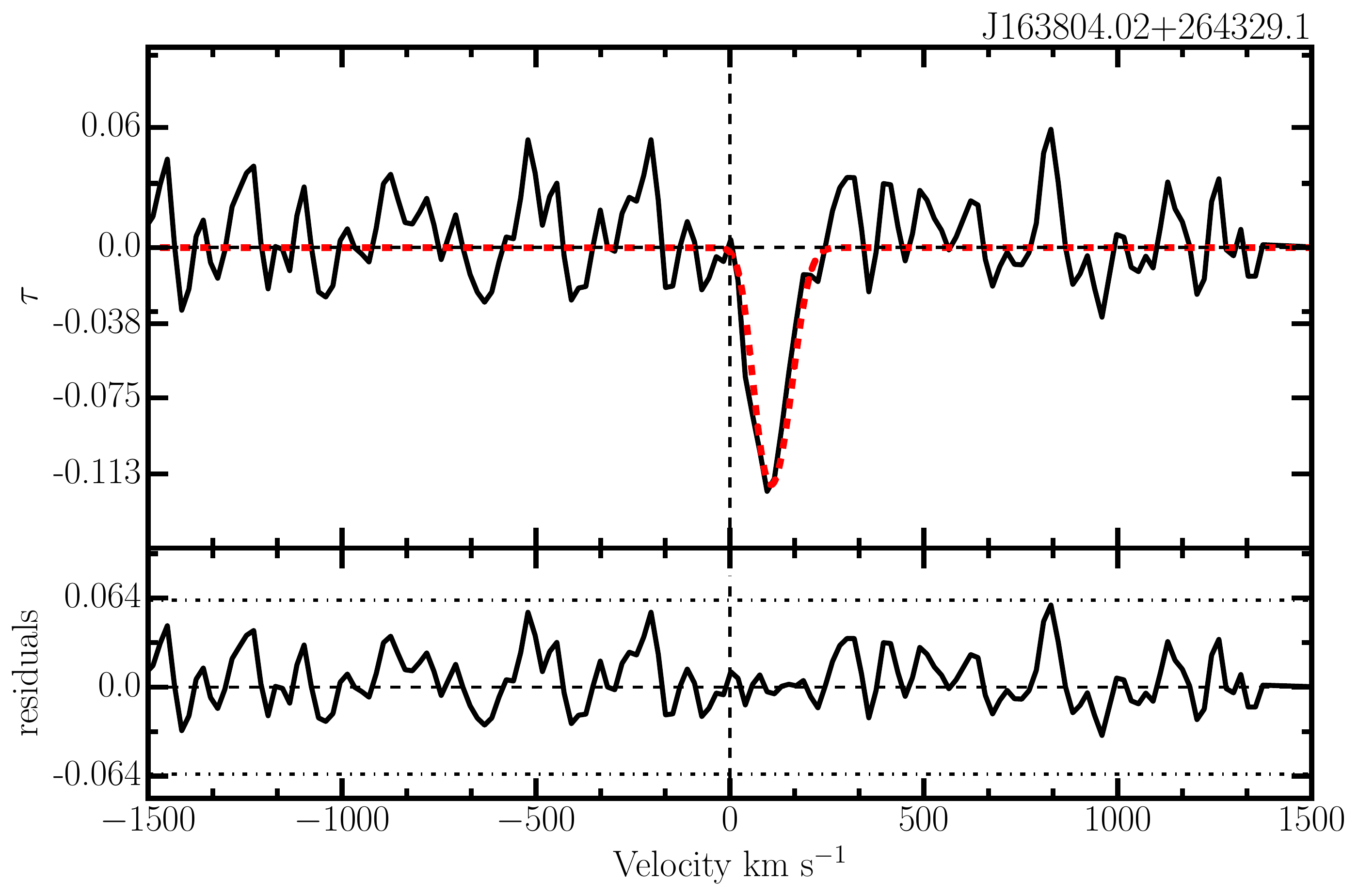}
                \includegraphics[trim = 0 0 0 0, clip,width=.33\textwidth]{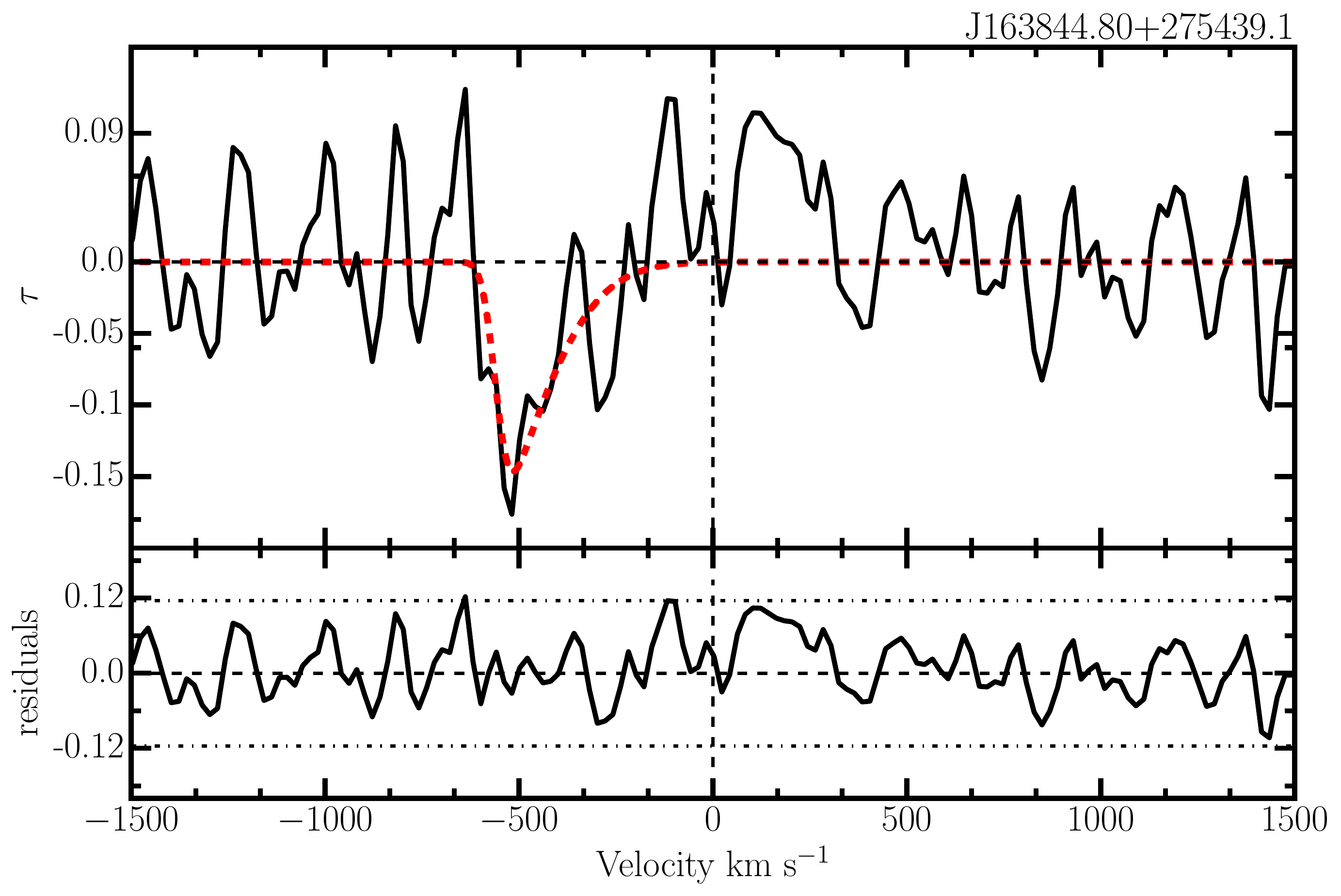}
                \includegraphics[trim = 0 0 0 0, clip,width=.33\textwidth]{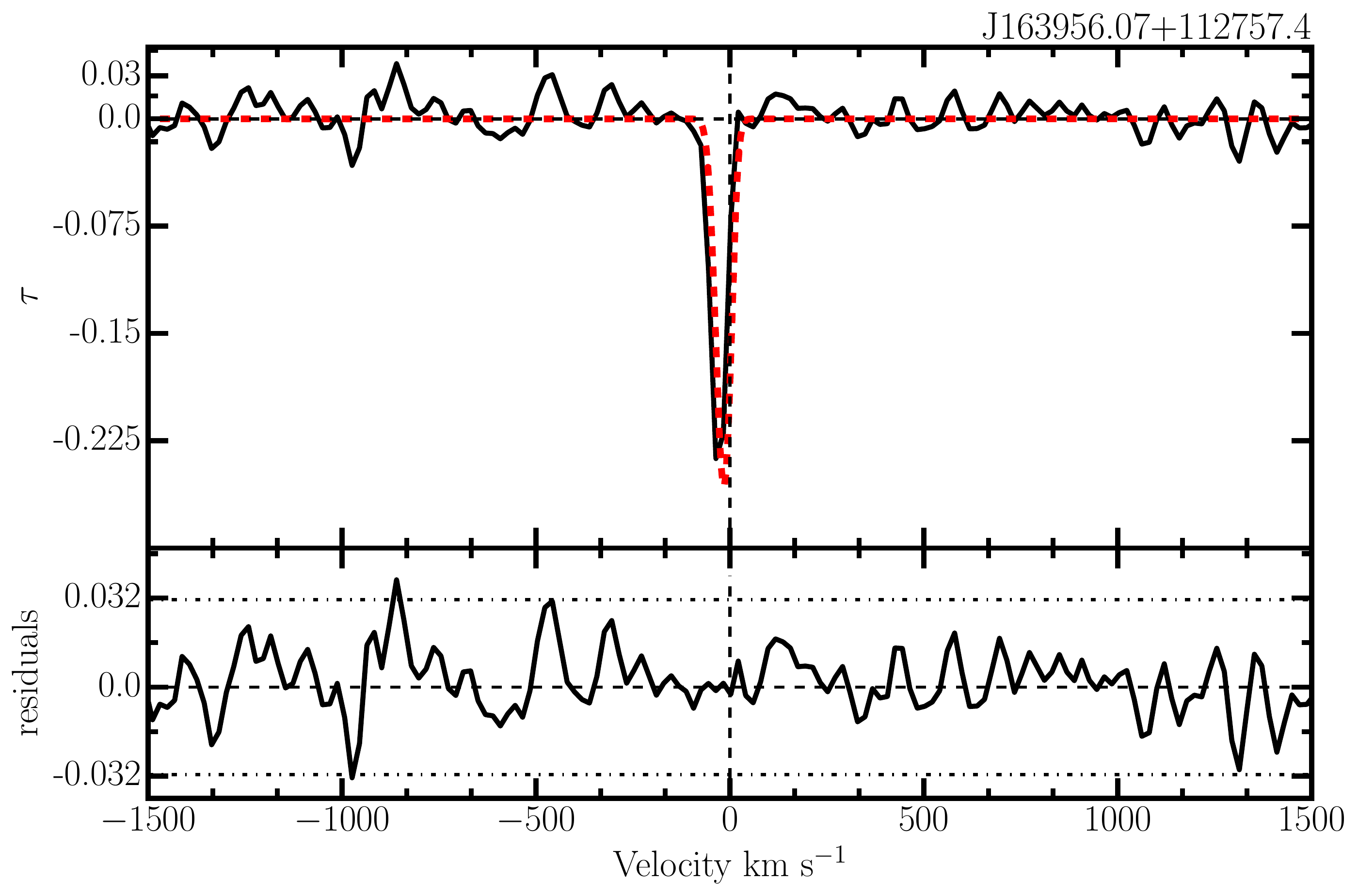}
                \includegraphics[trim = 0 0 0 0, clip,width=.33\textwidth]{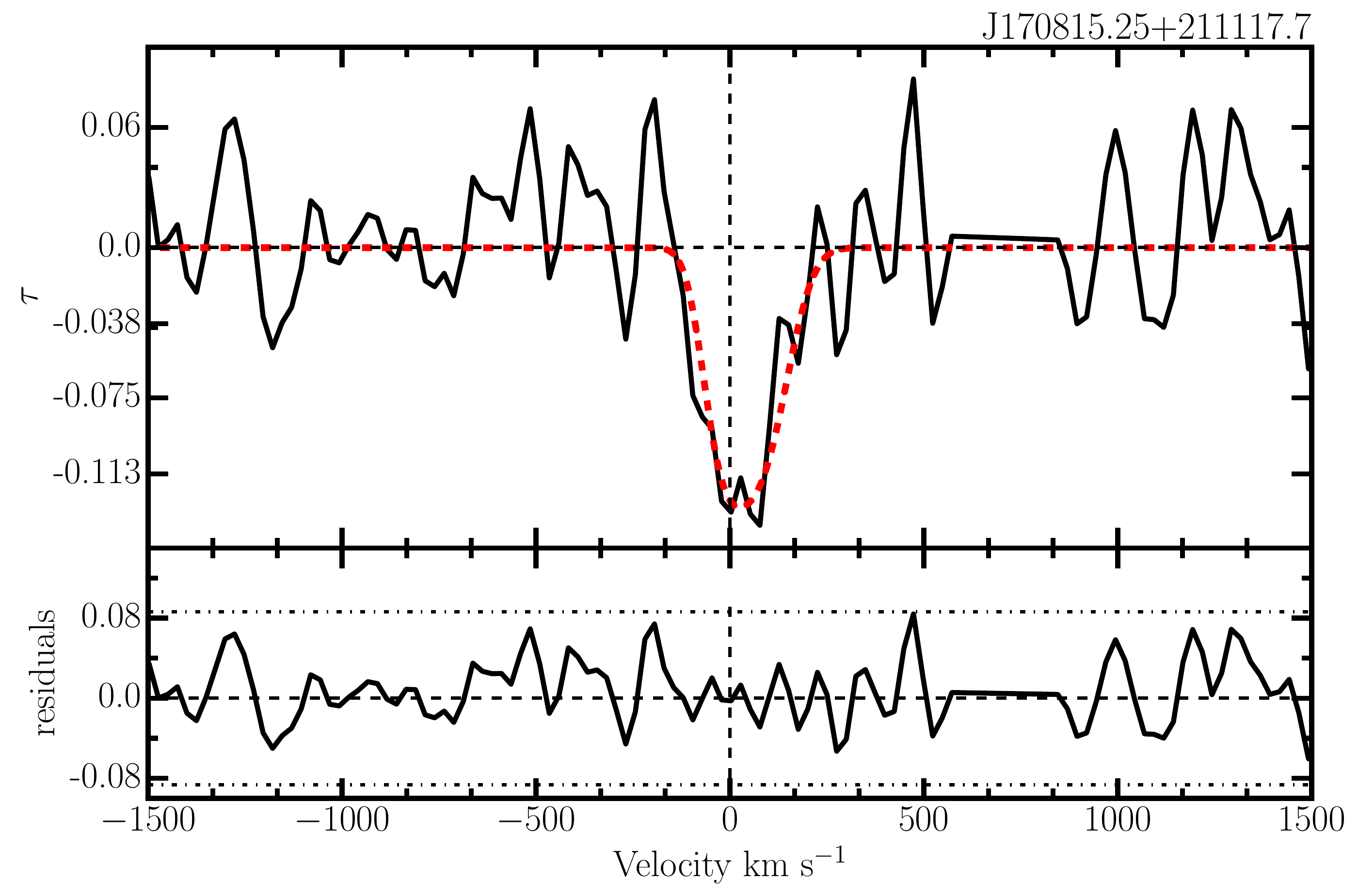}
                \caption{same as Fig.~\ref{fig:Profiles1}.}
\label{fig:Profiles3}
\end{center}
\end{figure*}

\clearpage

\longtab[1]{
\begin{landscape}
\begin{longtable}{l c c c c c c c c c c c}
\caption{\label{tab:long} Table of sources}\\
\hline\hline                                                         
Source  & z$_{\rm opt}$           & S$_{\rm{1.4 \ GHz}}$           &    P$_{\rm{1.4 \ GHz}}$     &    Radio type  & WISE type & Noise (3$\sigma$) &  $\tau_{\rm peak}$  & $\int \tau\cdot {\rm d}v$ & FWHM & FW20 & Centroid \\  
J2000   &  & [mJy] &[ \whz] & [C / E / I] & (6) & [mJy] & & [\kms] & [\kms] & [\kms] & [\kms]  \\
\hline
\endfirsthead
\caption{continued.}\\
\hline\hline
Source  & z$_{\rm opt}$           & S$_{\rm{1.4 \ GHz}}$           &    P$_{\rm{1.4 \ GHz}}$     &    Radio type  & WISE type & Noise (3$\sigma$) &  $\tau_{\rm peak}$  & $\int\tau\cdot {\rm d}v$ & FWHM & FW20 & Centroid \\  
J2000   &  & [mJy] &[ \whz] & [C / E / I] & (6) & [mJy] & & [\kms]& [\kms] & [\kms] & [\kms]  \\
\hline
\endhead
\hline
J075157.32+522309.2 & 0.1887 & 68.47 & 24.84 & C & $4.6\mu$m & <2.47 & - & - & - & - & - \\
J075244.19+455657.4 & 0.0517 & 55.54 & 23.55 & I & $12\mu$m & <1.85 & - & - & - & - & - \\
J075555.12+420756.7 & 0.15 & 57.56 & 24.54 & I & $12\mu$m & <5.58 & - & - & - & - & - \\
J075607.06+383401.0$^{\footnotesize (*)}$ & 0.2156 & 70.18 & 24.98 & C & dp & <2.82 & - & - & - & - & - \\
J075648.71+531256.2 & 0.0837 & 39.15 & 23.83 & C & dp & <1.78 & - & - & - & - & - \\
J075756.71+395936.1$^{\footnotesize (*)}$ & 0.0658 & 91.27 & 23.98 & C & $4.6\mu$m & 1.85 & 0.042 & 5.96 & 122.52$\pm$14.7 & 244.68$\pm$15.66 & -22.49$\pm$5.73 \\
J075828.10+374711.8$^{\footnotesize (*)}$ & 0.0408 & 243.0 & 23.98 & I & dp & <2.18 & - & - & - & - & - \\
J075846.99+270515.6$^{\footnotesize (*)}$ & 0.0987 & 69.05 & 24.23 & C & $4.6\mu$m & <3.76 & - & - & - & - & - \\
J075940.96+505023.9 & 0.0544 & 42.91 & 23.48 & C & $4.6\mu$m & <1.64 & - & - & - & - & - \\
J080041.98+321727.6$^{\footnotesize (*)}$ & 0.1872 & 104.0 & 25.01 & C & dp & <2.47 & - & - & - & - & - \\
J080601.51+190614.7$^{\footnotesize (*)}$ & 0.0979 & 142.0 & 24.54 & I & $12\mu$m & 3.08 & 0.099 & 18.17 & 189.89$\pm$9.44 & 266.22$\pm$9.54 & 104.05$\pm$4.53 \\
J080624.94+172503.7 & 0.1041 & 37.44 & 24.01 & I & dp & <2.65 & - & - & - & - & - \\
J080938.88+345537.2$^{\footnotesize (*)}$ & 0.0825 & 142.32 & 24.38 & I & $12\mu$m & 1.24 & 0.009 & 0.71 & 82.4$\pm$24.39 & 108.15$\pm$25.7 & -243.29$\pm$9.15 \\
J081040.29+481233.1 & 0.0775 & 73.18 & 24.03 & C & $4.6\mu$m & <5.88 & - & - & - & - & - \\
J081601.88+380415.4 & 0.1727 & 67.86 & 24.75 & I & dp & <2.99 & - & - & - & - & - \\
J081827.34+281402.8$^{\footnotesize (*)}$ & 0.2252 & 46.63 & 24.84 & C & $4.6\mu$m & <2.14 & - & - & - & - & - \\
J081854.09+224744.8$^{\footnotesize (*)}$ & 0.0958 & 194.0 & 24.65 & I & $12\mu$m & <2.12 & - & - & - & - & - \\
J082028.10+485347.4$^{\footnotesize (*)}$ & 0.1324 & 124.0 & 24.76 & I & dp & <3.18 & - & - & - & - & - \\
J082133.60+470237.3 & 0.128 & 1239.99 & 25.73 & I & $4.6\mu$m & 2.1 & 0.0038 & 0.56 & 41.47$\pm$3.33 & 47.24$\pm$4.35 & -254.33$\pm$12.0 \\
J082209.54+470552.9 & 0.1271 & 54.95 & 24.37 & C & dp & <2.94 & - & - & - & - & - \\
J082440.14+410305.6 & 0.058 & 75.93 & 23.79 & I & dp & <2.62 & - & - & - & - & - \\
J082814.20+415351.9 & 0.226 & 84.47 & 25.11 & I & dp & <2.57 & - & - & - & - & - \\
J082904.82+175415.8$^{\footnotesize (*)}$ & 0.0895 & 190.0 & 24.58 & I & $12\mu$m & <2.27 & - & - & - & - & - \\
J083138.83+223422.9$^{\footnotesize (*)}$ & 0.0869 & 92.56 & 24.24 & I & dp & <2.01 & - & - & - & - & - \\
J083139.79+460800.8$^{\footnotesize (*)}$ & 0.1311 & 123.0 & 24.75 & C & dp & <2.68 & - & - & - & - & - \\
J083411.09+580321.4$^{\footnotesize (*)}$ & 0.0934 & 46.31 & 24.01 & C & dp & <2.57 & - & - & - & - & - \\
J083548.14+151717.0 & 0.1684 & 45.48 & 24.55 & I & $4.6\mu$m & 4.5 & 0.1133 & 8.05 & 67.56$\pm$11.58 & 86.05$\pm$15.0 & -289.36$\pm$31.74 \\
J083637.84+440109.6$^{\footnotesize (*)}$ & 0.0554 & 134.0 & 23.99 & C & $12\mu$m & 1.93 & 0.016 & 1.37 & 77.92$\pm$17.38 & 206.4$\pm$26.8 & 25.67$\pm$13.27 \\
J083903.08+401545.6 & 0.1941 & 40.06 & 24.63 & I & dp & <1.96 & - & - & - & - & - \\
J083915.82+285038.7$^{\footnotesize (*)}$ & 0.079 & 126.0 & 24.29 & I & dp & <2.86 & - & - & - & - & - \\
J084307.11+453742.8$^{\footnotesize (*)}$ & 0.1919 & 331.0 & 25.54 & C & $12\mu$m & 5.8 & 0.273 & 25.86 & 79.16$\pm$37.17 & 126.75$\pm$19.98 & 57.85$\pm$59.03 \\
J084359.13+510524.9$^{\footnotesize (*)}$ & 0.1263 & 79.03 & 24.52 & I & dp & <3.17 & - & - & - & - & - \\
J084522.15+112550.4 & 0.0662 & 159.87 & 24.23 & I & dp & <3.72 & - & - & - & - & - \\
J084712.92+113350.1 & 0.1984 & 32.97 & 24.57 & C & $4.6\mu$m & <4.44 & - & - & - & - & - \\
J090100.09+103701.7 & 0.0295 & 45.96 & 22.96 & I & dp & <2.35 & - & - & - & - & - \\
J090105.25+290146.9$^{\footnotesize (*)}$ & 0.194 & 1670.0 & 26.25 & I & $12\mu$m & <1.67 & - & - & - & - & - \\
J090206.46+203037.6 & 0.0815 & 57.02 & 23.97 & I & dp & <1.87 & - & - & - & - & - \\
J090209.87+283042.9 & 0.0849 & 34.61 & 23.79 & C & dp & <1.82 & - & - & - & - & - \\
J090325.54+162256.0 & 0.1823 & 47.81 & 24.65 & C & $4.6\mu$m & 3.75 & 0.0949 & 8.98 & 163.02$\pm$43.22 & 187.67$\pm$41.56 & 2.97$\pm$36.58 \\
J090343.15+265022.5$^{\footnotesize (*)}$ & 0.0843 & 81.11 & 24.16 & C & dp & <2.0 & - & - & - & - & - \\
J090426.55+545805.6 & 0.1194 & 58.53 & 24.34 & I & dp & <5.85 & - & - & - & - & - \\
J090615.54+463619.0$^{\footnotesize (*)}$ & 0.0847 & 273.0 & 24.69 & C & $12\mu$m & <3.53 & - & - & - & - & - \\
J090652.79+412429.7$^{\footnotesize (*)}$ & 0.0274 & 56.34 & 22.98 & C & dp & <2.43 & - & - & - & - & - \\
J090734.91+325722.9 & 0.0491 & 44.53 & 23.4 & I & $4.6\mu$m & 2.8 & 0.4206 & 27.84 & 91.28$\pm$37.52 & 129.81$\pm$40.93 & 2.19$\pm$4.26 \\
J090937.44+192808.2$^{\footnotesize (*)}$ & 0.0278 & 63.33 & 23.05 & C & $12\mu$m & 2.05 & 0.119 & 14.82 & 118.93$\pm$5.68 & 181.98$\pm$6.25 & -38.79$\pm$2.52 \\
J091039.92+184147.7 & 0.0284 & 48.09 & 22.95 & I & dp & <3.0 & - & - & - & - & - \\
J091218.36+483045.1$^{\footnotesize (*)}$ & 0.1172 & 143.14 & 24.71 & I & dp & <3.95 & - & - & - & - & - \\
J091651.94+523828.3$^{\footnotesize (*)}$ & 0.1904 & 63.01 & 24.81 & I & dp & <3.31 & - & - & - & - & - \\
J092151.49+332406.7 & 0.0236 & 51.16 & 22.81 & I & dp & <1.83 & - & - & - & - & - \\
J092405.30+141021.4 & 0.1356 & 120.01 & 24.77 & C & dp & <4.41 & - & - & - & - & - \\
J092445.88+304933.0 & 0.2118 & 42.32 & 24.74 & I & dp & <4.71 & - & - & - & - & - \\
J092511.57+190713.1 & 0.1291 & 38.23 & 24.22 & C & $12\mu$m & <3.09 & - & - & - & - & - \\
J092740.64+554548.0 & 0.221 & 64.18 & 24.96 & I & $12\mu$m & <2.45 & - & - & - & - & - \\
J093004.05+341326.5 & 0.0421 & 30.43 & 23.1 & C & dp & <1.52 & - & - & - & - & - \\
J093414.30+241335.1 & 0.0504 & 30.55 & 23.26 & C & dp & <1.79 & - & - & - & - & - \\
J093551.59+612111.3$^{\footnotesize (*)}$ & 0.0394 & 148.0 & 23.73 & 2.0 & $4.6\mu$m & 1.82 & 0.073 & 33.08 & 536.43$\pm$9.97 & 825.84$\pm$10.68 & -78.08$\pm$4.49 \\
J093609.36+331308.3$^{\footnotesize (*)}$ & 0.0762 & 60.63 & 23.94 & C & dp & <1.79 & - & - & - & - & - \\
J094319.15+361452.1$^{\footnotesize (*)}$ & 0.0223 & 104.0 & 23.07 & C & dp & <1.86 & - & - & - & - & - \\
J094521.33+173753.2 & 0.1281 & 49.02 & 24.33 & I & $4.6\mu$m & <3.06 & - & - & - & - & - \\
J094542.23+575747.7$^{\footnotesize (*)}$ & 0.2289 & 89.3 & 25.14 & C & $4.6\mu$m & <2.64 & - & - & - & - & - \\
J100935.70+182601.5$^{\footnotesize (*)}$ & 0.1165 & 43.7 & 24.19 & I & dp & <3.68 & - & - & - & - & - \\
J101256.03+163853.0 & 0.118 & 37.56 & 24.13 & C & $4.6\mu$m & <8.88 & - & - & - & - & - \\
J101542.92+425803.6 & 0.1973 & 43.27 & 24.68 & C & $12\mu$m & <2.15 & - & - & - & - & - \\
J102053.67+483124.3$^{\footnotesize (*)}$ & 0.0532 & 81.85 & 23.74 & I & $12\mu$m & 2.54 & 0.05 & 6.51 & 124.8$\pm$3.62 & 179.13$\pm$11.06 & -67.23$\pm$4.92 \\
J102400.53+511248.1 & 0.2139 & 47.12 & 24.8 & C & dp & 2.02 & 0.0759 & 17.72 & 179.18$\pm$55.84 & 334.89$\pm$60.66 & -409.18$\pm$22.06 \\
J102544.22+102230.4 & 0.0457 & 92.83 & 23.66 & C & dp & 2.72 & 0.1222 & 9.62 & 66.95$\pm$4.9 & 81.26$\pm$6.18 & -15.43$\pm$6.68 \\
J102838.69+170211.2 & 0.1691 & 47.54 & 24.58 & I & dp & <3.66 & - & - & - & - & - \\
J103053.58+411316.0 & 0.0921 & 41.12 & 23.94 & C & dp & <1.85 & - & - & - & - & - \\
J103214.01+275601.6 & 0.0852 & 29.82 & 23.73 & I & $4.6\mu$m & <1.46 & - & - & - & - & - \\
J103653.01+444818.1 & 0.1274 & 32.52 & 24.14 & C & $12\mu$m & <2.71 & - & - & - & - & - \\
J103719.33+433515.3$^{\footnotesize (*)}$ & 0.0247 & 142.0 & 23.29 & C & $12\mu$m & <1.83 & - & - & - & - & - \\
J103932.12+461205.3 & 0.1861 & 30.81 & 24.48 & I & dp & 2.78 & 0.0859 & 14.75 & 86.1$\pm$9.25 & 127.26$\pm$12.78 & 3.23$\pm$6.0 \\
J104029.94+295757.7$^{\footnotesize (*)}$ & 0.0909 & 407.0 & 24.93 & C & $12\mu$m & <2.84 & - & - & - & - & - \\
J104609.61+165511.4$^{\footnotesize (*)}$ & 0.2069 & 63.21 & 24.89 & C & $4.6\mu$m & <2.42 & - & - & - & - & - \\
J104643.83+315301.1 & 0.1166 & 38.35 & 24.13 & C & dp & <3.72 & - & - & - & - & - \\
J104801.21+151438.4 & 0.2161 & 46.28 & 24.8 & I & $12\mu$m & <2.51 & - & - & - & - & - \\
J104931.69+232723.6 & 0.0631 & 81.23 & 23.89 & I & dp & <1.72 & - & - & - & - & - \\
J105327.25+205835.9$^{\footnotesize (*)}$ & 0.0526 & 79.07 & 23.72 & C & $12\mu$m & 1.96 & 0.023 & 2.59 & 156.05$\pm$15.46 & 189.77$\pm$20.9 & -58.39$\pm$6.85 \\
J105731.17+405646.1 & 0.0251 & 31.74 & 22.66 & C & dp & <2.66 & - & - & - & - & - \\
J110017.98+100256.8 & 0.036 & 125.77 & 23.58 & 2.0 & $4.6\mu$m & 2.87 & 0.2308 & 42.85 & 101.22$\pm$11.9 & 166.2$\pm$15.01 & -15.04$\pm$3.68 \\
J110305.78+191702.2$^{\footnotesize (*)}$ & 0.2143 & 98.95 & 25.12 & C & dp & <2.35 & - & - & - & - & - \\
J111113.19+284147.0 & 0.0287 & 36.42 & 22.84 & C & $12\mu$m & 2.47 & 0.0927 & 18.3 & 180.98$\pm$28.6 & 211.38$\pm$31.87 & 54.73$\pm$14.26 \\
J111622.70+291508.2$^{\footnotesize (*)}$ & 0.0453 & 72.88 & 23.54 & C & dp & <1.8 & - & - & - & - & - \\
J111834.85+614638.2 & 0.1924 & 38.52 & 24.61 & C & dp & <2.58 & - & - & - & - & - \\
J111836.00+313638.6 & 0.1185 & 52.89 & 24.29 & I & dp & <2.72 & - & - & - & - & - \\
J111916.54+623925.7 & 0.1102 & 32.28 & 24.0 & C & $12\mu$m & 3.63 & 0.2688 & 31.63 & 95.62$\pm$15.1 & 147.91$\pm$18.15 & -119.22$\pm$12.38 \\
J112030.04+273610.7$^{\footnotesize (*)}$ & 0.1125 & 177.0 & 24.76 & C & $4.6\mu$m & 4.39 & 0.147 & 10.93 & 62.48$\pm$1.25 & 95.5$\pm$3.63 & -107.64$\pm$2.03 \\
J112156.70+431456.9 & 0.1854 & 37.72 & 24.56 & C & $12\mu$m & <2.54 & - & - & - & - & - \\
J112332.04+235047.8 & 0.207 & 142.69 & 25.25 & I & dp & 3.75 & 0.1048 & 5.01 & 156.19$\pm$13.68 & 184.03$\pm$15.59 & 219.69$\pm$17.29 \\
J112349.91+201654.4$^{\footnotesize (*)}$ & 0.1304 & 102.0 & 24.66 & I & $12\mu$m & <3.51 & - & - & - & - & - \\
J113142.27+470008.6$^{\footnotesize (*)}$ & 0.1257 & 99.6 & 24.62 & I & dp & <2.65 & - & - & - & - & - \\
J113230.99+573109.3 & 0.1804 & 36.73 & 24.53 & C & dp & <2.68 & - & - & - & - & - \\
J113359.22+490343.4$^{\footnotesize (*)}$ & 0.0316 & 168.0 & 23.59 & I & dp & <1.84 & - & - & - & - & - \\
J113446.55+485721.9 & 0.0316 & 44.62 & 23.01 & I & dp & <2.25 & - & - & - & - & - \\
J113903.77+262142.2 & 0.0223 & 41.6 & 22.67 & C & $4.6\mu$m & <2.26 & - & - & - & - & - \\
J114505.01+193622.8$^{\footnotesize (*)}$ & 0.0216 & 777.0 & 23.91 & I & dp & <2.33 & - & - & - & - & - \\
J114520.25+642623.4 & 0.0616 & 73.94 & 23.83 & I & dp & <1.81 & - & - & - & - & - \\
J114722.13+350107.5$^{\footnotesize (*)}$ & 0.0629 & 276.0 & 24.42 & I & $12\mu$m & <1.65 & - & - & - & - & - \\
J115531.39+545200.4 & 0.0496 & 51.46 & 23.48 & C & dp & <1.78 & - & - & - & - & - \\
J115742.64+330810.4 & 0.0803 & 39.1 & 23.79 & C & $12\mu$m & <1.7 & - & - & - & - & - \\
J115954.66+302727.0 & 0.1064 & 23.44 & 23.83 & C & $12\mu$m & <2.54 & - & - & - & - & - \\
J120231.12+163741.8$^{\footnotesize (*)}$ & 0.1195 & 81.6 & 24.48 & C & $12\mu$m & 3.42 & 0.042 & 7.27 & 174.98$\pm$37.74 & 301.47$\pm$50.03 & -148.3$\pm$31.29 \\
J120255.33+261518.7 & 0.1936 & 36.39 & 24.59 & C & $4.6\mu$m & <1.52 & - & - & - & - & - \\
J120303.50+603119.1$^{\footnotesize (*)}$ & 0.0653 & 153.0 & 24.2 & C & $4.6\mu$m & <1.83 & - & - & - & - & - \\
J120320.81+131934.3 & 0.0584 & 107.17 & 23.94 & C & dp & <2.11 & - & - & - & - & - \\
J120551.46+203119.0$^{\footnotesize (*)}$ & 0.0238 & 80.0 & 23.01 & C & $12\mu$m & 2.03 & 0.034 & 3.23 & 90.26$\pm$8.7 & 122.3$\pm$9.81 & 17.71$\pm$3.45 \\
J120805.55+251414.2 & 0.0225 & 50.66 & 22.77 & C & dp & <1.84 & - & - & - & - & - \\
J120855.60+464113.8$^{\footnotesize (*)}$ & 0.101 & 68.38 & 24.25 & I & $12\mu$m & 2.29 & 0.052 & 2.48 & 32.09$\pm$0.81 & 68.99$\pm$1.49 & 33.84$\pm$1.1 \\
J121030.47+310518.6 & 0.0577 & 26.62 & 23.33 & C & dp & <1.12 & - & - & - & - & - \\
J121329.27+504429.3$^{\footnotesize (*)}$ & 0.0308 & 110.0 & 23.38 & C & dp & <2.07 & - & - & - & - & - \\
J121856.15+122643.0 & 0.0932 & 45.73 & 24.0 & C & dp & <1.58 & - & - & - & - & - \\
J122121.94+301037.2$^{\footnotesize (*)}$ & 0.1836 & 60.0 & 24.76 & C & $4.6\mu$m & <2.58 & - & - & - & - & - \\
J122513.09+321401.6 & 0.0592 & 50.84 & 23.63 & C & dp & 1.33 & 0.1493 & 10.49 & 54.51$\pm$11.08 & 125.9$\pm$13.92 & 168.95$\pm$10.0 \\
J122519.14+162104.6 & 0.197 & 28.33 & 24.5 & C & dp & <2.92 & - & - & - & - & - \\
J122622.51+640622.0 & 0.1102 & 47.1 & 24.17 & I & dp & <3.18 & - & - & - & - & - \\
J122823.09+162612.7$^{\footnotesize (*)}$ & 0.23 & 116.92 & 25.26 & I & $12\mu$m & <2.66 & - & - & - & - & - \\
J123011.85+470022.7$^{\footnotesize (*)}$ & 0.0391 & 93.76 & 23.52 & C & dp & <1.76 & - & - & - & - & - \\
J123200.55+331747.6$^{\footnotesize (*)}$ & 0.0788 & 93.83 & 24.16 & I & $4.6\mu$m & 1.68 & 0.034 & 3.76 & 145.75$\pm$7.25 & 174.74$\pm$9.48 & -49.83$\pm$4.28 \\
J123349.26+502622.7$^{\footnotesize (*)}$ & 0.2068 & 135.0 & 25.22 & I & $4.6\mu$m & <2.01 & - & - & - & - & - \\
J123905.13+174457.5 & 0.0654 & 66.4 & 23.84 & I & dp & 2.14 & 0.0477 & 2.72 & 78.49$\pm$14.24 & 103.59$\pm$15.73 & 40.05$\pm$7.72 \\
J124135.95+162033.6 & 0.0702 & 40.21 & 23.68 & I & dp & <2.51 & - & - & - & - & - \\
J124351.24+185025.9 & 0.228 & 30.19 & 24.67 & I & dp & <1.89 & - & - & - & - & - \\
J124428.54+331546.2 & 0.0843 & 71.48 & 24.1 & I & dp & <1.99 & - & - & - & - & - \\
J124707.32+490017.9$^{\footnotesize (*)}$ & 0.2069 & 1140.0 & 26.15 & C & $4.6\mu$m & 1.78 & 0.002 & 0.46 & 369.58$\pm$82.31 & 586.32$\pm$51.67 & -284.84$\pm$28.45 \\
J124709.68+324705.0 & 0.1351 & 32.21 & 24.19 & C & $4.6\mu$m & <2.48 & - & - & - & - & - \\
J125220.88+395100.9 & 0.2253 & 26.32 & 24.6 & C & $4.6\mu$m & <2.31 & - & - & - & - & - \\
J125236.90+285150.7$^{\footnotesize (*)}$ & 0.1951 & 430.0 & 25.67 & I & $12\mu$m & <2.14 & - & - & - & - & - \\
J125431.43+262040.6 & 0.0691 & 36.76 & 23.63 & I & dp & <2.2 & - & - & - & - & - \\
J125433.26+185602.2$^{\footnotesize (*)}$ & 0.1154 & 75.84 & 24.42 & C & $12\mu$m & 5.71 & 0.068 & 5.0 & 60.1$\pm$0.0 & 140.9$\pm$0.0 & 356.29$\pm$0.0 \\
J130125.26+291849.5 & 0.0234 & 35.64 & 22.65 & C & $4.6\mu$m & 1.52 & 0.0452 & 8.4 & 147.61$\pm$21.97 & 217.78$\pm$23.81 & 53.95$\pm$12.0 \\
J130132.61+463402.7$^{\footnotesize (*)}$ & 0.2055 & 97.0 & 25.07 & C & $12\mu$m & 2.01 & 0.018 & 4.24 & 172.41$\pm$41.9 & 584.1$\pm$85.29 & -308.84$\pm$40.27 \\
J130346.59+191617.4$^{\footnotesize (*)}$ & 0.0635 & 69.52 & 23.83 & I & dp & <2.59 & - & - & - & - & - \\
J130556.95+395621.5 & 0.1535 & 35.84 & 24.36 & C & dp & 4.08 & 0.0844 & 8.2 & 126.22$\pm$23.11 & 141.49$\pm$26.79 & -18.96$\pm$16.0 \\
J130619.24+111339.7 & 0.0857 & 118.97 & 24.34 & I & dp & <2.91 & - & - & - & - & - \\
J130621.72+434751.2$^{\footnotesize (*)}$ & 0.2026 & 136.54 & 25.21 & I & dp & <2.84 & - & - & - & - & - \\
J130837.91+434415.1$^{\footnotesize (*)}$ & 0.0358 & 60.81 & 23.26 & C & dp & <1.86 & - & - & - & - & - \\
J131424.68+621945.8$^{\footnotesize (*)}$ & 0.1308 & 72.0 & 24.51 & I & dp & <2.96 & - & - & - & - & - \\
J131535.10+620728.4 & 0.0308 & 44.72 & 22.99 & 2.0 & $4.6\mu$m & 1.83 & 0.0575 & 10.52 & 283.09$\pm$29.75 & 324.38$\pm$30.35 & 142.39$\pm$15.29 \\
J131739.20+411545.6$^{\footnotesize (*)}$ & 0.0661 & 246.0 & 24.41 & C & dp & 1.69 & 0.031 & 4.79 & 133.97$\pm$32.35 & 245.21$\pm$75.52 & 59.19$\pm$147.78 \\
J131941.39+162852.5 & 0.1587 & 12.18 & 23.92 & C & dp & <3.63 & - & - & - & - & - \\
J132035.40+340821.7$^{\footnotesize (*)}$ & 0.0231 & 96.6 & 23.07 & 2.0 & $4.6\mu$m & 1.8 & 0.162 & 46.64 & 272.0$\pm$4.29 & 416.47$\pm$4.67 & 26.57$\pm$1.94 \\
J132513.37+395553.2$^{\footnotesize (*)}$ & 0.0756 & 36.81 & 23.71 & C & dp & 1.75 & 0.053 & 9.46 & 138.16$\pm$52.7 & 210.53$\pm$80.31 & -133.97$\pm$20.58 \\
J132524.03+492022.7 & 0.1867 & 35.89 & 24.55 & C & $12\mu$m & <3.72 & - & - & - & - & - \\
J133455.94+134431.7 & 0.0231 & 26.31 & 22.5 & C & dp & 4.14 & 0.1544 & 15.68 & 130.14$\pm$8.44 & 150.73$\pm$18.68 & -101.17$\pm$14.06 \\
J133817.24+481629.7 & 0.0276 & 79.28 & 23.14 & 2.0 & $4.6\mu$m & 2.51 & 0.1014 & 7.26 & 139.61$\pm$15.7 & 234.59$\pm$23.93 & 176.44$\pm$16.81 \\J134035.20+444817.3$^{\footnotesize (*)}$ & 0.0654 & 36.22 & 23.57 & 2.0 & $4.6\mu$m & 1.69 & 0.26 & 12.3 & 43.43$\pm$8.54 & 62.65$\pm$2.45 & -23.25$\pm$1.44 \\
J134105.10+395945.4 & 0.1715 & 59.29 & 24.68 & C & $12\mu$m & <3.9 & - & - & - & - & - \\
J134111.14+302241.3 & 0.0404 & 38.72 & 23.17 & C & $4.6\mu$m & 1.98 & 0.0298 & 7.78 & 208.44$\pm$25.28 & 242.09$\pm$29.78 & -2.42$\pm$27.96 \\
J134442.16+555313.5$^{\footnotesize (*)}$ & 0.0373 & 132.0 & 23.63 & 2.0 & $4.6\mu$m & 1.79 & 0.091 & 45.28 & 570.07$\pm$2.12 & 638.38$\pm$2.11 & 84.89$\pm$0.08 \\
J134620.46+130501.6 & 0.0811 & 46.57 & 23.88 & C & $12\mu$m & <2.11 & - & - & - & - & - \\
J134649.45+142401.7 & 0.0216 & 162.0 & 23.23 & C & $4.6\mu$m & 2.4 & 0.0367 & 4.15 & 248.73$\pm$8.93 & 282.45$\pm$11.18 & 36.82$\pm$10.0 \\
J134808.76+304908.9 & 0.1692 & 49.82 & 24.6 & C & dp & <3.39 & - & - & - & - & - \\
J134840.10+181716.1 & 0.0731 & 36.63 & 23.68 & C & $4.6\mu$m & <2.18 & - & - & - & - & - \\
J135217.88+312646.4$^{\footnotesize (*)}$ & 0.0452 & 3530.0 & 25.23 & I & $4.6\mu$m & 2.48 & 0.057 & 45.28 & 99.94$\pm$1.01 & 230.82$\pm$0.41 & -141.6$\pm$0.18 \\
J135314.08+374113.9 & 0.2159 & 32.59 & 24.65 & C & dp & <3.15 & - & - & - & - & - \\
J135646.10+102609.0 & 0.1231 & 61.02 & 24.38 & C & $4.6\mu$m & 5.1 & 0.0614 & 10.41 & 128.68$\pm$21.1 & 147.99$\pm$25.04 & 169.32$\pm$15.0 \\
J135806.05+214021.1 & 0.0664 & 61.05 & 23.81 & I & dp & 2.09 & 0.1365 & 8.17 & 49.71$\pm$7.51 & 72.34$\pm$11.28 & 2.43$\pm$3.0 \\
J135908.74+280121.3 & 0.0645 & 34.06 & 23.53 & C & dp & <2.33 & - & - & - & - & - \\
J135942.61+124412.5 & 0.0392 & 44.4 & 23.2 & I & dp & <1.79 & - & - & - & - & - \\
J140026.40+175133.3$^{\footnotesize (*)}$ & 0.0506 & 87.7 & 23.72 & C & $12\mu$m & <1.91 & - & - & - & - & - \\
J140051.58+521606.5$^{\footnotesize (*)}$ & 0.1179 & 176.0 & 24.8 & C & $12\mu$m & <3.66 & - & - & - & - & - \\
J140810.47+524048.1$^{\footnotesize (*)}$ & 0.0829 & 176.0 & 24.48 & I & dp & <3.14 & - & - & - & - & - \\
J140935.47+575841.2$^{\footnotesize (*)}$ & 0.1799 & 114.0 & 25.01 & C & $12\mu$m & <3.92 & - & - & - & - & - \\
J141134.14+294914.1 & 0.1861 & 12.36 & 24.08 & C & dp & <2.01 & - & - & - & - & - \\
J141149.43+524900.1$^{\footnotesize (*)}$ & 0.0765 & 393.14 & 24.75 & I & dp & <1.96 & - & - & - & - & - \\
J141203.47+292801.7 & 0.1147 & 31.47 & 24.03 & I & $12\mu$m & <4.14 & - & - & - & - & - \\
J141557.25+495334.6 & 0.1854 & 37.87 & 24.57 & I & $4.6\mu$m & <2.15 & - & - & - & - & - \\
J141652.95+104826.7 & 0.0247 & 31.36 & 22.64 & I & dp & <2.65 & - & - & - & - & - \\
J142210.81+210554.1$^{\footnotesize (*)}$ & 0.1915 & 84.3 & 24.94 & C & dp & 2.22 & 0.048 & 8.36 & 179.82$\pm$14.28 & 274.93$\pm$22.61 & -196.05$\pm$6.23 \\
J142810.35+123711.7 & 0.0792 & 38.05 & 23.77 & I & dp & <1.68 & - & - & - & - & - \\
J142832.60+424021.0 & 0.1293 & 37.09 & 24.21 & I & $4.6\mu$m & <2.23 & - & - & - & - & - \\
J143418.19+242444.2 & 0.085 & 47.45 & 23.93 & C & dp & <1.62 & - & - & - & - & - \\
J143521.67+505122.9$^{\footnotesize (*)}$ & 0.0997 & 141.0 & 24.55 & C & $12\mu$m & 2.04 & 0.013 & 3.55 & 285.61$\pm$56.38 & 422.04$\pm$52.61 & -66.71$\pm$24.9 \\
J144104.37+532008.7 & 0.105 & 40.65 & 24.06 & 2.0 & $4.6\mu$m & <2.44 & - & - & - & - & - \\
J144433.70+192121.5 & 0.1905 & 121.02 & 25.1 & I & dp & <1.91 & - & - & - & - & - \\
J144557.78+173828.6 & 0.0653 & 270.89 & 24.45 & I & $12\mu$m & <3.21 & - & - & - & - & - \\
J144712.76+404744.9$^{\footnotesize (*)}$ & 0.1951 & 348.0 & 25.58 & I & $4.6\mu$m & <2.08 & - & - & - & - & - \\
J144921.58+631614.0$^{\footnotesize (*)}$ & 0.0417 & 2500.0 & 25.01 & I & $12\mu$m & 2.64 & 0.005 & 1.21 & 162.42$\pm$9.77 & 460.94$\pm$90.81 & 20.19$\pm$14.2 \\
J145049.40+100649.1 & 0.0545 & 48.22 & 23.53 & C & dp & <1.96 & - & - & - & - & - \\
J150034.56+364845.1$^{\footnotesize (*)}$ & 0.0661 & 60.66 & 23.81 & C & $12\mu$m & 2.11 & 0.19 & 20.3 & 100.97$\pm$15.38 & 161.26$\pm$54.0 & 17.96$\pm$24.06 \\
J150151.12+163705.9 & 0.1489 & 22.61 & 24.13 & I & $12\mu$m & <3.06 & - & - & - & - & - \\
J150457.12+260058.4$^{\footnotesize (*)}$ & 0.054 & 190.0 & 24.12 & C & dp & <2.27 & - & - & - & - & - \\
J150656.41+125048.6 & 0.0223 & 66.27 & 22.87 & I & $12\mu$m & <3.66 & - & - & - & - & - \\
J150721.87+101844.8 & 0.078 & 403.28 & 24.78 & C & $12\mu$m & 2.93 & 0.009 & 1.29 & 74.08$\pm$7.24 & 241.14$\pm$8.78 & -16.73$\pm$18.0 \\
J150950.99+155730.3$^{\footnotesize (*)}$ & 0.1874 & 424.0 & 25.62 & C & dp & <2.11 & - & - & - & - & - \\
J151319.23+343133.7 & 0.1272 & 34.59 & 24.17 & C & $12\mu$m & 3.12 & 0.2491 & 37.97 & 134.03$\pm$22.93 & 180.45$\pm$25.97 & 23.93$\pm$9.76 \\
J151641.59+291809.2$^{\footnotesize (*)}$ & 0.1299 & 72.07 & 24.51 & I & $12\mu$m & <4.13 & - & - & - & - & - \\
J151838.90+404500.2 & 0.0652 & 54.42 & 23.75 & C & dp & <1.41 & - & - & - & - & - \\
J152045.04+483922.9 & 0.078 & 47.01 & 23.85 & I & dp & <1.54 & - & - & - & - & - \\
J152115.79+151207.8$^{\footnotesize (*)}$ & 0.2148 & 405.0 & 25.74 & C & dp & <2.83 & - & - & - & - & - \\
J152326.91+283732.5 & 0.0824 & 89.41 & 24.18 & I & dp & <2.12 & - & - & - & - & - \\
J152349.34+321350.2$^{\footnotesize (*)}$ & 0.11 & 182.0 & 24.75 & C & $12\mu$m & <2.89 & - & - & - & - & - \\
J152446.01+230723.5 & 0.2157 & 40.76 & 24.74 & C & $12\mu$m & 1.95 & 0.0375 & 9.47 & 290.22$\pm$26.32 & 376.95$\pm$29.55 & 125.05$\pm$16.0 \\
J152500.83+332359.8$^{\footnotesize (*)}$ & 0.0816 & 58.73 & 23.98 & I & dp & <1.85 & - & - & - & - & - \\
J152650.94+101320.8 & 0.2243 & 45.15 & 24.83 & I & dp & <2.69 & - & - & - & - & - \\
J152922.49+362142.2$^{\footnotesize (*)}$ & 0.0989 & 37.8 & 23.97 & C & $12\mu$m & 2.36 & 0.075 & 18.35 & 231.54$\pm$31.63 & 360.19$\pm$38.04 & -29.27$\pm$17.78 \\
J153058.19+573625.2 & 0.1733 & 37.76 & 24.5 & I & $4.6\mu$m & <5.34 & - & - & - & - & - \\
J153202.23+301628.9$^{\footnotesize (*)}$ & 0.0653 & 32.94 & 23.53 & C & $12\mu$m & <2.04 & - & - & - & - & - \\
J153437.61+251311.4 & 0.0339 & 43.39 & 23.06 & C & $12\mu$m & 2.65 & 0.1317 & 10.92 & 40.88$\pm$6.26 & 141.08$\pm$8.11 & -85.83$\pm$7.0 \\
J153452.95+290919.8 & 0.201 & 48.57 & 24.75 & C & $12\mu$m & 1.72 & 0.0275 & 6.77 & 233.7$\pm$31.87 & 320.14$\pm$35.97 & 26.61$\pm$27.04 \\
J153535.08+134752.7 & 0.0524 & 25.47 & 23.22 & C & $12\mu$m & <2.57 & - & - & - & - & - \\
J153901.66+353046.0$^{\footnotesize (*)}$ & 0.0778 & 90.91 & 24.13 & C & $12\mu$m & <1.71 & - & - & - & - & - \\
J153935.60+553015.9 & 0.0517 & 24.18 & 23.18 & C & dp & <1.75 & - & - & - & - & - \\
J154144.30+472754.8 & 0.1104 & 82.92 & 24.41 & I & dp & <2.84 & - & - & - & - & - \\
J154146.55+455614.3 & 0.2023 & 44.06 & 24.72 & C & $12\mu$m & <2.75 & - & - & - & - & - \\
J154818.27+573549.3 & 0.0742 & 32.15 & 23.64 & C & $12\mu$m & <1.66 & - & - & - & - & - \\
J154912.33+304716.4$^{\footnotesize (*)}$ & 0.1116 & 914.0 & 25.47 & I & $12\mu$m & <3.65 & - & - & - & - & - \\
J155343.59+234825.4$^{\footnotesize (*)}$ & 0.1176 & 168.04 & 24.78 & I & $4.6\mu$m & <5.62 & - & - & - & - & - \\
J155424.12+201125.4 & 0.2223 & 53.52 & 24.89 & I & $12\mu$m & <2.24 & - & - & - & - & - \\
J155603.91+242652.8$^{\footnotesize (*)}$ & 0.0425 & 92.13 & 23.59 & I & dp & <1.91 & - & - & - & - & - \\
J155611.61+281133.3$^{\footnotesize (*)}$ & 0.2079 & 84.14 & 25.02 & I & dp & <2.0 & - & - & - & - & - \\
J155645.91+334248.9 & 0.2069 & 75.22 & 24.97 & I & dp & <2.89 & - & - & - & - & - \\
J155902.70+230830.4 & 0.1932 & 43.31 & 24.66 & C & $4.6\mu$m & 2.2 & 0.0449 & 17.72 & 441.28$\pm$22.82 & 520.07$\pm$25.77 & 96.76$\pm$18.4 \\
J155953.90+423339.9 & 0.2169 & 44.2 & 24.78 & C & dp & <1.72 & - & - & - & - & - \\
J155953.98+444232.4$^{\footnotesize (*)}$ & 0.0417 & 57.82 & 23.37 & C & dp & <2.04 & - & - & - & - & - \\
J160246.39+524358.3$^{\footnotesize (*)}$ & 0.1057 & 577.0 & 25.22 & C & $12\mu$m & 2.0 & 0.015 & 4.58 & 463.6$\pm$29.76 & 673.64$\pm$21.6 & -255.51$\pm$11.76 \\
J160332.08+171155.3$^{\footnotesize (*)}$ & 0.034 & 278.0 & 23.87 & I & dp & 2.26 & 0.066 & 3.5 & 46.98$\pm$2.59 & 71.7$\pm$2.85 & -15.75$\pm$1.25 \\
J160338.06+155402.5$^{\footnotesize (*)}$ & 0.1098 & 100.0 & 24.49 & C & $12\mu$m & 3.04 & 0.125 & 5.11 & 357.8$\pm$33.71 & 499.9$\pm$7.26 & -35.77$\pm$3.28 \\
J160426.51+174431.1$^{\footnotesize (*)}$ & 0.0409 & 72.96 & 23.45 & 2.0 & $12\mu$m & <2.86 & - & - & - & - & - \\
J160616.03+181459.8$^{\footnotesize (*)}$ & 0.0369 & 225.0 & 23.85 & I & dp & <2.24 & - & - & - & - & - \\
J160821.14+282843.2$^{\footnotesize (*)}$ & 0.0502 & 113.0 & 23.83 & I & dp & <1.79 & - & - & - & - & - \\
J160907.18+131908.2 & 0.1836 & 75.75 & 24.86 & C & dp & <2.66 & - & - & - & - & - \\
J160952.60+133148.0 & 0.0357 & 33.84 & 23.0 & C & $4.6\mu$m & 4.08 & 0.1596 & 22.33 & 281.43$\pm$26.2 & 308.18$\pm$28.99 & 37.59$\pm$30.15 \\
J161114.11+265524.2 & 0.0319 & 69.49 & 23.21 & I & dp & <2.66 & - & - & - & - & - \\
J161217.62+282546.4$^{\footnotesize (*)}$ & 0.0531 & 78.47 & 23.72 & C & dp & 2.58 & 0.061 & 5.0 & 77.16$\pm$11.19 & 114.96$\pm$12.2 & -20.64$\pm$4.55 \\
J161419.62+502756.2$^{\footnotesize (*)}$ & 0.0603 & 80.56 & 23.85 & I & dp & <1.99 & - & - & - & - & - \\
J161541.21+471111.7$^{\footnotesize (*)}$ & 0.1986 & 134.55 & 25.18 & I & $4.6\mu$m & <3.06 & - & - & - & - & - \\
J161740.53+350015.1 & 0.0298 & 141.18 & 23.46 & I & dp & 2.5 & 0.0376 & 1.55 & 113.76$\pm$16.41 & 134.27$\pm$17.37 & -189.92$\pm$9.48 \\
J162318.73+370547.5 & 0.2029 & 46.92 & 24.75 & I & dp & <14.97 & - & - & - & - & - \\
J162424.49+483142.3$^{\footnotesize (*)}$ & 0.0571 & 67.42 & 23.72 & C & dp & <1.99 & - & - & - & - & - \\
J162839.03+252755.9 & 0.2199 & 63.49 & 24.96 & I & $12\mu$m & <2.39 & - & - & - & - & - \\
J163124.69+250309.8 & 0.0623 & 39.31 & 23.56 & C & dp & <2.27 & - & - & - & - & - \\
J163804.02+264329.1 & 0.0652 & 40.64 & 23.62 & C & dp & 2.52 & 0.1196 & 13.6 & 110.78$\pm$13.82 & 159.24$\pm$15.02 & 95.88$\pm$6.74 \\
J163844.80+275439.1 & 0.1035 & 32.32 & 23.95 & I & dp & 3.54 & 0.1166 & 27.45 & 164.05$\pm$15.64 & 280.29$\pm$17.71 & -478.5$\pm$14.0 \\
J163956.07+112757.4 & 0.0792 & 159.39 & 24.39 & C & $4.6\mu$m & 4.98 & 0.1264 & 12.52 & 46.59$\pm$6.99 & 68.09$\pm$7.58 & -36.04$\pm$1.09 \\
J164331.91+304835.5 & 0.184 & 46.69 & 24.65 & C & $4.6\mu$m & <3.9 & - & - & - & - & - \\
J164332.24+254206.7 & 0.0571 & 54.43 & 23.63 & I & dp & <2.69 & - & - & - & - & - \\
J164419.97+454644.4 & 0.2246 & 89.39 & 25.12 & C & $12\mu$m & <1.82 & - & - & - & - & - \\
J164516.33+132130.2 & 0.1934 & 78.15 & 24.92 & I & $12\mu$m & <3.57 & - & - & - & - & - \\
J165240.83+231847.3 & 0.1623 & 53.8 & 24.59 & I & dp & <3.15 & - & - & - & - & - \\
J170528.99+221604.8 & 0.0496 & 62.65 & 23.56 & I & dp & <1.24 & - & - & - & - & - \\
J170727.45+260957.9 & 0.1125 & 42.74 & 24.14 & C & dp & <1.86 & - & - & - & - & - \\
J170735.95+353949.6 & 0.164 & 103.91 & 24.89 & I & dp & <2.86 & - & - & - & - & - \\
J170815.25+211117.7 & 0.2241 & 34.36 & 24.71 & C & $12\mu$m & 2.92 & 0.1569 & 28.38 & 211.14$\pm$17.49 & 284.89$\pm$20.99 & 3.23$\pm$13.0 \\
J171056.30+394131.2 & 0.0622 & 77.57 & 23.86 & I & dp & <1.8 & - & - & - & - & - \\
J171522.97+572440.3 & 0.0273 & 45.84 & 22.89 & C & dp & <2.1 & - & - & - & - & - \\
J171523.73+302824.1 & 0.1111 & 30.66 & 23.99 & C & dp & <2.56 & - & - & - & - & - \\
J172223.65+320128.2 & 0.227 & 68.38 & 25.02 & I & dp & <3.24 & - & - & - & - & - \\
\end{longtable}
\tablefoot{(1): J2000 coordinates of the sources; (2) redshift measured from the SDSS spectrum; (3) flux at 1.4 GHz; (4) radio power; (5) radio classification as compact (C), extended (E) and interacting (I) sources; (6) WISE colour-colour classification as dust poor (dp), $12$~\um\ bright ($12\mu$m) and $4.6$~\um\ bright ($4.6\mu$m) galaxies (see Sect.~\ref{sec:sample}); (7) $3$--$\sigma$ detection limit of the \HI\ spectrum; (8) peak optical depth of the \HI\ detections; (9) full width at half maximum; (10) full width at $20\%$ of the peak flux; (11) position of the centroid of the line. Galaxies denoted by (*) have been presented in \ps\ and \pz.} \end{landscape} }
\end{appendix}

\end{onecolumn}

\end{document}